\documentclass[twoside,11pt,hyphens]{article}

\usepackage[]{melba}
\usepackage{hhline}
\usepackage{placeins}
\usepackage{pdflscape}
\usepackage{rotating}
\usepackage{hyperref}
\usepackage{url}

\usepackage{amsmath,amsfonts}

\usepackage{amsmath,amsfonts}

\usepackage{mathrsfs}
\usepackage{colortbl}
\usepackage{tabularx}
\usepackage{stfloats}
\usepackage{multirow}

\melbaheading{007}{https://www.melba-journal.org/article/22514}{2021}{1-32}{09/2020}{04/2021}{Caliv\'a, Cheng, Shah and Pedoia}{Medical Imaging with Deep Learning (MIDL) 2020}{Marleen de Bruijne, Tal Arbel, Ismail Ben Ayed, Hervé Lombaert}

\ShortHeadings{Adversarial Robust Training in MRI Reconstruction}{Caliv\'a, Cheng, Shah and Pedoia}
\firstpageno{1}

\title{Adversarial Robust Training of Deep Learning MRI Reconstruction Models}

\author{\name Francesco Caliv\'a$^{1}$\thanks{Contributed equally} \email francesco.caliva@ucsf.edu 
\AND
\name Kaiyang Cheng$^{1,2 *}$ \email victorcheng21@berkeley.edu
\AND
\name Rutwik Shah$^{1}$ \email rutwik.shah@ucsf.edu 
\AND
\name Valentina Pedoia$^{1}$ \email valentina.pedoia@ucsf.edu
\\
\addr $^{1}$ Center for Intelligent Imaging (CI$^{2}$), Department of Radiology and Biomedical Imaging, University of California, San Francisco\\
\addr $^{2}$ Department of Electrical Engineering and Computer Sciences, University of California, Berkeley
}

\begin{document}

\maketitle

\begin{abstract}
Deep Learning (DL) has shown potential in accelerating Magnetic Resonance Image acquisition and reconstruction. Nevertheless, there is a dearth of tailored methods to guarantee that the reconstruction of small features is achieved with high fidelity. In this work, we employ adversarial attacks to generate small synthetic perturbations, which are difficult to reconstruct for a trained DL reconstruction network. Then, we use robust training to increase the network's sensitivity to these small features and encourage their reconstruction. Next, we investigate the generalization of said approach to real world features. For this, a musculoskeletal radiologist annotated a set of cartilage and meniscal lesions from the knee Fast-MRI dataset, and a classification network was devised to assess the reconstruction of the features. Experimental results show that by introducing robust training to a reconstruction network, the rate of false negative features (4.8\%) in image reconstruction can be reduced. These results are encouraging, and highlight the necessity for attention to this problem by the image reconstruction community, as a milestone for the introduction of DL reconstruction in clinical practice. To support further research, we make our annotations and code publicly available at~\url{https://github.com/fcaliva/fastMRI\_BB\_abnormalities\_annotation}.
\end{abstract}


Magnetic Resonance Imaging (MRI) is a widely used screening modality. However, long scanning time, tedious post processing operations and lack of standardized acquisition protocols make automated and faster MRI desirable. Deep learning research in the domain of accelerated MRI reconstruction is an active area of research. In a recent retrospective study, \cite{recht2020using} suggests that clinical and (up to $4\times$) DL-accelerated images can be interchangeably utilized. However, results of that study demonstrate that clinically relevant features can lead to discordant clinical opinions. Focusing on imaging the knee joint, small structures- which are clinically relevant not only for grading the severity of lesions, but also for deciding further course of treatment \citep{mcalindon2014oarsi}- are often difficult to reconstruct, and their reconstruction quality is generally overlooked by standard image fidelity metrics. 
This adds to the concerning results presented at the Fast-MRI challenge, which was held during NeurIPS 2019 (\url{https://slideslive.com/38922093/medical-imaging-meets-neurips-4}): top-performing deep learning models failed in reconstructing some relatively small abnormalities such as meniscal tear and subchondral osteophytes. A similar outcome was observed during the second Fast-MRI challenge in 2020: multiple referee radiologists raised that none of the submitted 8$\times$ accelerated reconstructed images would be clinically acceptable \citep{muckley2020state}. We argue that in the current state, it must be acknowledged that irrespective of a remarkable improvement in image quality of accelerated MRI, the false negative reconstruction phenomenon is still present. Such phenomenon closely relates to the instability of deep learning models for image reconstruction, which was discussed in \cite{antun2020instabilities}. Instabilities can manifest in the form of severe reconstruction artefacts caused by imperceptible perturbations in the sampling and/or image domains. Otherwise, instabilities can lead to what \cite{cheng2020addressing} defined as false negative features: small perceptible features, which in spite of their presence in the sampled data, have disappeared upon image reconstruction. Instances of this phenomenon can be tumors or lesions that disappeared in the reconstructed images. Arguably, this could be a direct consequence of the limited training data often available in medical imaging, which might not accurately represent certain pathological phenotypes. This can cause a dangerous outcome as shown in \cite{cohen2018distribution}, where reconstructed images were hallucinated, in the sense that important clinical features were not reconstructed simply because they were not available in the training distribution.
\begin{figure}[!tb]
 \centering
 \includegraphics[width=0.99\columnwidth]{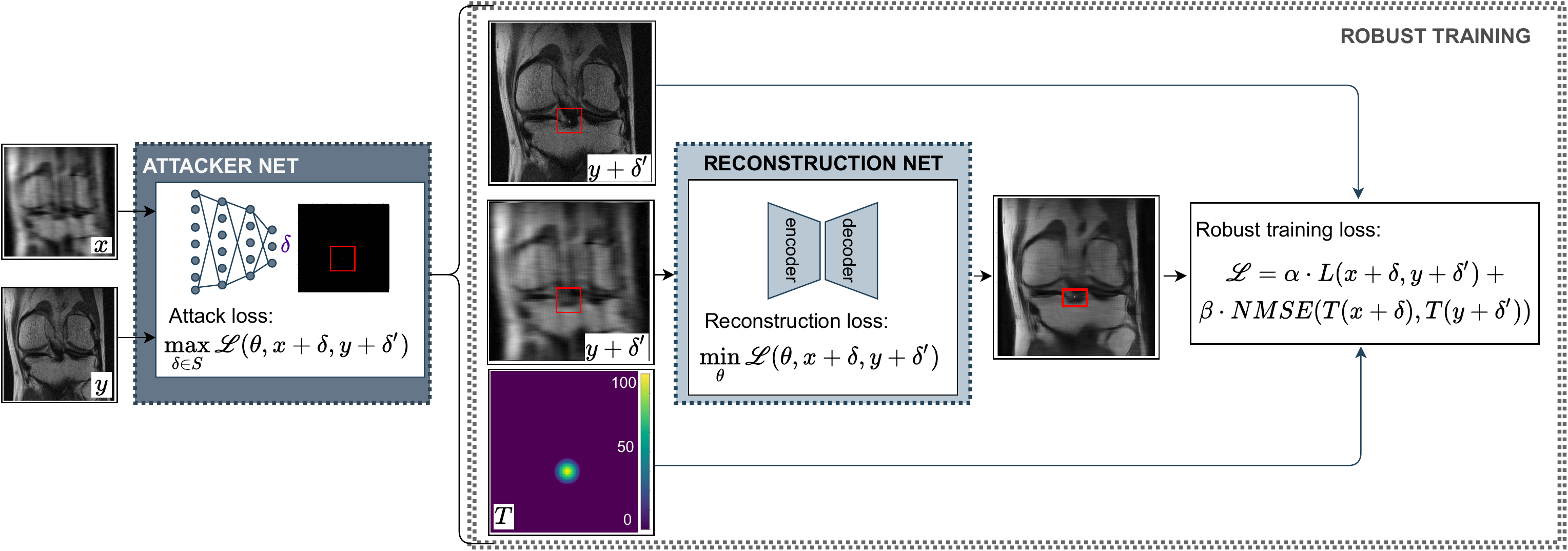}
 \caption{Overview of the proposed approach. At each training iteration, an attacker identifies which small features are difficult to be reconstructed, given the current network parameters and the input MRI. The network parameters are then updated so that the reconstruction of the small features is encouraged.}
 \label{fig:FNAF_training}
\end{figure}
In an attempt to better explain this false negative phenomenon, in \cite{cheng2020addressing}, we investigated two hypotheses: \textit{i)} under-sampling is the cause for loss of small abnormality features; and \textit{ii)} the small abnormality features are not lost during under-sampling; albeit they result to be attenuated and potentially rare. A graphical overview of the approach is shown in Fig.~\ref{fig:FNAF_training}.

Based on these assumptions, if an under-sampling procedure had removed the information related to a small feature available in the fully-sampled signal, a reconstruction algorithm would be able to recover that information only through other confounded structural changes, if present. This led to a realization for the necessity of new loss functions or metrics to be adopted in image reconstruction. In practice, deep learning reconstruction models trained to only maximize image quality can fail when it comes to reconstructing small and infrequent structures, as experimentally shown in many studies \citep{cheng2020addressing,zbontar2018fastmri,antun2020instabilities,muckley2020state}. With respect to the second hypothesis, if the (albeit rare) features were still available, a learning-based reconstruction method should be able to reconstruct those features; yet, this is contingent upon placing the right set of priors during training and to achieve this result, adversarial training is a valid solution. 
\begin{figure}[!tb]
 \centering
 \includegraphics[width=0.99\columnwidth]{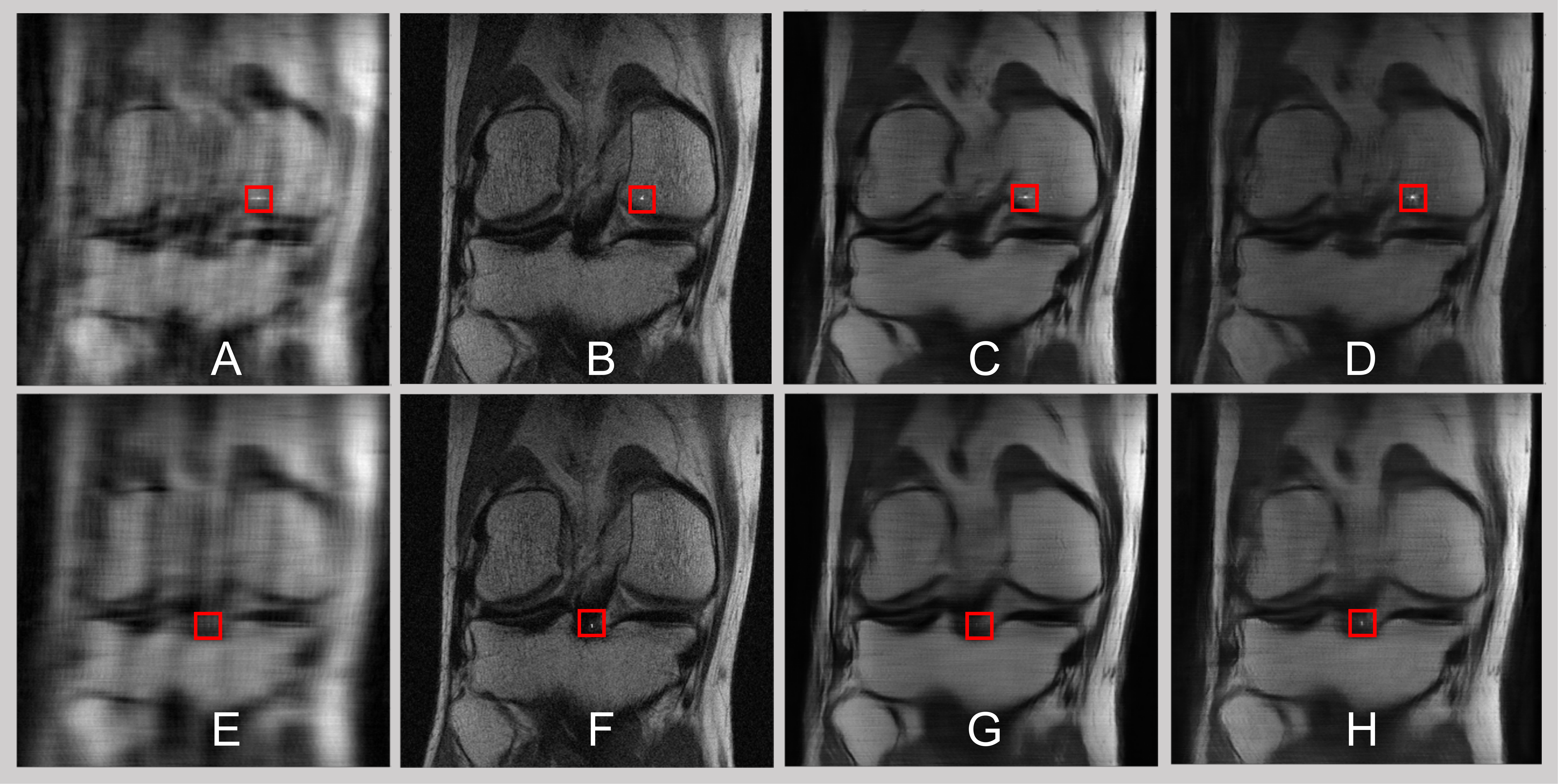}
 \caption{The top row (A-D) shows a `failed' FNAF attack. The bottom row (E-H) shows a `successful' FNAF attack. Column 1 contains the under-sampled zero-filled images. Column 2 contains the fully-sampled images. Column 3 contains U-Net reconstructed images. Column 4 contains FNAF-robust U-Net reconstructed images. (C-G-D-H)}
 \label{fig:FNAF_attack}
\end{figure}
To investigate these hypotheses, in \cite{cheng2020addressing}, False Negative Adversarial Features (FNAF) were introduced. Examples of FNAF are shown in Fig.~\ref{fig:FNAF_attack}; these are minimally perceivable features synthetically injected in the input data- through adversarial attacks- to mislead the reconstruction network and make it unable to reconstruct the features. Subsequently, in the pursuance of reconstructing these small features, a robust training approach was adopted, where the loss function over-weighted the reconstruction importance of such features. 
\newpage
This paper extends \cite{cheng2020addressing} leading to the following contributions:
\begin{itemize}
 \item A musculoskeletal (MSK) radiologist manually annotated features that are relevant in the diagnosis and monitoring of MSK related diseases, including bone marrow edema and cartilage lesions. Bounding boxes were placed in relevant regions of the Fast-MRI knee dataset. The manual annotations can be accessed at \url{https://github.com/fcaliva/fastMRI_BB_abnormalities_annotation}. 
 \item We investigated the effects of robust training with FNAF on reconstructing real-world abnormalities present in the Fast-MRI knee dataset, using the available bounding boxes.
 \item We quantitatively investigated the effects of training using real abnormalities, utilizing the available bounding boxes.
 \item We investigated the effects of robust training on a downstream task, such as feature abnormality classification. The experimental study showed that when a FNAF-based robust training is employed, the false negative rate in abnormality classification is reduced.
 \item We further highlight the need for using features that better represent real-world features as part of the robust training procedure, speculating that this would reduce network instability and improve reliability and fidelity in image reconstruction. 
\end{itemize}

\section{Related works}
\subsection{Clinical background}
During the past decade, non-invasive imaging has played a big role in the discovery of biomarkers in support of diagnosis, monitoring and assessment of knee joint degeneration. Osteoarthritis (OA) is a degenerative joint disease, and a leading cause of disability for 303 million people worldwide \citep{kloppenburg2020osteoarthritis}. Scoring systems, such as the MRI Osteoarthritis Knee Score (MOAKS)~\citep{hunter2011evolution} and the Whole-Organ Magnetic Resonance Imaging Score (WORMS)~\citep{peterfy2004whole} have been applied for grading cartilage lesions in subjects with OA, and comparing lesion severity with other findings such as meniscal defects, presence of bone marrow lesions, as well as additional radiographic and clinical scores \citep{link2003osteoarthritis,felson2001association,felson2003bone}. Morphological abnormalities which are detectable from MRIs have been considered to be associated with incident knee pain \citep{joseph2016mri}, 
and predictors of total knee replacement surgery \citep{roemer2015can}. 
\cite{roemer2015can} observed a significant knee replacement risk when knees had exhibited, among other phenotypes, severe cartilage loss, bone marrow lesions, meniscal maceration, effusion or synovitis. 
Small abnormalities are fundamental for clinical management, and current image reconstruction techniques should be reliable in recovering such features.

\subsection{MRI reconstruction with deep learning}
MRI is a first-choice imaging modality when it comes to studying soft tissues and performing functional studies. While it has been widely adopted in clinical environments, MRI has limitations, one of which depends on the data collection: it is a sequential and progressive procedure where data points are acquired in the \textit{k-}space. The higher the resolution, the more data points are needed to be sampled to satisfy the desired image quality. A strategy to reduce the scanning time is to acquire a reduced number of phase-encoding steps. However, this results in aliased raw data, which- in parallel imaging for example- are resolved by exploiting the knowledge about the utilized coil geometry and spatial sensitivity \citep{glockner2005parallel}. The new paradigm in image reconstruction is that DL-based approaches are capable of converting under-sampled data to images that include the entire information content \citep{recht2020using}. According to \cite{liang2019deep} and \cite{hammernik2020machine}, this is key in MRI reconstruction. There exists multiple approaches to accomplish MRI reconstruction by means of DL. \cite{liang2019deep} suggest three main categories of methods: data-driven, model-driven or integrated. Data-driven approaches do not require particular domain knowledge and mainly rely on the availability of a large amount of data. As a result, they tend to be data-hungry in their efforts to learn the mapping between \textit{k-}space and the reconstructed MRI. Conversely, model-based approaches mitigate the need for big data by restricting the solution space through the injection of physics-based prior knowledge. Those methods, which reproduce the iterative approach of compressed sensing~\citep{lustig2007sparse}, belong to the model-based set. Integrated approaches can combine positive aspects of both previous solutions.

\subsection{Adversarial attacks}
Adversarial attacks are small, imperceptible perturbations purposefully added to input images with the aim to mislead machine learning models. To apply adversarial attacks to MRI reconstruction with deep learning, it is important to understand the most studied forms of adversarial attacks. There exists a vast literature which attempts to explain adversarial examples \citep{goodfellow, bubeck2018adversarialconstraints, gilmer2018adversarialsphere, mahloujifar2019curse, shafahi2018adversarialinevitable}. One notable theory by \cite{more_data} states that adversarial examples are a consequence of data scarcity, and is linked to the fact that the true data distribution is not captured in small datasets. Another profound explanation is provided by \cite{bug_feature}, which shows that adversarial successes are mainly supported by a model's ability to generalize on a standard test set by using non-robust features. In other words, adversarial examples are more likely a product of datasets rather than that of machine learning models. To make a model resistant to adversarial attacks without additional data, one could employ adversarial training and provide the model with a prior that remarks the fact that non-robust features are not useful as demonstrated in \cite{goodfellow} and \cite{madry2018towards}. These findings are orthogonal to the second hypothesis investigated in this paper: if we interpret the distribution of FNAF as the distribution of robust features, we may attribute FNAF reconstruction failure to the dataset's inability to capture FNAF's distribution.

While the majority of adversarial attacks focus on discriminative models, \cite{generative} proposes a framework to attack variational autoencoders (VAE) and VAE-Generative Adversarial Networks (GAN). Specifically, input images are imperceptibly perturbed so that the generative models synthesize target images that belong to a different class. Although reconstruction models can be seen as generative, we differ from this body of work, mainly because we focus on generating perceptible features that perform non-targeted attacks.

Going beyond small perturbations, a set of more realistic attacks produced by 2D and 3D transformations are proposed in \cite{xiao2018spatially} and \cite{synthesizing}. Similarly to our work, these studies perform perceptible attacks. Arguably, the most realistic attacks are physical attacks, which in \cite{physical} are achieved by altering the physical space before an image is digitally captured. \cite{Kgler2018PhysicalAI} propose a physical attack on dermatoscopy images by drawing on the skin around selected areas. Although these attacks could more easily translate to real world scenarios, it would be nearly impossible to perform physical attacks to MRIs. Recently, \cite{chen2020realistic} leveraged adversarial attacks to introduce realistic intensity inhomogeneities to cardiac MRIs, which in turn boosted the performance in a segmentation task.

\subsection{Adversarial attacks and robustness in medical imaging}
Adversarial robustness can improve performance in medical imaging tasks. \cite{liu2019robustifying} investigated the robustness of convolutional neural networks (CNNs) for image segmentation to visually-subtle adversarial perturbations, and showed that in spite of a defense mechanism in place, the segmentation on unperturbed images was generally of higher quality than that obtained on perturbed images. Adversarial robustness has mainly been associated with security, where an adversary is assumed to attack a machine learning system. In \cite{Kgler2018PhysicalAI} and \cite{Finlayson1287}, attackers are malicious users that aim to fool medical systems by applying physical attacks. Cases like these are hardly found in real life, and if we rely on the genuine assumption that no malicious entities try to mislead diagnoses or treatments \citep{Finlayson1287}, it is easy to understand why ideas from the robustness community have found limited application in medical imaging. Although \cite{Finlayson1287} voiced concerns of insurance fraud, it has been argued that robustness goes beyond security and has implications in other domains including interpretability of DL models \citep{engstrom2019learning, kaur2019perceptually, santurkar2019image, tsipras2018robustness, fong2017interpretable}. 
In this context, the work from \cite{fong2017interpretable} aimed to identify image parts that were responsible and could explain the classifier's decisions. \cite{uzunova2019interpretable_medical} used VAE to generate perturbations which were employed to explain medical image classification tasks. Furthermore, while independent and identically distributed machine learning focuses on improving the prediction outcome in average cases while neglecting edge cases, robust machine learning focuses on finding the worst case scenario and edge cases to potentially patch them. This perspective is very useful in medical imaging analyses and applications, as the danger for each misprediction can be high and abnormalities can be edge cases.

\section{Materials and Methods}
This study aims to enhance the quality of accelerated MRI that are reconstructed with deep learning approaches. A complete overview of the proposed approach is shown in Fig.~\ref{fig:FNAF_training}. In the interest of accomplishing such task, we adversarially attack a reconstruction network to identify small (synthetic) features which are difficult for that network to reconstruct. We refer to them using the acronym `FNAF', which stands for `False Negative Adversarial Features'. Simultaneously, we apply ideas of robust training to make the reconstruction network less sensitive to the adversarial features (\textit{i.e.} FNAF), and reduce their reconstruction error. Next, we test whether the approach has generalized to real world features by analyzing reconstructed MRIs from the knee Fast-MRI dataset \citep{zbontar2018fastmri}. To do so, MR images are fed into a classification network, purposefully trained to establish whether upon reconstruction, these real small features are still present in the MRI.
\begin{table}[!b]
\centering
 \caption{Summary of the imaging data used in the experimental study.}
 \label{tab:dataset}%
 \resizebox{0.99\columnwidth}{!}{
 \begin{tabular}{c|c|c|m{0.6\textwidth}}
 \hline\hline
Dataset & Train & Validation & Comments\\
\hline
MRI reconstruction & 973 (34,742) & 199 (7,135) & 3D Volumes (2D slices). Single-coil emulated \textit{k}-space data, from \cite{zbontar2018fastmri}. Details in Section~\ref{sec:MRI reconstruction dataset}.\\
\hline
Abnormality annotation & 468 & 97 & Coronal proton density-weighted with fat suppression volumes, annotated for abnormalities based on WORMS scale. Details in Section~\ref{section: bbox-labeling}. Annotations publicly available \url{https://github.com/fcaliva/fastMRI\_BB\_abnormalities\_annotation}{here}. \\
\hline
Abnormality classification ($32\times32$) & 103,652 & 2,552 & \multirow{2}{0.6\columnwidth}{Image patches extracted from the volumes with abnormality annotations available. Details in Section~\ref{sec:abnormality dataset}.} \\
Abnormality classification ($64\times64$) & 7,664 & 2,110 & \\
& & & \\
\hline
\hline
\end{tabular}
}
\end{table}
\subsection{Imaging data}
\subsubsection{MRI reconstruction dataset}\label{sec:MRI reconstruction dataset}
The Fast-MRI dataset for the single-coil reconstruction task was utilized \citep{zbontar2018fastmri}. This portion of the dataset is comprised of emulated single-coil (ESC) \textit{k-}space data, which are combined \textit{k-}space data derived from multi-coil \textit{k-}space data to approximate single-coil acquisitions \citep{tygert2020simulating}. ESC uses responses from multiple coils and linearly combines them in a complex-valued formulation, which is fitted to the ground-truth root-sum-of-squares reconstruction in the least-squares sense. The dataset comprises proton-density weighted with and without fat suppression sequences; we refer interested readers to the original paper for details about the acquisition. We used the original train/validation data splits which include 973/199 volumes (34,742/7,135 slices) respectively. Details are summarized in Table~\ref{tab:dataset}.
\begin{figure}[!t]
 \centering
 \includegraphics[width=0.99\columnwidth]{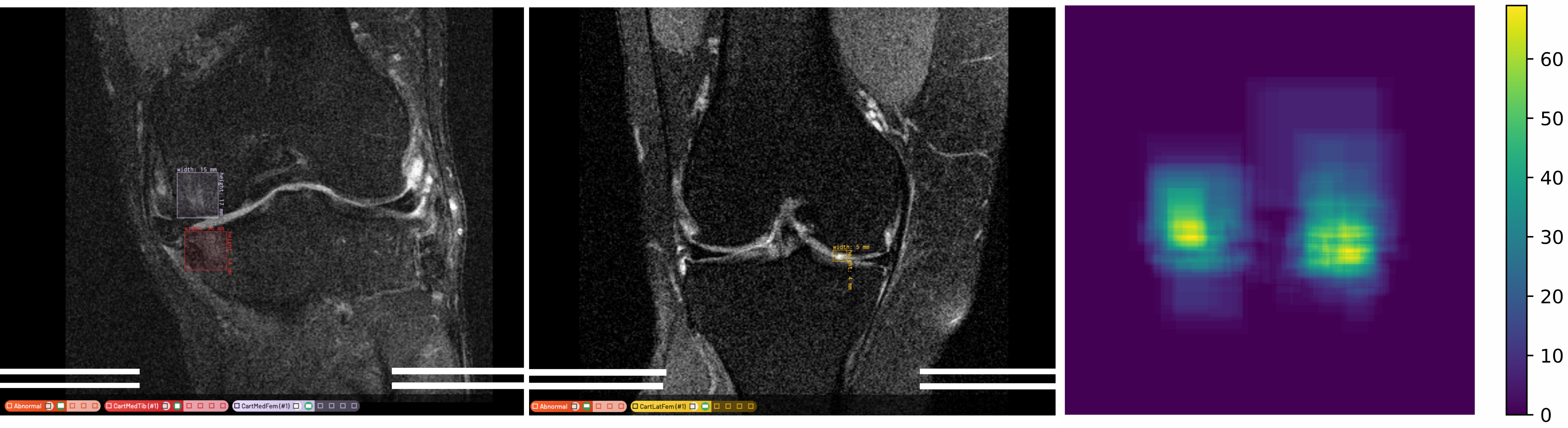} 
\caption{Depiction of the bounding box-based annotations. Left: Cartilage lesions visible in the medial tibia and femur. Middle: Cartilage lesion in the lateral femur. Right: Regional distribution of abnormalities in the training dataset.}
\label{fig:Example_annotation}
\end{figure}
\subsubsection{Fast-MRI dataset abnormality annotation}\label{section: bbox-labeling}
To further assess the generalization ability of the proposed reconstruction approach to real features, coronal proton density-weighted with fat suppression knee sequences from the aforementioned dataset were annotated. These include a total of 515 MR exams, 468 and 97 exams from the training and validation sets respectively. All MRI volumes were examined for cartilage, meniscus, bone marrow lesions (BML), cysts, and based on the WORMS scale \citep{peterfy2004whole} by the MSK radiologist involved in the study. Bounding boxes were placed for cartilage and BML in 4 sub-regions of the knee at the tibio-femoral joint; medial and lateral femur, medial and lateral tibia. Cartilage lesion were defined as partial or full thickness defects observed in one or more slices extending to include any breadth. Annotated bone marrow lesions included any increased marrow signal abnormality adjacent to articular cartilage in one or more slices, at least encompassing 5\% of the articular marrow regions.
Similar bounding boxes were placed in the two sub-regions of the meniscus: medial and lateral. Meniscal lesions were defined to include intrasubstance signal abnormality, simple tear, complex tear or maceration. Given the sparse occurrence of cysts, a single bounding box label was used to encompass all encountered cystic lesions in any of the sub-regions. Fig.~\ref{fig:Example_annotation} is exemplary of the performed annotations. Specifically, the MRI on the left-hand side displays the presence of cartilage lesions in the medial tibial (red) and femoral (purple) compartments. In the middle, a cartilage lesion in the lateral femur is marked by a golden color-coded bounding box. On the right-hand side, a map representative of the distribution location of abnormalities in the training set is shown.

\subsubsection{Abnormality presence classification dataset}\label{sec:abnormality dataset}
Performing anomaly classification at the image level can hardly discriminate whether all clinical abnormalities have correctly been reconstructed. This is especially true when an image includes more than one abnormality, as per the case in the left image of Fig.~\ref{fig:Example_annotation}. A good way to provide a less coarse analysis is to perform it in a patch-based fashion. This has the drawback of introducing severe class imbalance; therefore we performed adaptive patch sampling based on the distribution of abnormality in the training set as shown in the right of Fig.~\ref{fig:Example_annotation}. Two configurations of patch size were explored, $32\times32$ and $64\times64$. With respect to the $32\times32$ patches, these were created by utilizing a sliding window with stride 2, whereas for the $64\times64$ patches a stride of 8 pixels was adopted. The available annotations allowed to define whether a patch comprised abnormalities or not. 

To further enforce class balance, the number of normal patches was capped to the number of abnormal patches, by randomly selecting normal patches within the pool of normal patches. As a result, the two datasets comprised of: 103,652 train/2,552 validation data points in the $32\times32$ patch dataset, and 7,644 train/2,110 valid data samples in the $64\times64$ patch dataset. Importantly, since we do not have access to patient information in the Fast-MRI dataset, to prevent any form of data leakage across data splits, we pertained the original dataset's training and validation splits, refraining from creating a test split, as potential data-leakage could lead to an overoptimistic classification performance. Details about the dataset are summarized in Table~\ref{tab:dataset}.

\subsection{False negative adversarial attack on reconstruction networks}
Adversarial attacks aim to maximize the loss $\mathscr{L}$ of a machine learning model, parameterized by $\theta$. This can be achieved by changing a perturbation parameter $\delta$ within the set $S\subseteq R^d$ of the allowed perturbation distribution \citep{madry2018towards}. Here, $S$ was restricted to be a set of visible small features in all locations of an inputted MR image. Formally, this is defined as:
\begin{equation}\label{eq. standard_attack}
 \max_{\delta \in S} {\mathscr{L}}(\theta, x+\delta, y)
\end{equation}
In principle, $\mathscr{L}$ could be any arbitrary loss function. 
For traditional image reconstruction, the reconstruction loss is minimized so that all the features including the introduced perturbation (\textit{i.e.} small features) are reconstructed. Conversely, the attacker aims to maximize Eq.~\ref{eq. standard_attack} and identify a perturbation, which the network is not capable of reconstructing. 

Next, let $\delta$ be an under-sampled perturbation which is added to an under-sampled image and $\delta^{\prime}$ the respective fully-sampled perturbation, the objective function becomes:
\begin{equation}\label{eq. fn_attack}
 \max_{\delta \in S} {\mathscr{L}}(\theta, x+\delta, y+\delta^{\prime})
\end{equation}
with:
\begin{equation}\label{eq. delta_prime}
 \delta = U(\delta^{\prime})
\end{equation}
$U$ can be any under-sampling function, which is comprised of an indicator function $M$ that acts as a mask in the \textit{k-}space domain, and an operator that allows for a conversion from image to \textit{k-}space and vice-versa such as the Fast Fourier Transform $\mathcal{F}$ and its inverse $\mathcal{F}^{-1}$. The under-sampling as well as the \textit{k-}space mask $M$ functions are the same as in the implementations provided by \cite{zbontar2018fastmri}.
\begin{equation}\label{eq. under-sampling}
 U(y) = \mathcal{F}^{-1}(M(\mathcal{F}(y)))
\end{equation}
As we synthetically construct the small added features, we can measure the loss value within the area where the introduced features are located to assess whether they have been reconstructed. To do so, a mask was placed over the reconstructed image and the perturbed target image, so that only the area of the small feature was highlighted. The area was relaxed to also include a small region at a distance $d$ from the feature border. The motivation for the mask accounting for boundaries is; if only the loss of the FNAF’s foreground was measured, it might not capture cases where the FNAF had blended in with the background. Therefore, the loss was computed in a 5 pixels distance range from the boundary of the FNAF.
The loss is defined as
\begin{equation}\label{eq. attack_loss}
 \mathscr{L}=\alpha \cdot L(x,y) + \beta \cdot NMSE(T(x),T(y)) 
\end{equation}
where x and y are the under-sampled and the fully-sampled (\textit{i.e.} target) reconstructed MRIs, respectively. $L$ is the original whole-image objective function reported by image reconstruction methods in their original papers, such as L1 loss for the U-Net in \cite{zbontar2018fastmri} and SSIM loss for I-RIM in \cite{putzky2019rim}. $T$ is an indicator function which masks over the FNAF in the fully-sampled and under-sampled reconstructed images. Weights $\alpha$ and $\beta$ are hyper-parameters, which were set to 1 and 10 during adversarial training (details in Section~\ref{section:Training implementation details}). This allows one to better preserve both the image quality and robustness of FNAF. During the evaluation of the attack, $\alpha$ and $\beta$ were set to 0 and 1 respectively, to guarantee that the loss value was representative of the FNAF reconstruction. The loss can be maximized by either random search (RS) or finite-difference approximation gradient ascent (FD). 

With random search, random shapes of features $\delta$ are placed at random locations in the image, and the $\delta$ which maximizes the loss in Eq.~\ref{eq. fn_attack} is identified. As demonstrated in \cite{random_search} and \cite{engstrom2017exploring}, random search is an effective optimization technique. The location of the $\delta$ feature is a crucial factor in finding FNAF but the ($c, r$) coordinates of $\delta$ are non-differentiable. This limitation can be addressed by employing the finite central difference reported in Eq.~\ref{eq. partialL}, with step size $h$, which makes it possible to approximate the partial derivatives for each location parameter ($p$) and optimize the low-dimensional non-differentiable parameters to update $p$ and maximize Eq.~\ref{eq. fn_attack} via gradient ascent.
\begin{equation}\label{eq. partialL}
 \frac{\partial L}{\partial p}= \frac{L\left(p+\frac{h}{2}\right)-L\left(p-\frac{h}{2}\right)}{h}
\end{equation} 

\subsection{Under-sampling information preservation verification} \label{sec: information preservation}
A benefit of having a synthetic feature generator is that one can produce unlimited features, but also quantify the amount of preserved information after \textit{k-}space under-sampling. To guarantee that the information of $\delta$ was preserved irrespective of under-sampling in the \textit{k-}space, the following condition needs to be satisfied:
\begin{equation}\label{eq. epsilon}
 D(x+\delta, x) > \epsilon
\end{equation}
where $D$ is a distance function, and $\epsilon$ is a noise error tolerance threshold; $x+\delta$ and $x$ obey: 
\begin{equation}\label{eq. Udelta_prime}
 U(y+\delta^{\prime}) = U(y)+U(\delta^{\prime}) \\ 
 = x+\delta
\end{equation} as $U$ is linear and closed under addition. MSE is used for $D$.

\subsection{FNAF implementation details}\label{section:Training implementation details}
FNAF were constrained to include 10 connected pixels for the evaluation of the attacks. Attack masks were placed within the center of a 120$\times$120 crop of the image, to ensure that the added features were small enough and placed in reasonable locations. For random search, 11 randomly-shaped FNAF were generated at random locations for each sample in the validation set, and the highest adversarial loss were recorded. 

\subsection{Robust training with False Negative Adversarial Features}\label{section:FNAF-robust training}
Our attack formulation, which is expressed in Eq.~\ref{eq. attack_loss}, allows the reconstruction models to simultaneously undergo standard and adversarial training. This allows one to do FNAF-robust training on a pre-trained model and in turn, speed up convergence. In contrast, small perturbations-based adversarial training approaches would require training only on robust features \citep{madry2018towards}. To accelerate training, we adopted ideas from \cite{adv_free}. Briefly, for FNAF-robust training, we used a training set which included original and adversarial examples. The inner maximization can be performed by either random search or finite-difference approximation gradient ascent, as described above. In our experiments, we opted for random search as in our previous study \citep{cheng2020addressing}, it proved to be a more effective form of attack. RS reduced the implementation to be a data augmentation approach. 

In Eq.~\ref{eq. attack_loss}, $\beta$ was set to 10 to encourage the reconstruction of the introduced features. To prevent from overfitting on the FNAF attack successes, the selection of the model (to be attacked) was done by choosing the model which showed the best reconstruction loss on the validation set, while the adversarial component of the loss was ignored. The models trained in this experiment are named `FNAF U-Net' and `FNAF I-RIM'.

During training of FNAF I-RIM, FNAF attacks were constrained to include 10 to 1000 uniformly sampled connected pixels. This was to relax the constraint, as advised in \cite{cheng2020addressing}. We intended to do the same for FNAF U-Net but the training could not reach convergence, so during training of FNAF U-Net, FNAF was restricted to 10 connected pixels.

\subsection{Training with real abnormalities}\label{section:training-bbox}
Using the annotations described in Section~\ref{section: bbox-labeling}, we implemented robust training on the reconstruction models using real abnormalities. Same as in the experiment described in Section~\ref{section:FNAF-robust training}, Eq.~\ref{eq. attack_loss} was optimized, by altering the term $T$ which here represents annotations of bounding box masks. This experiment is an upper-bound for FNAF robust training. The models trained in this experiment are named `B-Box U-Net' and `B-Box I-RIM'.
\begin{figure}[!tbh]
 \centering
 \includegraphics[width=0.99\columnwidth]{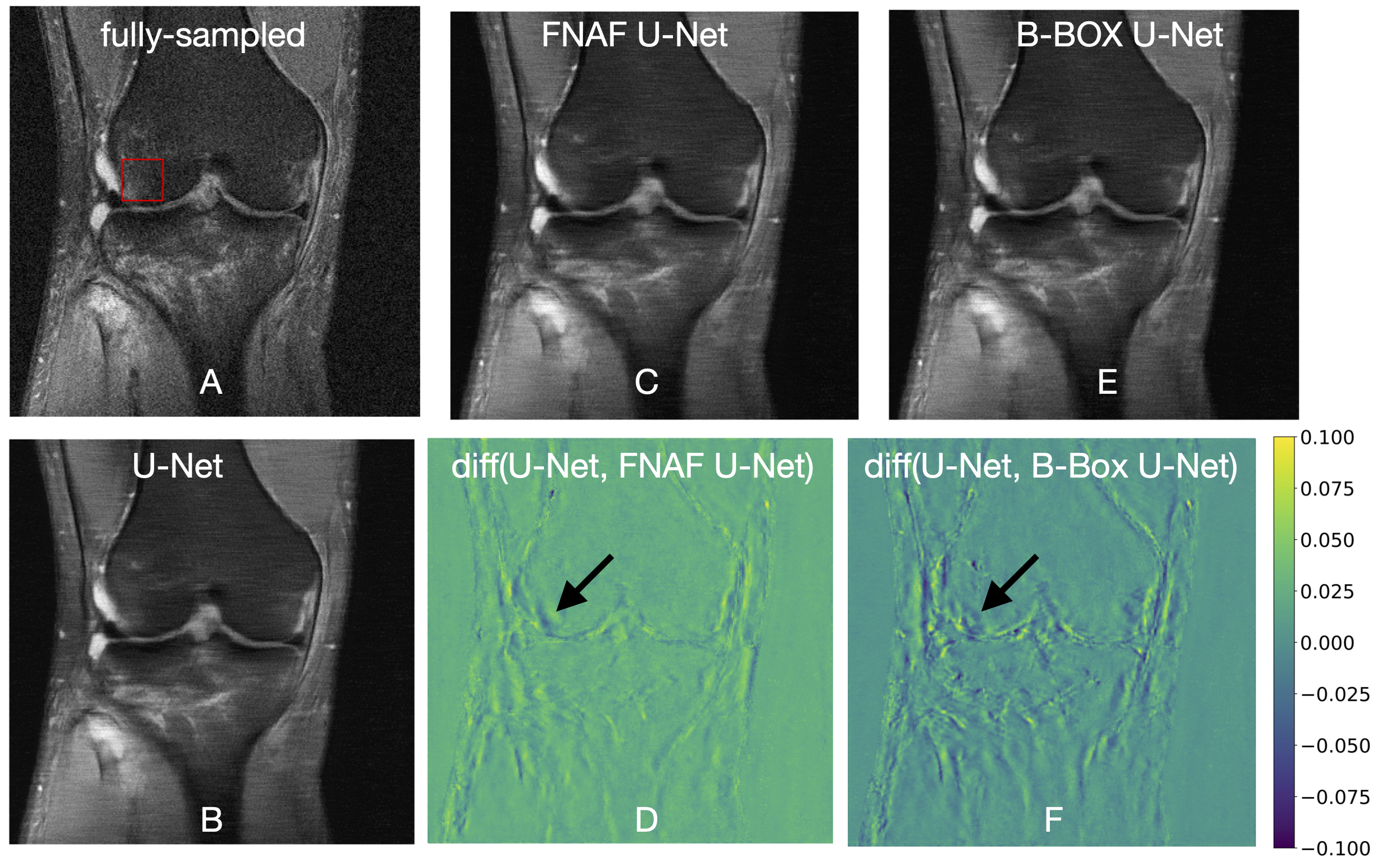}
 \caption{Example a reconstructed 4$\times$ accelerated MRI. The abnormality within the bounding box in (A) was not perfectly reconstructed by the baseline U-Net (B), and detected as a false negative feature by the abnormality classifier. The abnormality was better reconstructed after robust training with FNAF (C) and real abnormalities (E). Reconstruction differences between the images reconstructed with the baseline U-Net and the two robust trained networks are shown in (D) and (F). The arrows point to regions where signal was better preserved upon reconstruction, which in turn might have helped the abnormality classifier. Baseline U-Net trained with settings reported in \cite{zbontar2018fastmri}.}
 \label{fig:classification_comments_UNET}
\end{figure}
\subsection{Evaluation of generalization to real-world abnormalities}
We aimed to answer the following question: How well does a robust training approach- which encourages the reconstruction of FNAF- improve the network's performance on reconstructing real (rare or common) small features?

An MSK radiologist inspected the reconstructed MRI slices from the validation set which include abnormalities, as per the annotations described in Section~\ref{section: bbox-labeling} and observed only slight improvements in the reconstruction of the abnormalities, which can be attributable to the robust training. An example is pictured in Fig.~\ref{fig:classification_comments_UNET} (more details in Section~\ref{Measuring the presence of real-world abnormalities in the reconstruction}). We understand that this investigative approach could be biased by human cautiousness, and it is also practically impossible to confine the radiologist's analysis to the regions which include the abnormalities. Lastly, the opinion of a single radiologist is not enough to fairly investigate the reconstruction due to individual bias, so a formal analysis with multiple readers is needed to confirm our preliminary observations. Therefore, we conducted a quantitative analysis of the reconstruction by automating abnormality detection in the reconstructed images.

\subsubsection{Abnormality presence classification}
Abnormalities present in the Fast-MRI dataset were manually annotated by means of bounding boxes, as described in Section~\ref{section: bbox-labeling}. The benefits of producing such annotations are two-fold. Bounding boxes allow for restricting the region in which the reconstruction error can be calculated. This overcomes the limitation of global evaluation metrics which merely provide an `average' evaluation of the reconstruction quality, which in the first place, might not be representative of the reconstruction accuracy in the small regions that contain abnormalities. Furthermore, bounding boxes allow for implementing an unbiased tool (\textit{e.g.} a classifier) capable of assessing the presence or absence of the abnormalities in the reconstructed image, making it possible to fairly compare multiple reconstruction approaches. 

\subsection{Experimental setup}
\label{section:Experimental setup}
We conducted our experiments with single-coil setting, including 4$\times$ and 8$\times$ acceleration factors \citep{zbontar2018fastmri}. We evaluated our method with two deep learning based reconstruction methods; a modified U-Net \citep{zbontar2018fastmri}- here considered as baseline for MRI reconstruction- and invertible Recurrent Inference Machines (I-RIM) \citep{putzky2019invert}, which was the winner model of the single-coil knee Fast-MRI challenge. For U-Net, we followed the training procedures described in \cite{zbontar2018fastmri}. For I-RIM, we followed the training procedures described in \cite{putzky2019rim} and used the official released pre-trained model. 

\subsubsection{Attack evaluation metrics}
The average attack loss for the validation set and the attack hit rate were computed. The average attack loss is defined in Eq.~\ref{eq. attack_loss}. An attack is considered a hit when the loss is higher than a threshold value $\gamma$, which was empirically set to 0.25, as we observed the FNAF disappeared when the loss value was greater than 0.25. The hit rate was set to be conservatively low. To achieve this effect, $\gamma$ was set at a high value, and as a consequence there may be cases in which, despite $\mathscr{L}\leq \gamma$, the FNAF were lost. We speculate that the actual hit rate is likely to be higher than the value reported in the current work. 

\subsubsection{Abnormality classification}
Squeezenet \citep{iandola2016squeezenet} pretrained on the Imagenet dataset was the network of choice. The selection was driven by two factors: while providing excellent classification performance, Squeezenet has a small footprint by design and is fully convolutional. This means that it can accept, as its input, images of any size, and without requiring major architectural modifications. Furthermore, the limited memory footprint and flexibility to inputs of various dimensions are particularly appealing for us, for the prospective future combination of image reconstruction and abnormality detection in an end-to-end framework. Two fine-tuning strategies were tested; in one, all layers pre-trained on Imagenet with exception for the last classification convolutional layer were frozen. The last- namely the classification layer- was replaced by two $1\times1$ convolutional layers; the first produced 512 feature channels, the second provided two feature channels, which were activated with a ReLU activation function and fed into a Global Average Pooling to provide the logits subsequently inputted to a softmax activation function. The last is a function that converts an $N$-dimensional vector of arbitrary real values to a $N$-dimensional vector of real values in the range [0, 1] summed to 1. In this case, $N=2$ and the obtained values can be interpreted as probability values that a particular image includes an abnormality or not. In the second fine-tuning strategy, all the weights in the modified version of the Squeezenet were fine-tuned. Ultimately, for completeness we also trained a Squeezenet from scratch. 

During the learning phase, classification networks were trained and validated on the patches extracted from the original Fast-MRI dataset, as described in Section~\ref{sec:abnormality dataset}. Trained models were inferred on the same patches, this time extracted from the original Fast-MRI images as well as from images reconstructed with U-Net and I-RIM, and the versions robust trained with FNAF (FNAF U-Net and FNAF I-RIM) and with real abnormalities (B-Box U-Net and B-Box I-RIM). The performance on the patches extracted from the fully sampled images is an upper bound for abnormality presence classification performance.

For fine tuning, a binary cross-entropy loss was minimized, using stochastic gradient descent as the optimizer with learning rate as $1\times10^{-3}$ (0.1 decay rate), weight decay $5\times10^{-4}$ and 128 samples per batch. Early stopping was implemented when no classification improvement in the validation set were observed for 60 epochs, with the maximum number of epochs set to 100.
\begin{table}[!b]
\centering
 \caption{FNAF attack evaluations, when the loss was maximized using random search. Metrics computed on (A) 4$\times$ and (B) 8$\times$ retrospectively undersampled MRIs.}
 \label{tab:attack_models}%
 \resizebox{0.95\columnwidth}{!}{
 \begin{tabular}{c|c|c c c|c|c}
\hhline{===~===}
AF=$4\times$ & Attack Rate (\%) & NMSE & & AF=$8\times$ & Attack Rate (\%) & NMSE \\
\hhline{---~---}
 U-Net & 88.89 & 0.5798 & &U-Net & 92.96 & 0.7928 \\
 FNAF U-Net & 3.406 & 0.0422 & &FNAF U-Net & 13.38 & 0.0781 \\
 B-Box U-Net & 88.09 & 0.5793 & & B-Box U-Net & 92.98 & 1.072 \\
 I-RIM & 10.62 & 0.1948 & & I-RIM & 91.00 & 0.3305 \\
 FNAF I-RIM & 0.014 & 0.0072 & & FNAF I-RIM & 2.859 & 0.0245 \\
 B-Box I-RIM & 17.39 & 0.2056 & & B-Box I-RIM & 92.81 & 0.3489 \\
\hhline{===~===}
\multicolumn{1}{c}{} \\[-2.2ex]
\multicolumn{3}{c}{(A)} & \multicolumn{3}{c}{(B)} \\
\end{tabular}}
\end{table}

\section{Results and Discussion}
The results of the attack are reported in Table~\ref{tab:attack_models} and confirm that hypothesis 2 is true in many cases; small features are not lost during under-sampling. 
The high success rate of the random search method for both models showed that it is fairly easy to find a FNAF in the search space that was heuristically defined. Although I-RIM was more resilient to the attacks as opposed to the baseline U-Net, the attack rate was still fairly high. This is concerning, yet expected since deep learning methods are not explicitly optimized for such objective, so these FNAF are at the tail-end of the distribution or even out-of-distribution with respect to the training distribution. Fortunately, we can modify the objective as specified in Section~\ref{section:FNAF-robust training} to produce a FNAF-robust model which appears to be resilient to the attacks. Furthermore, this shows a minimal effect in the standard reconstruction quality, which is reported in Table~\ref{tab:standard_eval}. B-BOX training seems to worsen the network's vulnerability to FNAF attacks which could mean the FNAF constructed might not be representative of the real abnormalities.

\begin{table}[!tbh]
\centering
 \caption{Image reconstruction performance comparison. Evaluation conducted using structural similarity index measure (SSIM), peak signal-to-noise-ration (PSNR) and normalized mean-squared error (NMSE). Metrics computed on the (A) 4$\times$ and (B) 8$\times$ retrospectively undersampled MRIs from the Fast-MRI knee validation set.}
 \label{tab:standard_eval}%
 \begin{tabular}{c|c|c|c}
\hline\hline
\multicolumn{4}{c}{ AF=$4\times$}\\
Method & NMSE & PSNR & SSIM \\
\hline
U-Net & 0.0345 $\pm$ 0.050 & 31.85 $\pm$ 6.53 & 0.7213 $\pm$ 0.2614 \\
FNAF U-Net & 0.0349 $\pm$ 0.050 & 31.78 $\pm$ 6.45 & 0.7192 $\pm$ 0.2621 \\
B-Box U-Net & 0.0351 $\pm$ 0.050 & 31.76 $\pm$ 6.46 & 0.7181 $\pm$ 0.2623 \\
\hline
I-RIM & 0.0341 $\pm$ 0.058 & 32.45 $\pm$ 8.08 & 0.7501 $\pm$ 0.2539 \\
FNAF I-RIM & 0.0338 $\pm$ 0.057 & 32.48 $\pm$ 8.01 & 0.7496 $\pm$ 0.2541 \\
B-Box I-RIM & 0.0331 $\pm$ 0.056 & 32.57 $\pm$ 8.04 & 0.7493 $\pm$ 0.2548 \\
\hline\hline
\multicolumn{1}{c}{} \\[-2.2ex]
\multicolumn{4}{c}{(A)} \\
\multicolumn{1}{c}{}\\[-2.2ex]
\hline\hline
\multicolumn{4}{c}{ AF=$8\times$}\\
Method & NMSE & PSNR & SSIM \\
\hline
U-Net & 0.0493 $\pm$ 0.057 & 29.85 $\pm$ 5.22 & 0.6548 $\pm$ 0.2935 \\
FNAF U-Net & 0.0495 $\pm$ 0.057 & 29.84 $\pm$ 5.21 & 0.6539 $\pm$ 0.2931 \\
BBOX U-Net & 0.0496 $\pm$ 0.057 & 29.81 $\pm$ 5.19 & 0.6520 $\pm$ 0.2935 \\
\hline
I-RIM & 0.0444 $\pm$ 0.068 & 30.95 $\pm$ 7.11 & 0.6916 $\pm$ 0.2933 \\
FNAF I-RIM & 0.0439 $\pm$ 0.0672 & 30.98 $\pm$ 7.03 & 0.6909 $\pm$ 0.2933 \\
B-Box I-RIM & 0.0430 $\pm$ 0.066 & 31.07 $\pm$ 7.06 & 0.6904 $\pm$ 0.2944 \\
\hline\hline
\multicolumn{1}{c}{} \\[-2.2ex]
\multicolumn{4}{c}{(B)}\\
\end{tabular}
\end{table}

\subsection{Under-sampling information preservation verification}
To investigate whether small features were preserved upon under-sampling (\textit{i.e.} our second hypothesis), the acceptance rate of the adversarial examples was measured and a high acceptance rate ($>$99\%) was seen across all the tested settings. This proves that in most cases, the small feature's information was not completely lost through under-sampling. Recall the information preservation (IP) loss, which was introduced in Sec.~\ref{sec: information preservation}, is a MSE loss. When compared to the FNAF loss, the IP loss showed a small negative correlation ($r=-0.14$). This supports the hypothesis that the availability of richer information weakens the attacks. Nevertheless, such negative correlation is weak, indicating that there is not a strong association. Therefore, the preservation of information alone cannot predict the FNAF-robustness of the model. Thus the loss of information due to under-sampling is a valid but insufficient explanation for the existence of FNAF.

\begin{table}[!t]
\caption{Image reconstruction normalized mean-squared error computed for each reconstruction approach within the abnormality regions. The metric is reported per compartment and was computed on the (A) 4$\times$ and (B) 8$\times$ retrospectively undersampled MRIs from the Fast-MRI knee validation set. `All' refers to the average NMSE on all the bounding boxes. \textbf{Bold} values represent statistically significant difference when a model's performance was compared against the relative baseline reconstruction method (e.g. (FNAF U-Net vs U-Net), (FNAF I-RIM vs I-RIM)). A two-sided Wilcoxon rank-sum test was conducted using $p\leq \alpha$, with $\alpha$ set to $0.05$ as the significance level. Per-abnormality summaries of the NMSE computed in all the samples are reported in Fig.~\ref{fig:boxplot-4x}-~\ref{fig:boxplot-8x}. Column 2- titled N- reports the count of occurrences of each abnormality, in the validation-set.}
\label{table: real_world_recon} 
\centering
\resizebox{0.95\columnwidth}{!}{\begin{tabular}{c|c|c|c|c|c|c|c}
\hline\hline
\multicolumn{7}{c}{AF=4$\times$}\\
Compartment & N & U-Net & FNAF U-Net & B-Box U-Net & I-RIM & FNAF I-RIM & B-Box I-RIM \\
\hline
Cart Med Fem& 19 & 0.04720 & 0.04771 & 0.04695 & 0.04003 & 0.03989 & \textbf{0.03685} \\
Cart Lat Fem& 20 & 0.05818 & 0.06046 & 0.06353 & 0.04588 & 0.04574 & \textbf{0.04365} \\
Cart Med Tib& 5 & 0.08705 & 0.08843 & 0.09137 & 0.09260 & 0.08821 & 0.08289 \\
Cart Lat Tib& 7 & 0.06148 & 0.06145 & \textbf{0.06959} & 0.05064 & 0.05035 & \textbf{0.04595} \\
BML Med Fem & 24 & 0.04153 & 0.04142 & \textbf{0.04298} & 0.03880 & \textbf{0.03790} & \textbf{0.03440} \\
BML Lat Fem & 29 & 0.05803 & 0.05847 & \textbf{0.06009} & 0.05245 & \textbf{0.05099} & \textbf{0.04690} \\
BML Med Tib & 10 & 0.06031 & 0.06091 & 0.06315 & 0.05905 & \textbf{0.05744} & \textbf{0.05201} \\
BML Lat Tib & 16 & 0.04905 & 0.04920 & 0.04730 & 0.04777 & \textbf{0.04661} & \textbf{0.04389} \\
Med Men & 11 & 0.04902 & 0.04964 & 0.05014 & 0.04139 & 0.04151 & \textbf{0.0392} \\
Lat Men & 11 & 0.04314 & 0.04421 & \textbf{0.04838} & 0.03321 & 0.03335 & \textbf{0.03158} \\
Cyst & 5 & 0.03557 & 0.03585 & 0.03665 & 0.03494 & 0.03496 & 0.02945 \\
\hline
All & & 0.05213 & 0.05277 & 0.05437 & 0.04649 & 0.04569 & 0.04232 \\
\hline\hline
\multicolumn{1}{c}{} \\[-2.2ex]
\multicolumn{7}{c}{(A)} \\
\multicolumn{1}{c}{}\\[-2.2ex]
\hline\hline
\multicolumn{7}{c}{AF=8$\times$}\\
Compartment & N& U-Net & FNAF U-Net & B-Box U-Net & I-RIM & FNAF I-RIM & B-Box I-RIM \\
\hline
Cart Med Fem & 19 & 0.08435 & 0.08399 & \textbf{0.07794} & 0.06181 & 0.06418 & \textbf{0.05611} \\
Cart Lat Fem & 20 & 0.08809 & 0.08767 & 0.08742 & 0.06375 & 0.06402 & 0.06216 \\
Cart Med Tib & 5 & 0.15370 & 0.15350 & 0.13490 & 0.13850 & 0.13500 & 0.11910 \\
Cart Lat Tib & 7 & 0.11830 & 0.11790 & 0.13470 & 0.09268 & 0.09446 & \textbf{0.08259} \\
BML Med Fem & 24 & 0.08110 & 0.08178 & 0.07576 & 0.06020 & 0.06065 & \textbf{0.05374} \\
BML Lat Fem & 29 & 0.10280 & 0.10050 & \textbf{0.09199} & 0.08084 & 0.08059 & \textbf{0.06936} \\
BML Med Tib & 10 & 0.10990 & \textbf{0.10610} & 0.10840 & 0.08304 & \textbf{0.08070} & 0.07396 \\
BML Lat Tib & 16 & 0.07505 & 0.073440 & 0.07519 & 0.06799 & 0.06768 & \textbf{0.06196} \\
Med Men & 11 & 0.0828 & 0.08043 & 0.08265 & 0.06721 & 0.06654 & 0.06524 \\
Lat Men & 11 & 0.09097 & 0.08793 & 0.09225 & 0.05334 & 0.05439 & 0.05393 \\
Cyst & 5 & 0.07787 & 0.07586 & 0.07092 & 0.05993 & 0.06199 & 0.04969 \\
\hline
All & & 0.09228 & 0.091 & 0.08852 & 0.07085 & 0.07108 & 0.06417 \\
\hline\hline
\multicolumn{1}{c}{} \\[-2.2ex]
\multicolumn{7}{c}{(B)} \\
\end{tabular}}
\end{table}

\subsection{Real-world abnormality reconstruction: direct pixel evaluation}
To measure how well the annotated abnormalities were reconstructed, the bounding box regions in the reconstructed images were compared with those in the respective fully-sampled MRI from the Fast-MRI dataset. Normalized mean-squared error (NMSE) was the adopted metric, and was preferred over other metrics such as structural similarity index measure (SSIM) due to the variety of bounding box sizes (in number of pixels) that characterized the abnormality regions. Table~\ref{table: real_world_recon} shows that for 8$\times$ acceleration, the FNAF U-Net outperformed the U-Net baseline for the majority of the classes. This holds true for the total average value. Nevertheless, FNAF U-Net showed no improvements in the 4$\times$ setting. FNAF I-RIM showed improvements over I-RIM in the 4$\times$ setting but not the 8$\times$ setting. This shows that FNAF training may provide better abnormality reconstruction, but not consistently, as FNAF might still be too different from the real-world abnormalities. B-BOX training showed improvements over FNAF-robust training and the baselines in all but the U-Net 4$\times$ undersampling. This may be motivated by imperfect training and lack of abnormalities in the training set. Results were tested for statistical significance by means of a two-sided Wilcoxon rank-sum test ($p\leq0.05)$. As reported in Table~\ref{table: real_world_recon}, the use of I-RIM with robust training focused on real abnormalities (\textit{i.e.} B-Box I-RIM), significantly improved the reconstruction quality of all the lesions (bold values), but those located at the medial tibial cartilage and cyst (at 4$\times$ AF), lateral femoral cartilage, medial tibial cartilage, menisci and cyst (at 8$\times$ AF) when compared against the baseline U-Net and I-RIM based reconstructions. A complete summary of NMSE computed on every abnormality with the various reconstruction approaches is reported in Fig.~\ref{fig:boxplot-8x} included in the Appendix. 

\begin{figure}[!tbh]
 \centering
 \includegraphics[width=0.99\columnwidth]{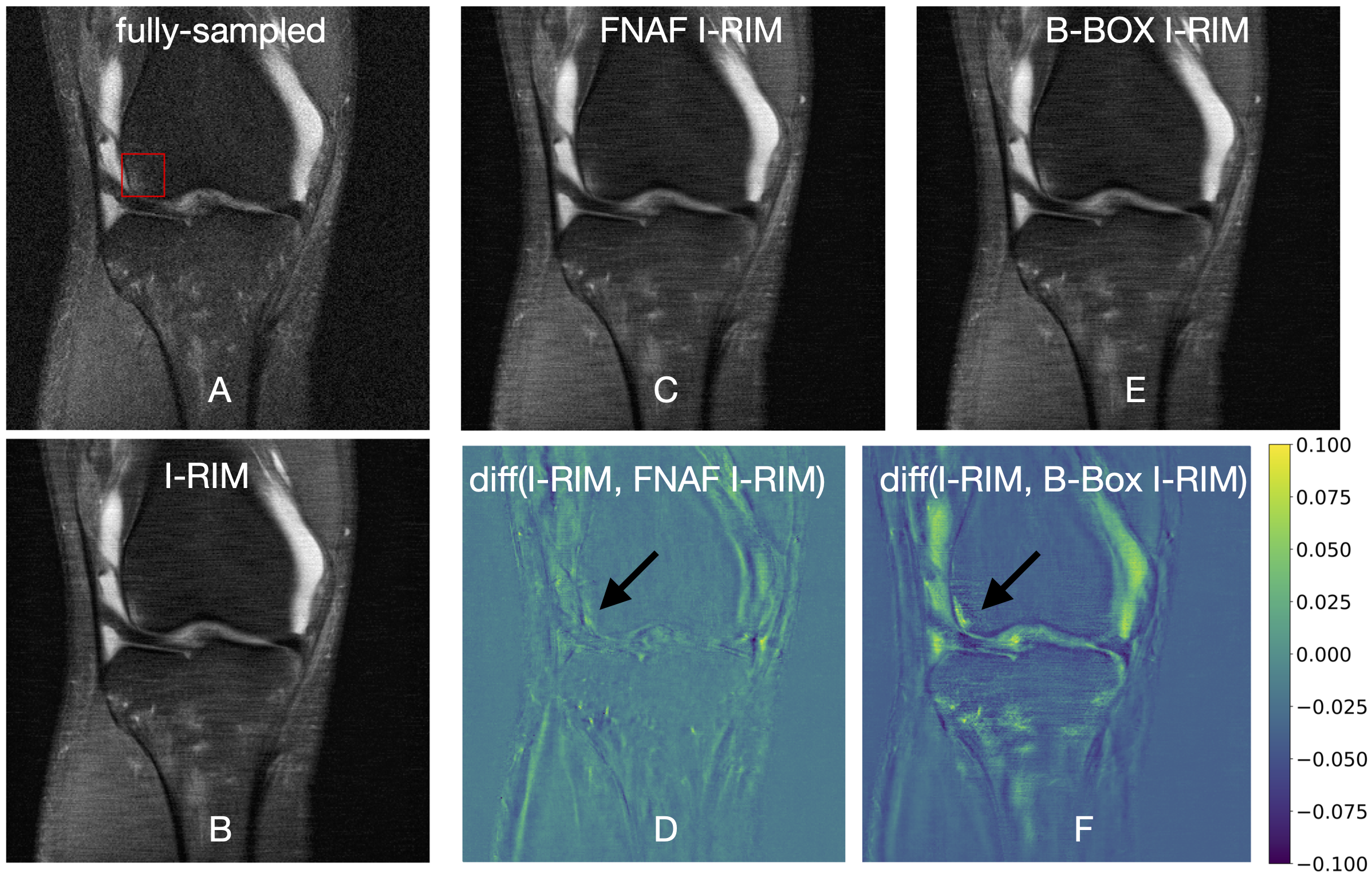}
 \caption{Example of a reconstructed 4$\times$ accelerated MRI. The abnormality within the bounding box in (A) was not perfectly reconstructed by the baseline I-RIM (B), and detected as a false negative feature by the abnormality classifier. The abnormality was better reconstructed after robust training with FNAF (C) and real abnormalities (E). Reconstruction differences between the images reconstructed with the baseline I-RIM and the two robust trained networks are shown in (D) and (F). The arrows point to regions where signal was better preserved upon reconstruction, which in turn might have helped the abnormality classifier. Baseline I-RIM is the trained model released by \cite{putzky2019rim}.}
 \label{fig:classification_comments_IRIM}
\end{figure}

\subsection{Measuring the presence of abnormalities in the reconstruction} \label{Measuring the presence of real-world abnormalities in the reconstruction}
With respect to the classification of abnormalities within the patches that were extracted from the reconstructed MRIs, a Squeezenet where only the final classification layer was fine-tuned showed superior performance compared to the other two training settings, namely fine-tuning of all layers and training from scratch. The results for the better performing classification model are reported in Table~\ref{table:feature_extractor_32x32} and Table~\ref{table:feature_extractor_64x64} where it is shown that the FNAF robust training approach for U-Net contributed to the enhancement in the reconstruction quality of clinically relevant features. This is represented by a reduction in the number of false negative features in the classification of real abnormalities. However, the classifications obtained on the U-Net and the robust trained versions disagree to the same amount (McNemar's test $p>0.05$). We inspected the prediction probabilities (normal/abnormal patch) on false negative cases in images reconstructed with U-Net and the proposed approach, and interestingly observed that for the network, which consumed 32$\times$32 patches, in 61\% of the cases the classifier's predictions were less confident when the reconstruction was performed using FNAF robust training as opposed to the baseline U-Net. We observed a similar behavior when analyzing the classification on patches from images reconstructed using I-RIM and FNAF I-RIM (54\% of the cases). Our interpretation for such decrease in classification confidence is that FNAF robust training proved an added value to the reconstruction. Furthermore, we suggest that this decrease in confidence is more prevalent on smaller patches as opposed to the 64$\times$64 patches, because the small abnormalities are mainly local features, which when analyzed on larger patches, they might be smoothed out. With respect to the experiment where robust training was done using real abnormalities, we observed higher sensitivity with B-Box U-Net, in contrast to B-Box I-RIM. A possible explanation of such result is that when robust training is implemented on synthetic features, these can inherently be unlimited as opposed to the real abnormalities, and this might cause slight overfitting on the real features. For this reason, a network with a smaller capacity like U-Net might better leverage robust training using real abnormalities.

\begin{table}[!tb]
\caption{Abnormality classification performance on patches of size $32\times32$. Metrics computed on fully-sampled images, (A) 4$\times$ and (B) $8\times$ undersampled images, which were reconstructed using the various methods studied in this paper. The reported metrics are Sensitivity (Sn), Specificity (Sp), F1 score, Kappa and Area Under the ROC curve (AUROC).}
\label{table:feature_extractor_32x32}
\centering
\begin{tabular}{ c| c | c | c | c |c }
\hline\hline
\multicolumn{6}{c}{AF=4$\times$}\\
Input & Sn & Sp & F1 score & Kappa & AUROC\\
\hline
Fully-sampled & 86.12 & 65.90 & 78.25& 52.04& 86.24\\
\hline
U-Net & 89.74 & 62.93 & 79.17 & 52.69& 86.24\\
FNAF U-Net & 90.23 & 63.51 & 79.64 & 53.76 & 86.37\\
B-Box U-Net & 91.13 & 62.11 & 79.63 & 53.27 & 86.71\\
\hline
I-RIM & 94.50 & 52.31 & 78.09 & 46.84 & 85.70\\
FNAF I-RIM & 94.09 & 54.20 & 78.49 & 48.32 & 85.79\\
B-Box I-RIM & 91.38 & 55.77 & 77.62 & 47.17 & 84.89\\
\hline\hline
\multicolumn{1}{c}{} \\[-2.2ex]
\multicolumn{6}{c}{(A)} \\
\multicolumn{1}{c}{}\\[-2.2ex]
\hline\hline
\multicolumn{6}{c}{AF=8$\times$}\\
Input & Sn & Sp & F1 score & Kappa & AUROC\\
\hline
Fully-sampled & 86.12 & 65.90 & 78.25& 52.04& 86.24\\
\hline
U-Net & 85.30 & 68.95 & 78.89 & 54.26 & 86.04\\
FNAF U-Net & 85.71 & 69.19 & 79.21 & 54.92 & 86.28\\
B-Box U-Net & 87.11 & 68.78 & 79.83 & 55.91 & 87.04\\
\hline
I-RIM & 92.28 & 55.93 & 78.14 & 48.24 & 85.43\\
FNAF I-RIM & 91.71 & 56.67 & 78.08 & 48.41 & 85.34\\
B-Box I-RIM & 87.03 & 59.06 & 76.40 & 46.11 & 84.20 \\
\hline\hline
\multicolumn{1}{c}{} \\[-2.2ex]
\multicolumn{6}{c}{(B)} \\
\multicolumn{1}{c}{}\\[-2.2ex]
\end{tabular}
\end{table}

\begin{table}[!tb]
\caption{Abnormality classification performance on patches of size $64\times64$. Metrics computed on fully-sampled images, (A) 4$\times$ and (B) $8\times$ undersampled images, which were reconstructed using the various methods studied in this paper.}
\label{table:feature_extractor_64x64}
\centering
\begin{tabular}{ c| c | c | c | c |c }
\hline\hline
\multicolumn{6}{c}{AF=4$\times$}\\
Input & Sn & Sp & F1 score & Kappa & AUROC\\
\hline
Fully-sampled & 78.43 & 80.47 & 79.31 & 58.89 & 87.93\\
\hline
U-Net & 69.87 & 83.12& 74.90 & 52.62 & 86.51 \\
FNAF U-Net & 70.65 & 83.32 & 75.49 & 53.93 & 86.68\\
B-Box U-Net & 73.08 & 82.73 & 76.85 & 55.78 & 87.00 \\
\hline
I-RIM & 81.54 & 76.45 & 79.60 & 58.00 & 86.58 \\
FNAF I-RIM & 81.44 & 76.45 & 79.54 & 57.90 & 86.63 \\
B-Box I-RIM & 78.91 & 78.31 & 78.61 & 57.22 & 86.49 \\
\hline\hline
\multicolumn{1}{c}{} \\[-2.2ex]
\multicolumn{6}{c}{(A)} \\
\multicolumn{1}{c}{}\\[-2.2ex]
\hline\hline
\multicolumn{6}{c}{AF=8$\times$}\\
Input & Sn & Sp & F1 score & Kappa & AUROC\\
\hline
Fully-sampled & 78.43 & 80.47 & 79.31 & 58.89 & 87.93\\
\hline
U-Net & 60.06 & 88.62 & 70.11 & 48.61 & 86.10\\
FNAF U-Net & 62.00 & 88.32 & 71.44 & 50.26 & 86.36 \\
B-Box U-Net &67.93 & 84.79 & 74.24 & 52.68 & 86.85 \\ 
\hline
I-RIM & 75.61 & 78.61 & 76.84 & 54.21 & 85.21 \\
FNAF I-RIM & 74.44 & 78.70 & 76.14 & 53.13 & 85.31 \\
B-Box I-RIM & 72.01 & 81.45 & 75.65 & 53.44 & 85.27 \\
\hline\hline
\multicolumn{1}{c}{} \\[-2.2ex]
\multicolumn{6}{c}{(B)} \\
\end{tabular}
\end{table}

It is worth mentioning that a drop in classification performance was expected when the input image was DL reconstructed. However, we assumed that such drop in performance would mainly be associated to a covariate shift which comes from subtle changes in the pixel intensity values caused by the image reconstruction. With that in mind, we assumed that the network with the smallest performance drop (meaning that more diagnostic features are still identifiable) was the network that more reliably reconstructed fine details in the MRI. This is hard to prove, mainly because errors propagate from the under-sampling and zero-filling in \textit{k-}space to the reconstruction error, up to the actual classification error per se, which might in turn not exclusively depend on errors in the reconstruction.

Overall, we suggest that further reconstruction improvements could be obtained by reducing the semantic difference between FNAF and real-world abnormalities, although it certainly requires further investigation. Ideally, we want to construct the space of FNAF to be representative of not only the size but also the semantics of real-world abnormalities. Possible direction for improving FNAF and making them more realistic include: \textit{i)} relaxing the pixel constraint more so that the FNAF space can include real-world abnormalities; \textit{ii)} modeling the abnormality features by introducing domain knowledge. In a recently published work, \cite{chen2020realistic} suggest to introduce domain knowledge in adversarial attacks to improve the reconstruction performance and boost performance in downstream tasks. 

Examples of the effect of robust training using FNAF as well as the real abnormalities on U-Net and I-RIM reconstruction networks are displayed in Fig.~\ref{fig:classification_comments_UNET} and Fig.~\ref{fig:classification_comments_IRIM} respectively. In Fig.~\ref{fig:classification_comments_UNET}, a coronal section of knee MR (fully-sampled MRI) from the validation set is depicted in (A). This is an MRI from the Fast-MRI dataset which was utilized as image reconstruction ground truth, as well as an input in the validation of the abnormality classification network. In (A), the bounding box highlights the presence of a bone marrow lesion at the lateral edge of the femur, based on the radiologist's manual annotation. From an accurate analysis, it can be observed that the bounding box includes a portion of the adjacent joint fluid with high signal to test for confounding. In the reported experiment, prior to reconstruction, MRIs were undersampled at a $4\times$ AF. In (B), the reconstruction result by using a baseline U-Net is displayed. This shows that U-Net recreated parts of the BML features in the corresponding region of the femur. Despite that, the lesion was not detected by our lesion classification model, making this a false negative case. In (C), the resulting reconstructed image with the proposed robust training approach appears to have better reconstructed BML features in the corresponding region and, in turn, it is identified by the lesion detection model, correctly making it a true positive case. In (D), a difference map between U-Net and FNAF reconstructions is shown. Radiologist's opinion propounds that FNAF robust training allowed for a finer reconstruction of particular features, which better mimic the ground truth and is crucial for a precise clinical interpretation.
In (E), the reconstruction is obtained post-robust training with real abnormalities (\textit{i.e.} B-Box U-Net reconstruction); the associated difference map is shown in (F). It is clear that when using real features, the network proved more sensitive to finer anatomical structures. Similar consideration can be done about the reconstruction obtained with I-RIM and robust trained versions, as shown in Fig.~\ref{fig:classification_comments_IRIM}. The difference map in Fig.~\ref{fig:classification_comments_IRIM}-(F), between an image reconstructed with I-RIM and B-Box I-RIM further credits the hypothesis that more realistic FNAF would enhance the reconstruction quality as demonstrated by the more prominent difference observed in bone marrow lesion. This would allow radiologists to make accurate diagnoses and avoid missing out on small details, often of utmost importance for image interpretation.

To summarize, our experiments showed that by training on our imperfect FNAF, one can force convolution filters to be more sensitive to small features, and this would be especially powerful if FNAF were more realistic. This was supported by the significant reconstruction improvements which were observed when robust training was implemented using the abnormality bounding boxes. We speculate that this is indicative of a promising direction to move forward to improve clinical robustness.

\section{Limitations}
We showed that the introduction of small abnormalities has potential to improve MRI reconstruction using DL techniques; however, this was only shown on a single-coil reconstruction task, based on emulated single-coil \textit{k-}space data. A multi-channel reconstruction approach would better suit clinical applications. In practice, the current formulation involved the introduction of FNAF in the image space, which forced us to treat the reconstruction as a single-coil problem. Further studies are needed to experimentally prove the effect of this method on high resolution 3D sequences, different anatomies, and scanners from multiple vendors. It will also be necessary to introduce FNAF directly in \textit{k-}space, to make it suitable to multi-channel reconstruction. The current formulation of the method requires knowledge of the forward function of MRI reconstruction, which is not explicitly available in the single-coil formulation. Further efforts to simulate the forward function or other approaches are required. Robust adversarial training is only empirically effective, which might explain the limited improvements over the baseline for real-world abnormalities. Theoretically robust methods might help resolve this issue. We believe that the evaluation metrics for abnormality reconstruction are still imperfect when compared to radiologist evaluations. Better metrics that capture clinical relevance are certainly needed. Ultimately, regarding the abnormality classification framework, we assumed that a drop in performance was associated to a covariate shift resulting from different image reconstruction techniques; it will be interesting to analyze whether adversarial domain adaptation techniques would mitigate the drop in performance we observed.

\section{Conclusions}
The connection between FNAF to real-world abnormalities is analogous to the connection between lp-bounded adversarial perturbations and real-world natural images. In the natural images sampled by non-adversaries, lp-bounded perturbations most likely do not exist. But their existence in the pixel space goes beyond security, as they reveal a fundamental difference between deep learning and human vision \citep{bug_feature}. Lp-bounded perturbations violate the human prior: humans see perturbed and original images the same. FNAF violate the reconstruction prior: an algorithm should recover (although it may be impossible) all features. We relaxed this prior to only small features, which often are the most clinically relevant. Therefore, the failure of deep learning reconstruction models to reconstruct FNAF is important even if FNAF might not be representative of the real-world abnormalities. Lp-bounded perturbations inspired works that generate more realistic attacks, and we hope to bring the same interest in the domain of MRI reconstruction. 

This work expanded \cite{cheng2020addressing} to show the possibility of translating adversarial robust training with FNAF to real world abnormalities.
For this, an MSK radiologist annotated the Fast-MRI dataset by placing bounding boxes on top of abnormalities observed in the menisci, cartilage, and bone in knee MRIs. Subsequently, a fully convolutional neural network was employed to assess whether the abnormalities had been reconstructed by deep learning reconstruction methods. This was addressed as a patch-based image classification problem, and demonstrated that our robust training approach contributes to a reduction in the number of false negative features.

We investigated two hypotheses for the false negative problem in deep-learning-based MRI reconstruction. By developing the FNAF adversarial robustness framework, we show that this problem is difficult, but not impossible. Within this framework, there is potential to bring the extensive theoretical and empirical ideas from the adversarial robustness community, especially in the area of provable defenses \citep{wong2018provable, mirman2018differentiable, raghunathan2018certified, balunovic2020adversarialgap} to tackle the problem. 
We suggest that our work gives further credit for the necessity of new loss functions or metrics to be adopted in image reconstruction, which would be more aware of finer details, as per the clinically relevant small features. Furthermore, we hope this will be a starting point for more research in the field, so that a more realistic search space for the FNAF could be identified. This would certainly enhance the generalization capabilities of FNAF to real-world abnormalities. We made the bounding box annotations publicly available at \url{https://github.com/fcaliva/fastMRI\_BB\_abnormalities\_annotation} so that they can serve as a validation set for future bodies of work.


\acks{This work was supported by the NIH/NIAMS R00AR070902 grant. \\We would like to thank Fabio De Sousa Ribeiro (University of Lincoln) for making a distributed Pytorch boilerplate implementation publicly available at \url{https://github.com/fabio-deep/Distributed-Pytorch-Boilerplate}. This was adapted to our abnormality classification problem. We would like to thank Radhika Tibrewala (New York University) for the support with exporting the annotation; Sharmila Majumdar and Madeline Hess (UCSF) for the fruitful discussions. Ultimately, we would like to thank the Reviewers and Editors for their thoughtful comments and effort which were instrumental to improve our manuscript.}

%
\ethics{The work follows appropriate ethical standards in conducting research and writing the manuscript, following all applicable laws and regulations regarding treatment of animals or human subjects.}

\coi{We declare that we do not have conflicts of interest.}

\bibliography{sample}

\newpage
\appendix 
\section*{Appendix A}
\subsection*{Correlation Analysis}
\begin{figure}[!hb]
 \centering
 \includegraphics[width=0.24\columnwidth]{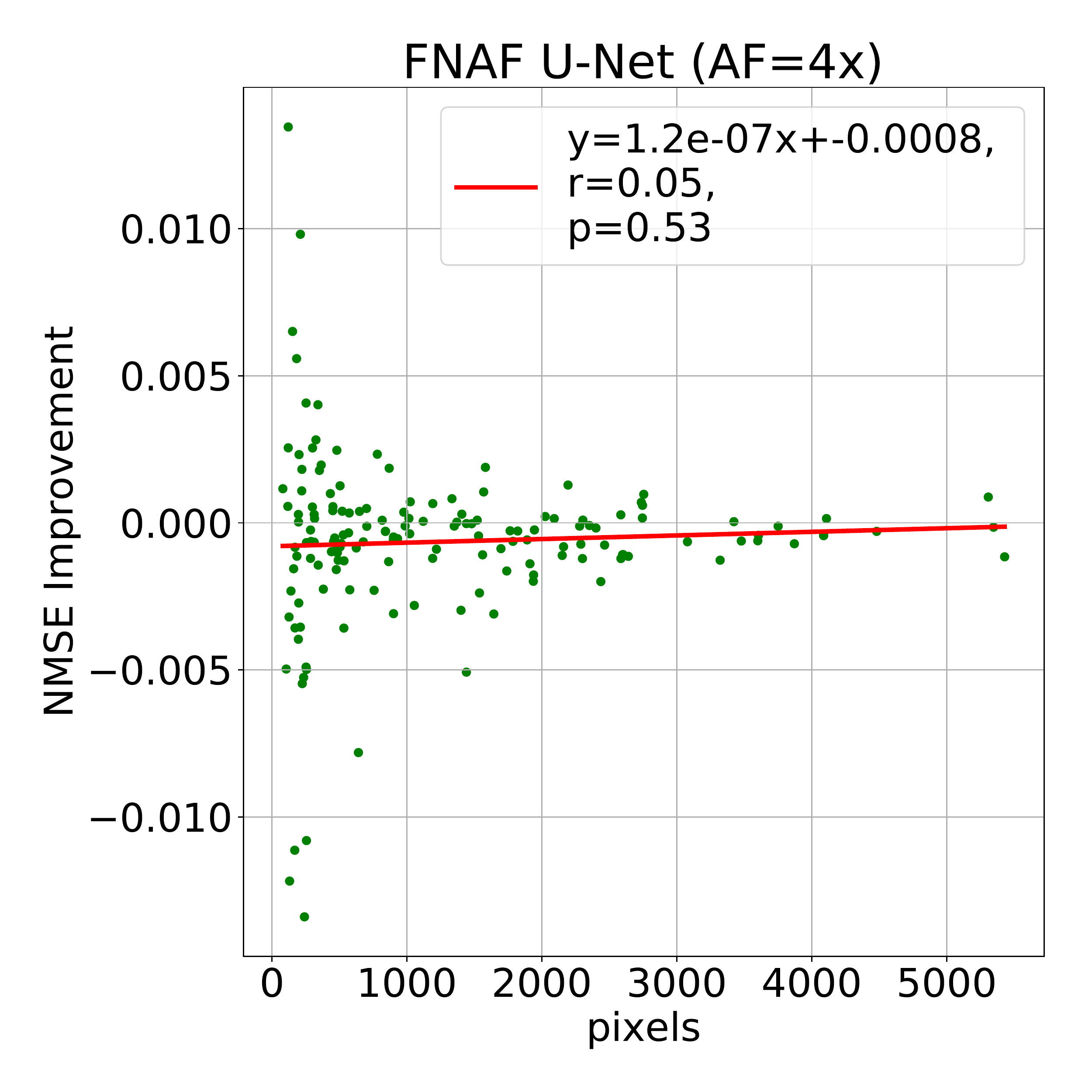}
 \includegraphics[width=0.24\columnwidth]{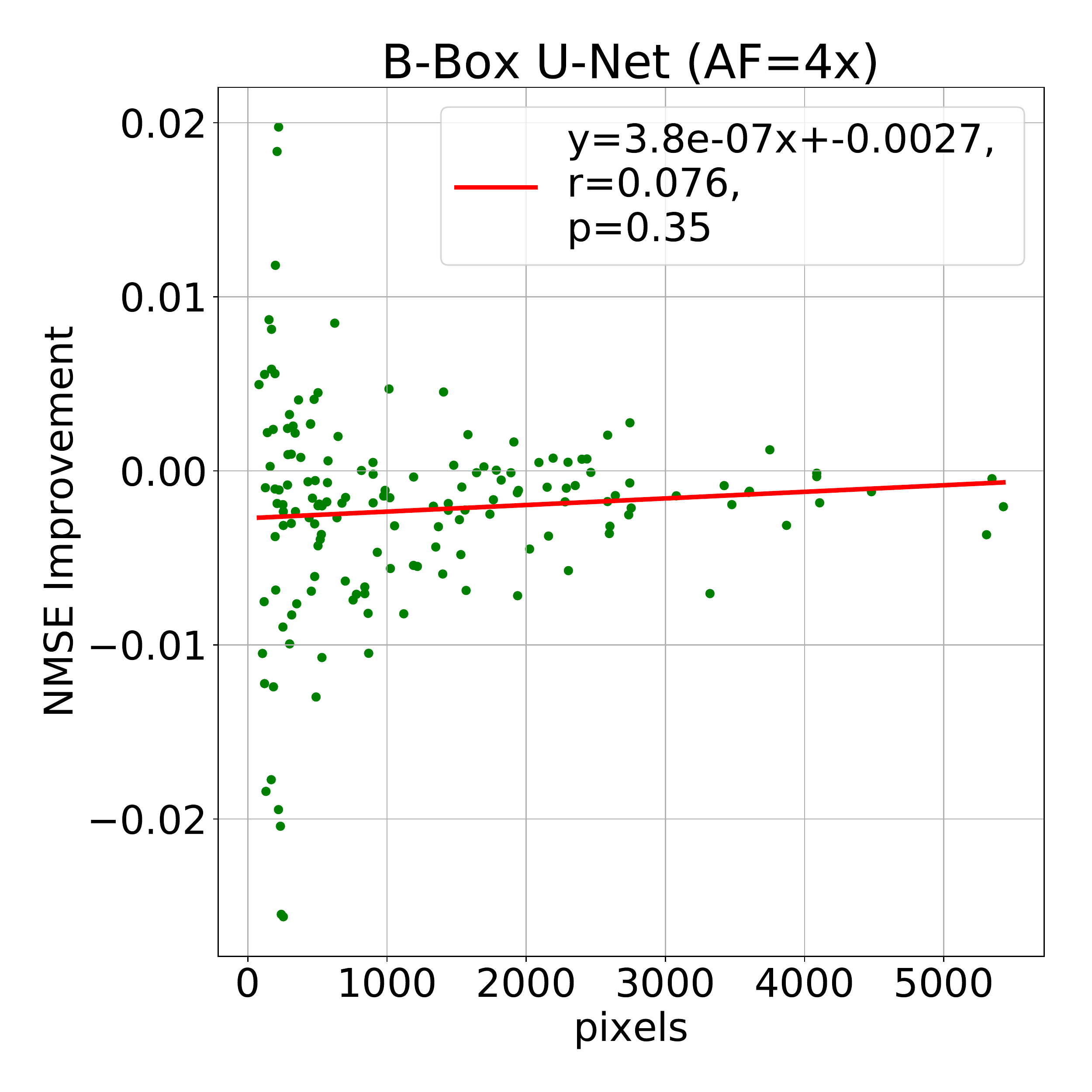}
  \includegraphics[width=0.24\columnwidth]{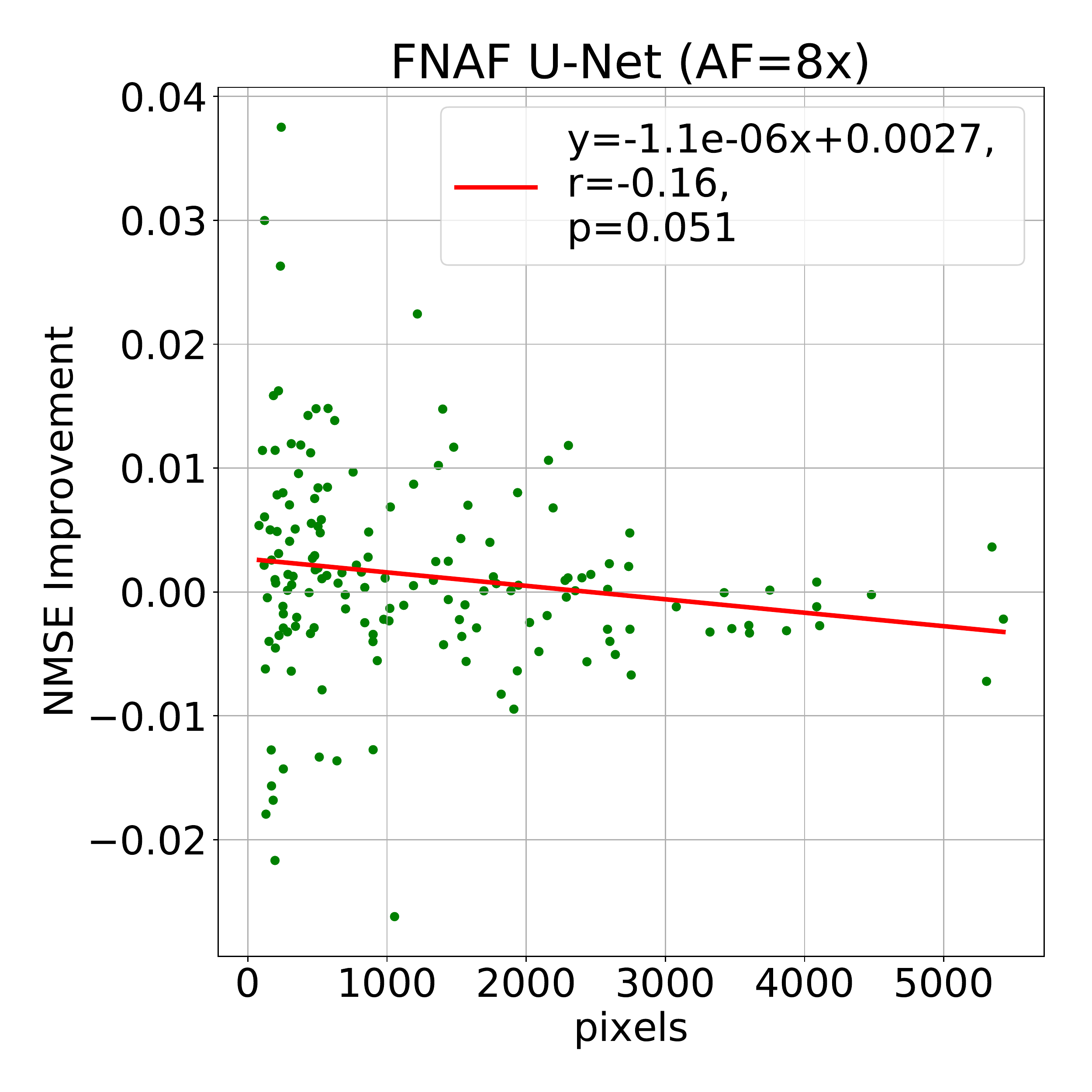}
 \includegraphics[width=0.24\columnwidth]{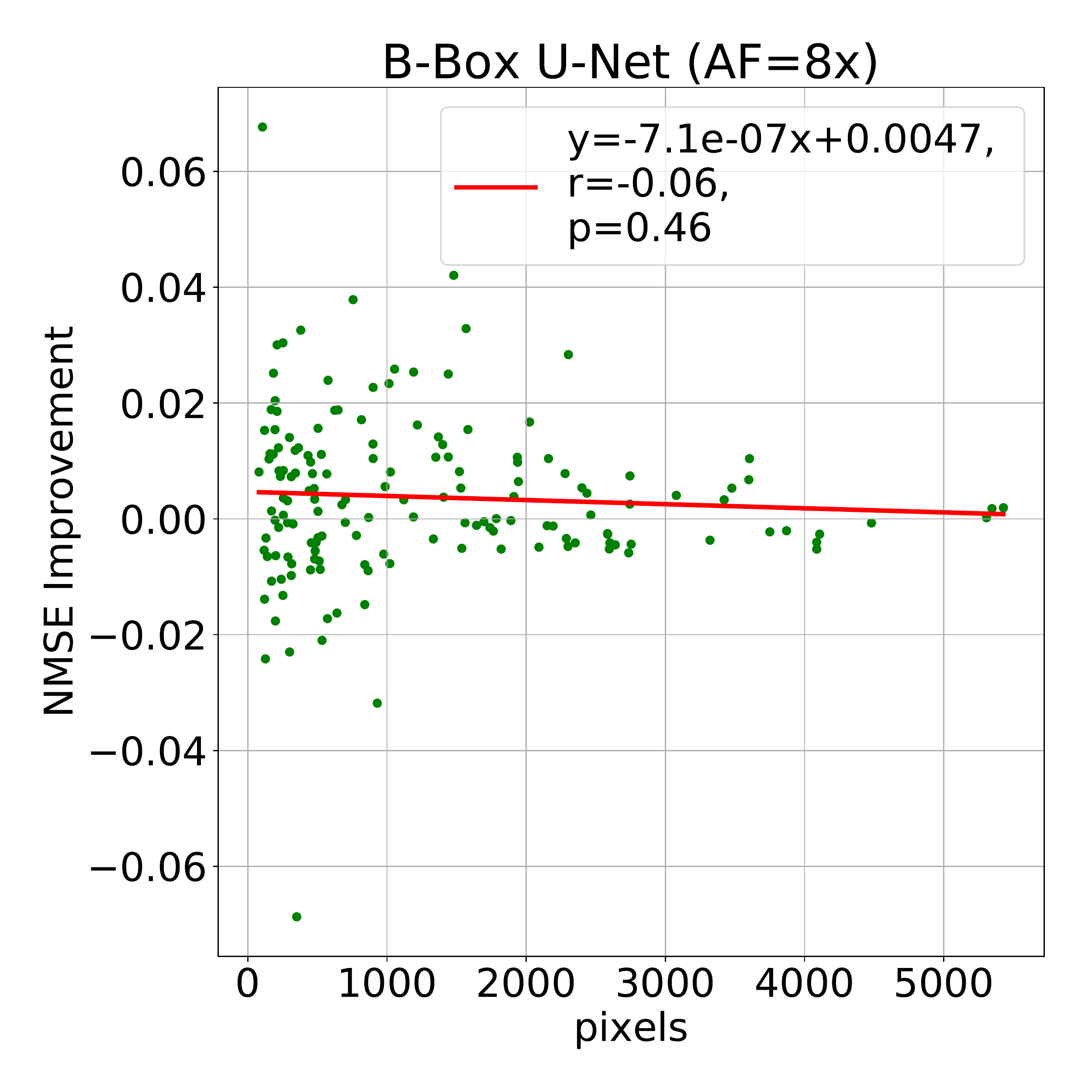}
 \includegraphics[width=0.24\columnwidth]{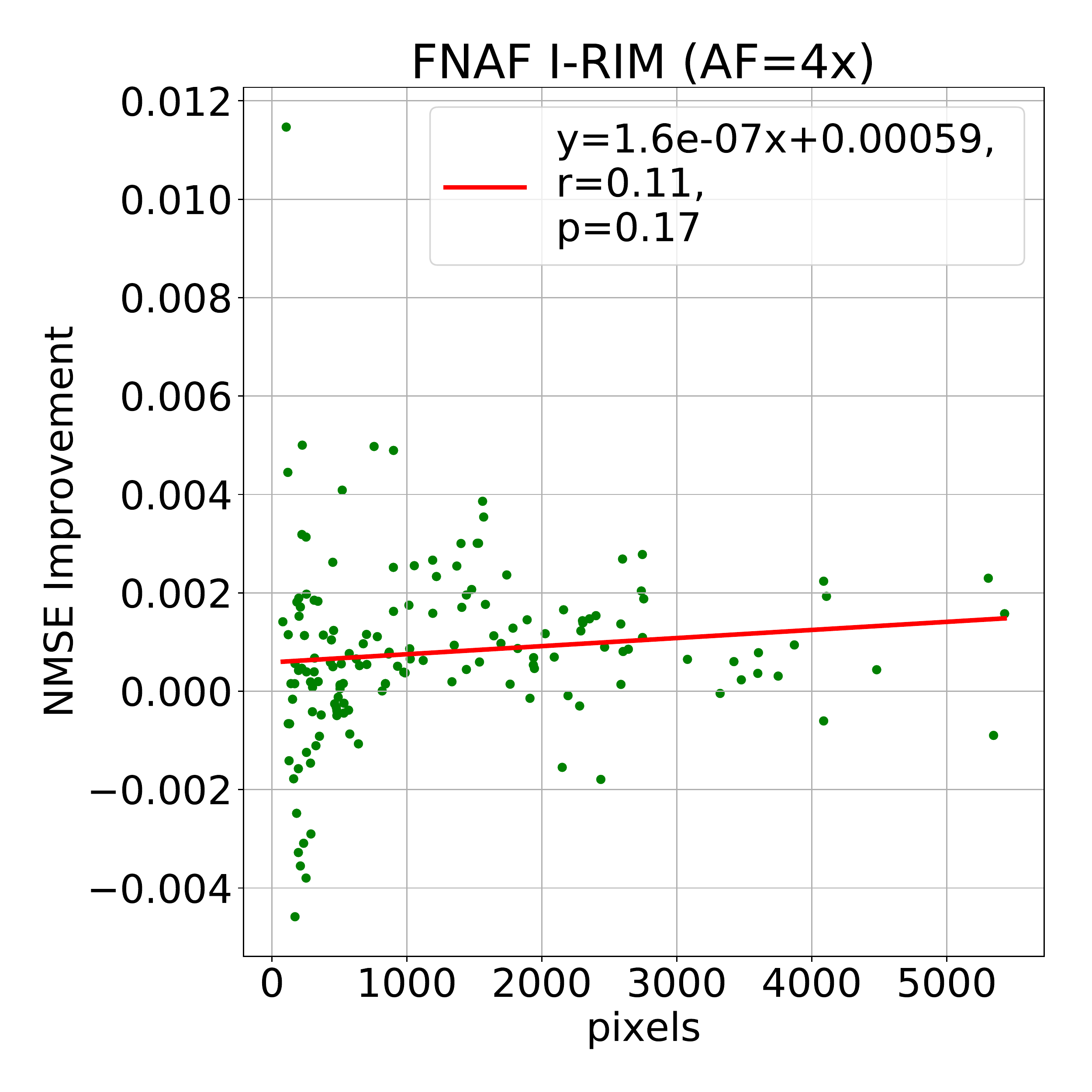}
 \includegraphics[width=0.24\columnwidth]{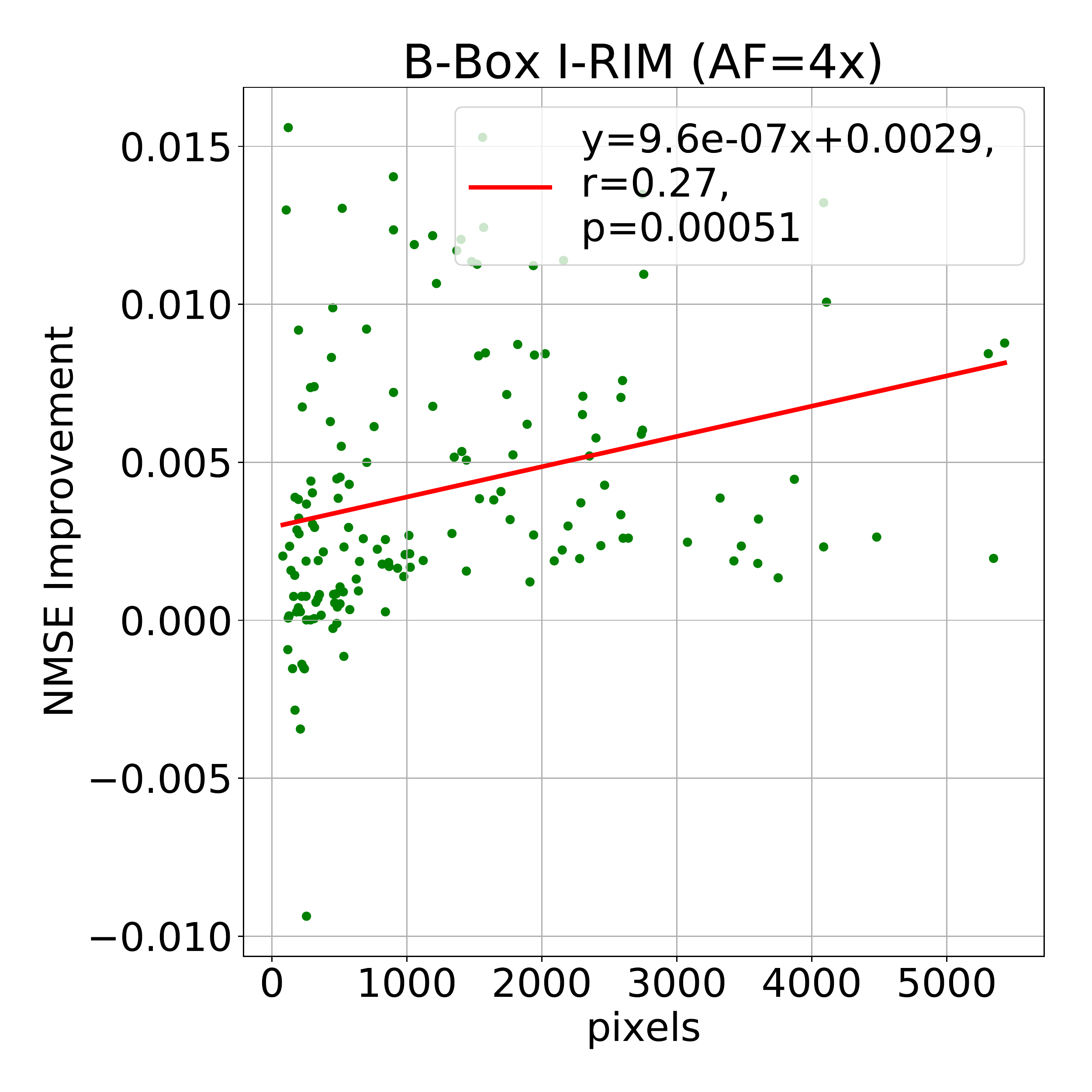}
  \includegraphics[width=0.24\columnwidth]{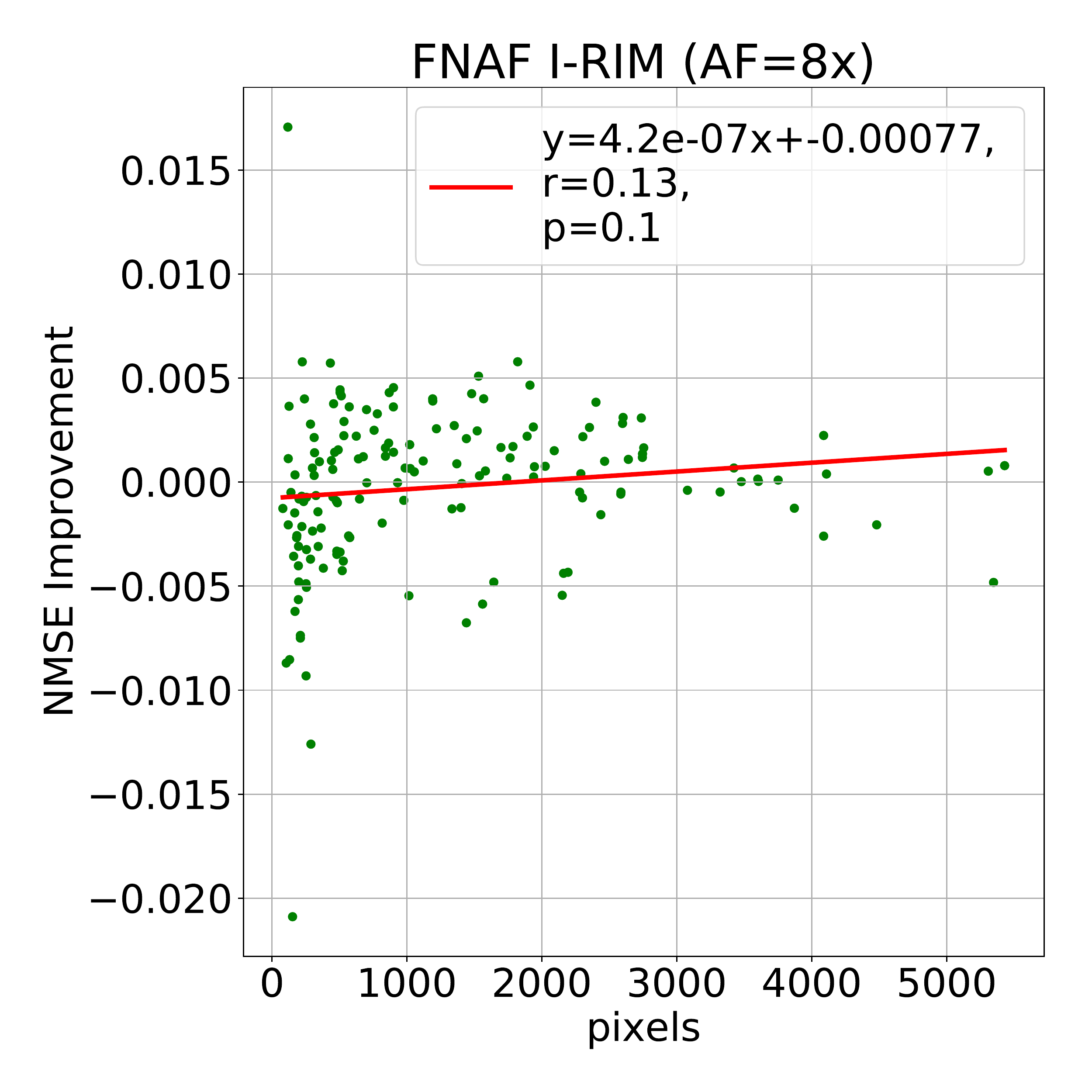}
 \includegraphics[width=0.24\columnwidth]{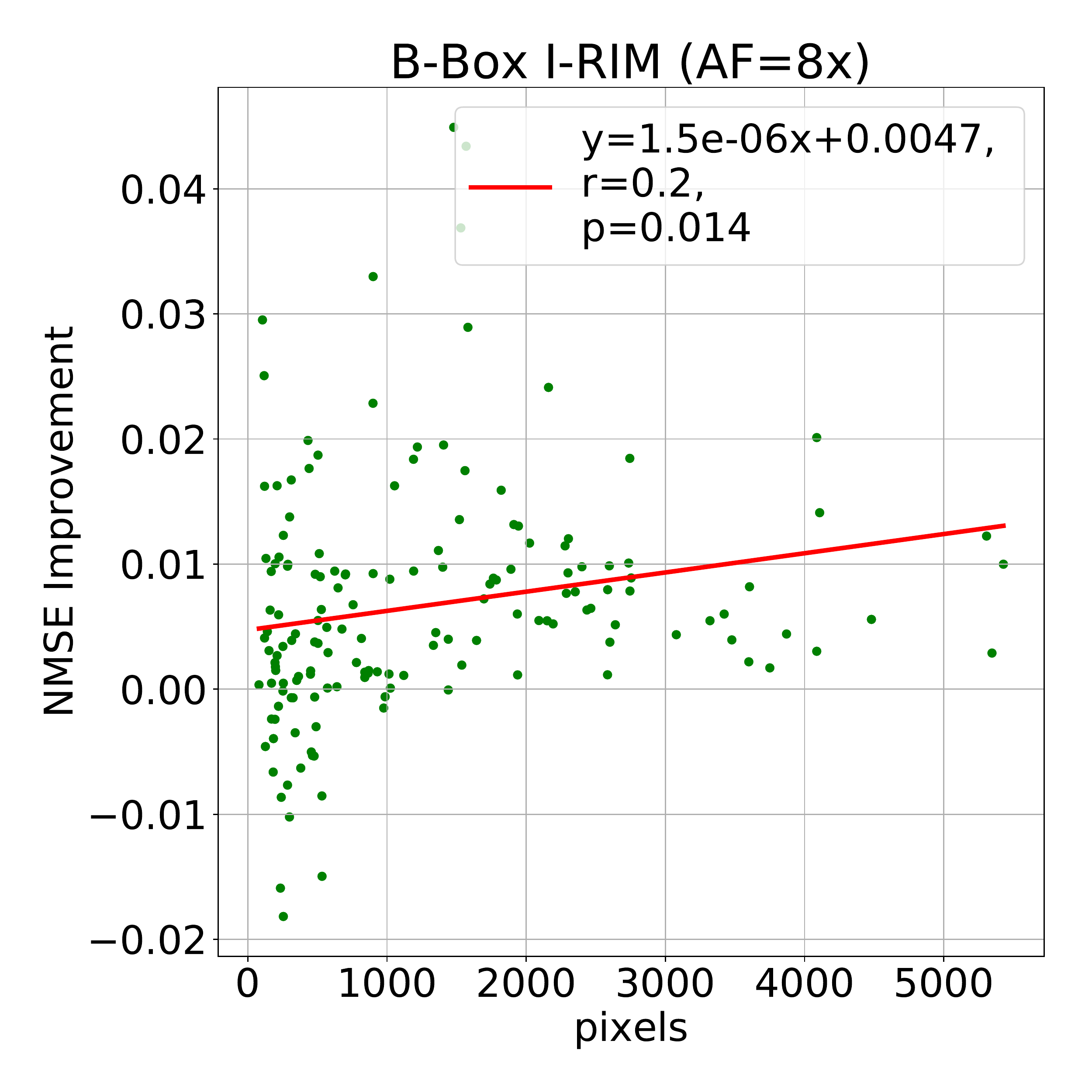}
 \caption{Correlation analysis between NMSE reconstruction improvements and the size of abnormalities (measured in terms number of occupied pixels)- when implementing robust training with FNAF or real abnormalities over the baseline models. Top: FNAF U-Net and B-Box U-Net improvements over the baseline U-Net. Bottom:
 FNAF I-RIM and B-Box I-RIM improvements over the baseline I-RIM. Input MRIs undersampled with a 4$\times$ acceleration factor (Column 1 and 2) and a 8$\times$ acceleration factor (Column 3 and 4).}
\label{fig:improvements_UNET}
\end{figure}

\begin{figure}[!tb]
 \centering
 \includegraphics[width=0.30\columnwidth]{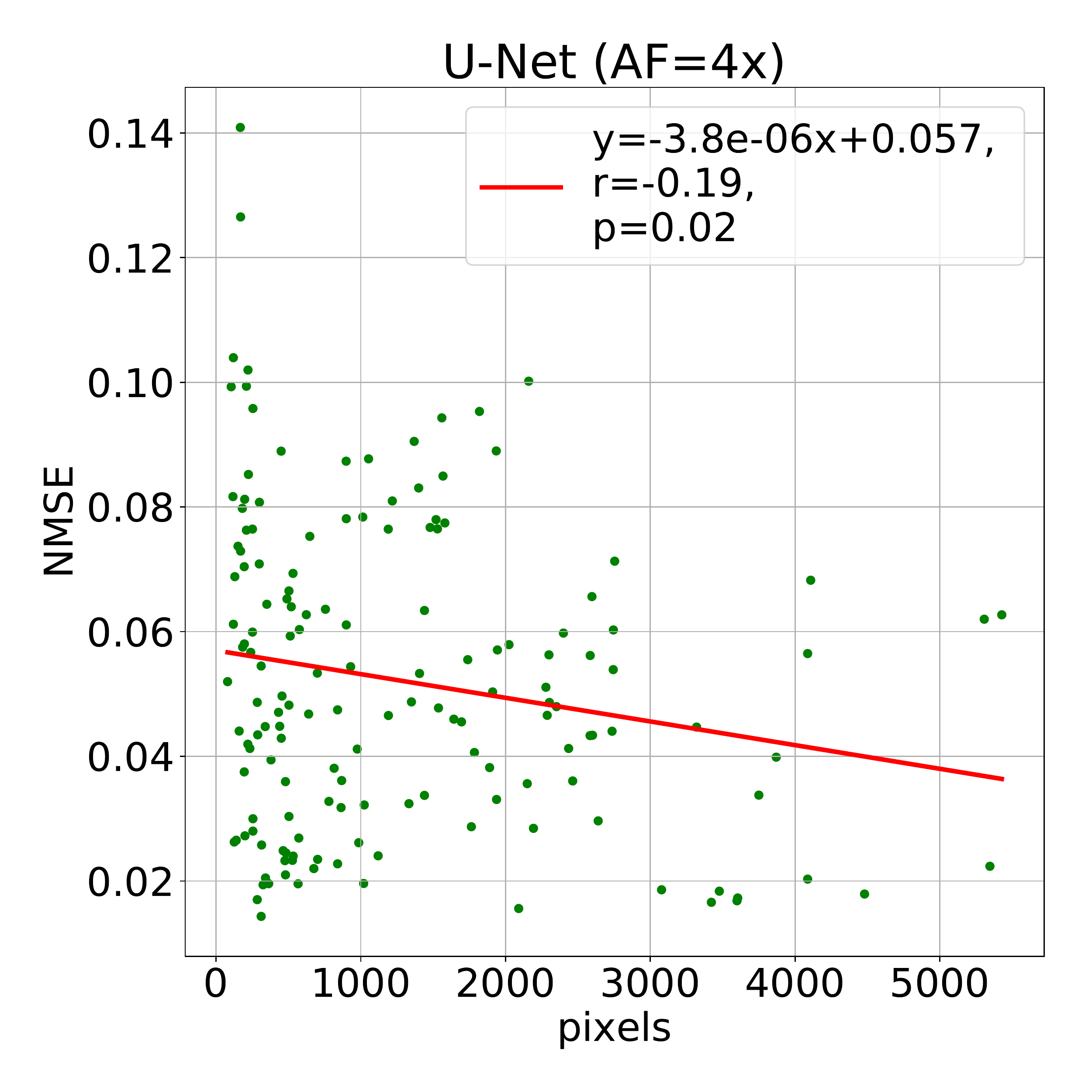} 
 \includegraphics[width=0.30\columnwidth]{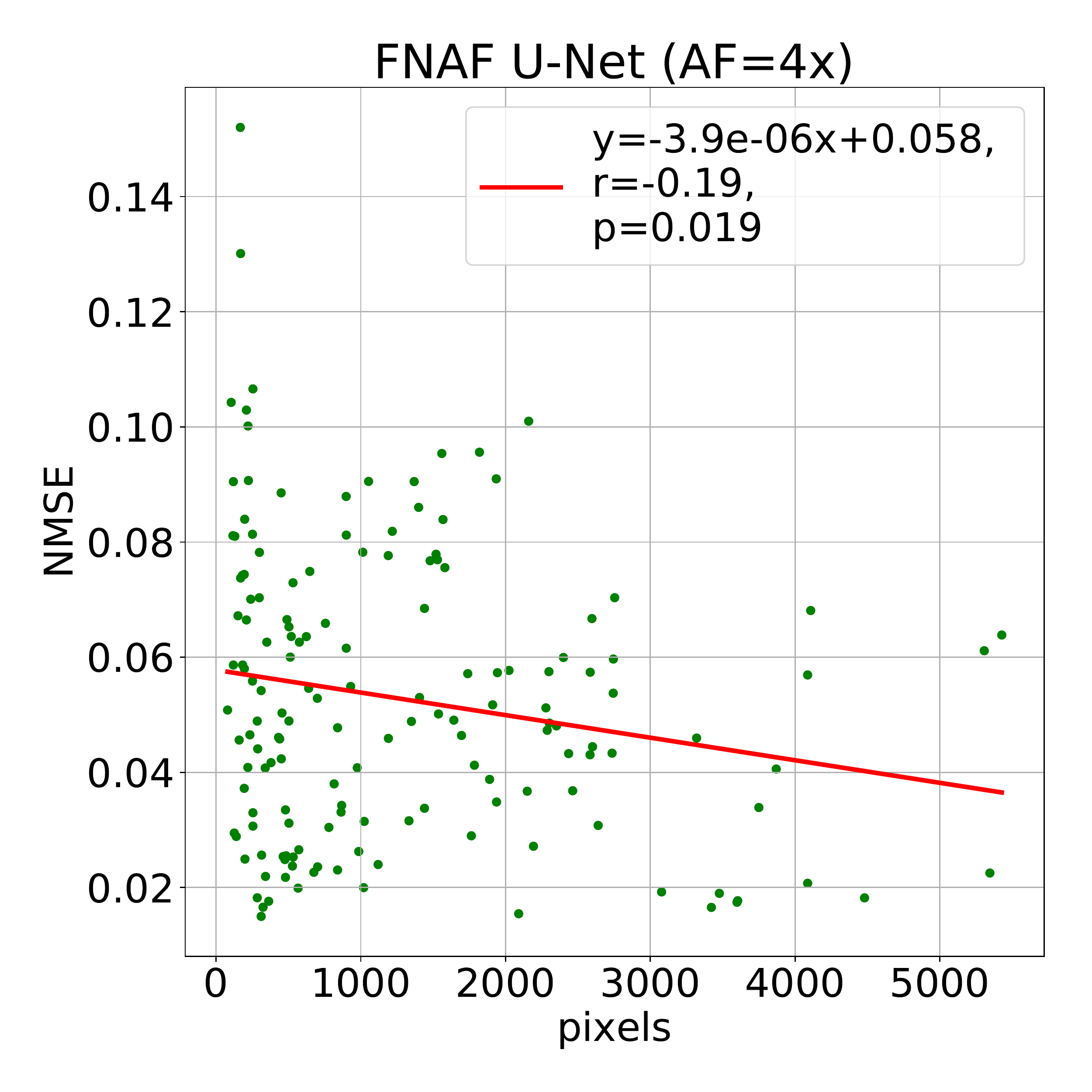} 
 \includegraphics[width=0.30\columnwidth]{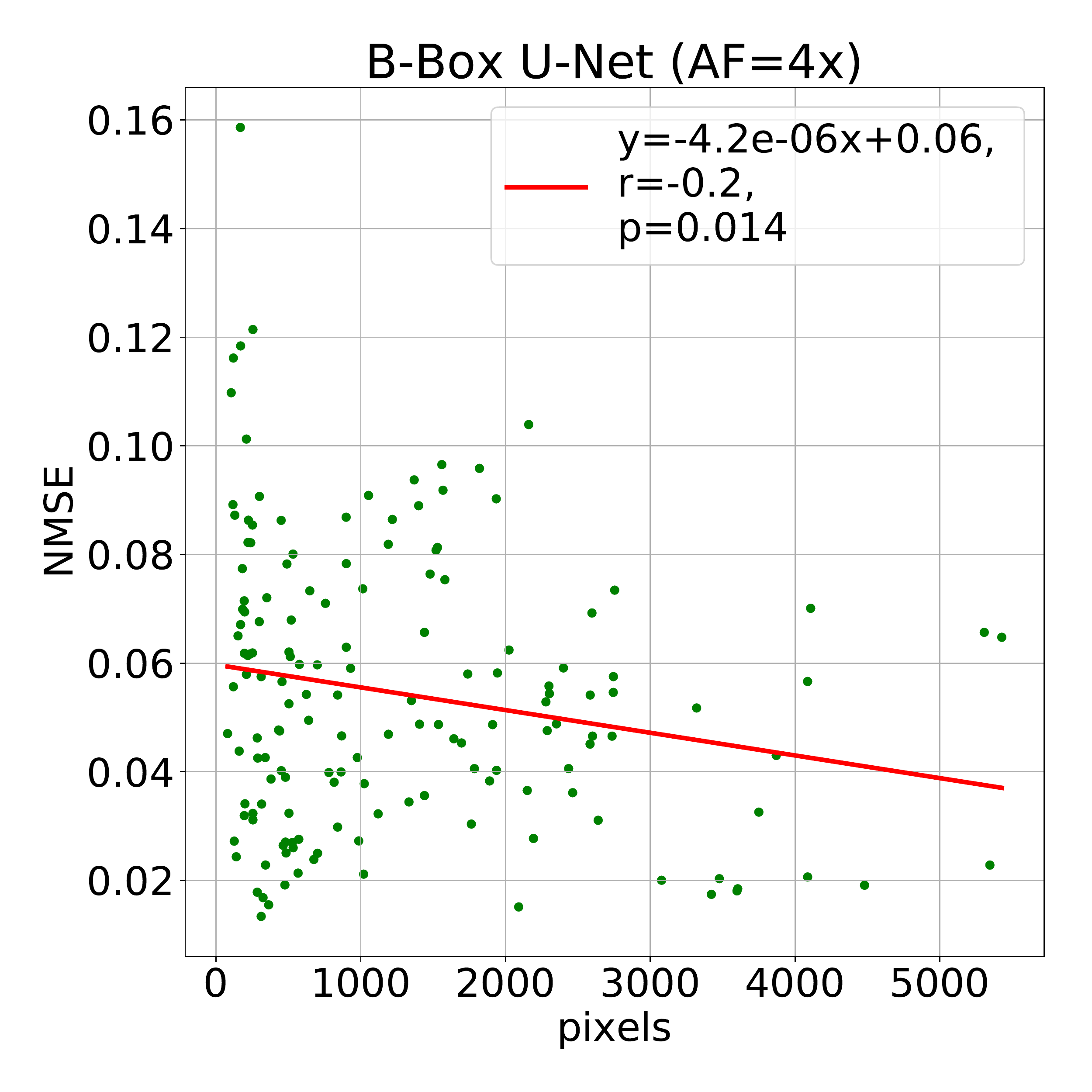}

 \includegraphics[width=0.30\columnwidth]{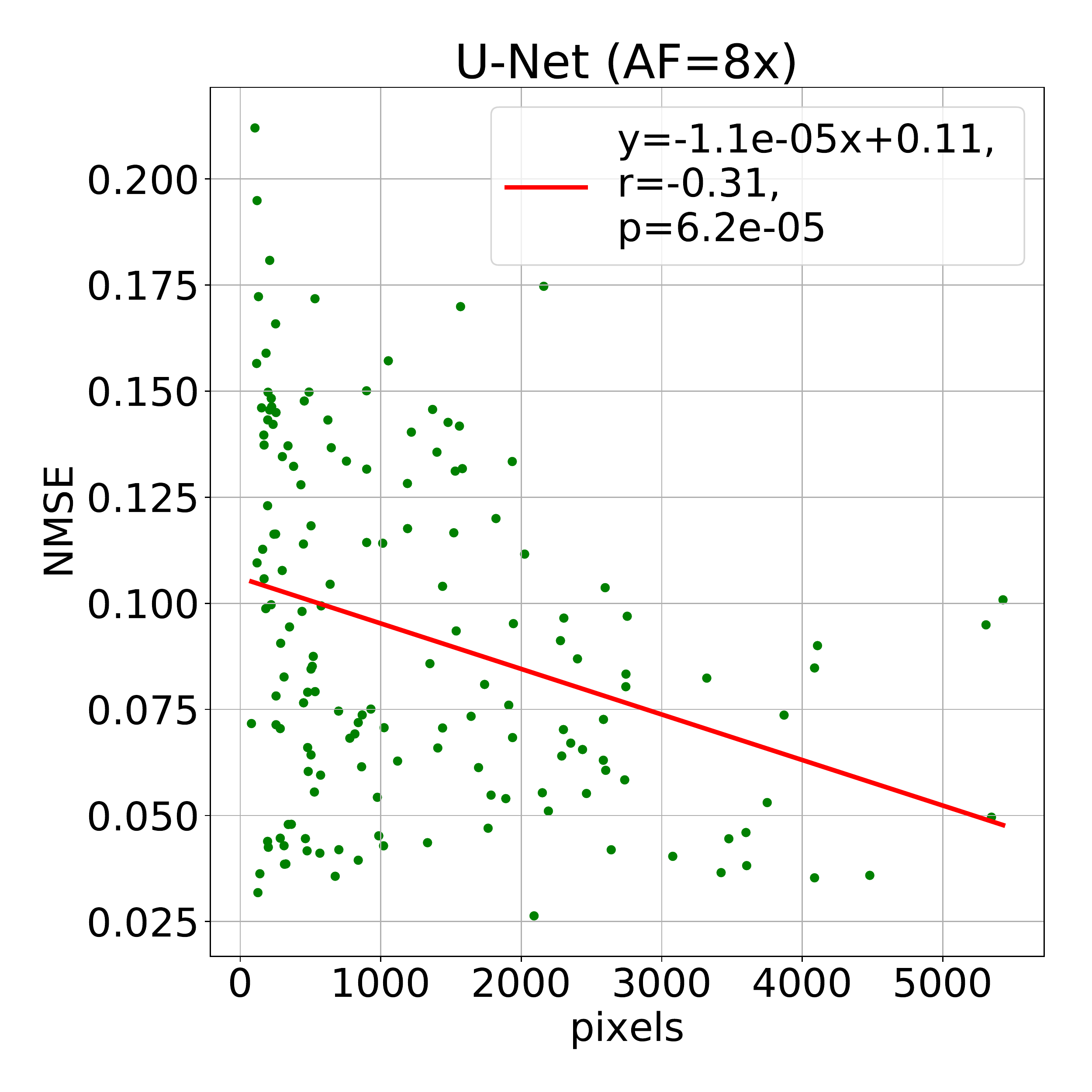} 
 \includegraphics[width=0.30\columnwidth]{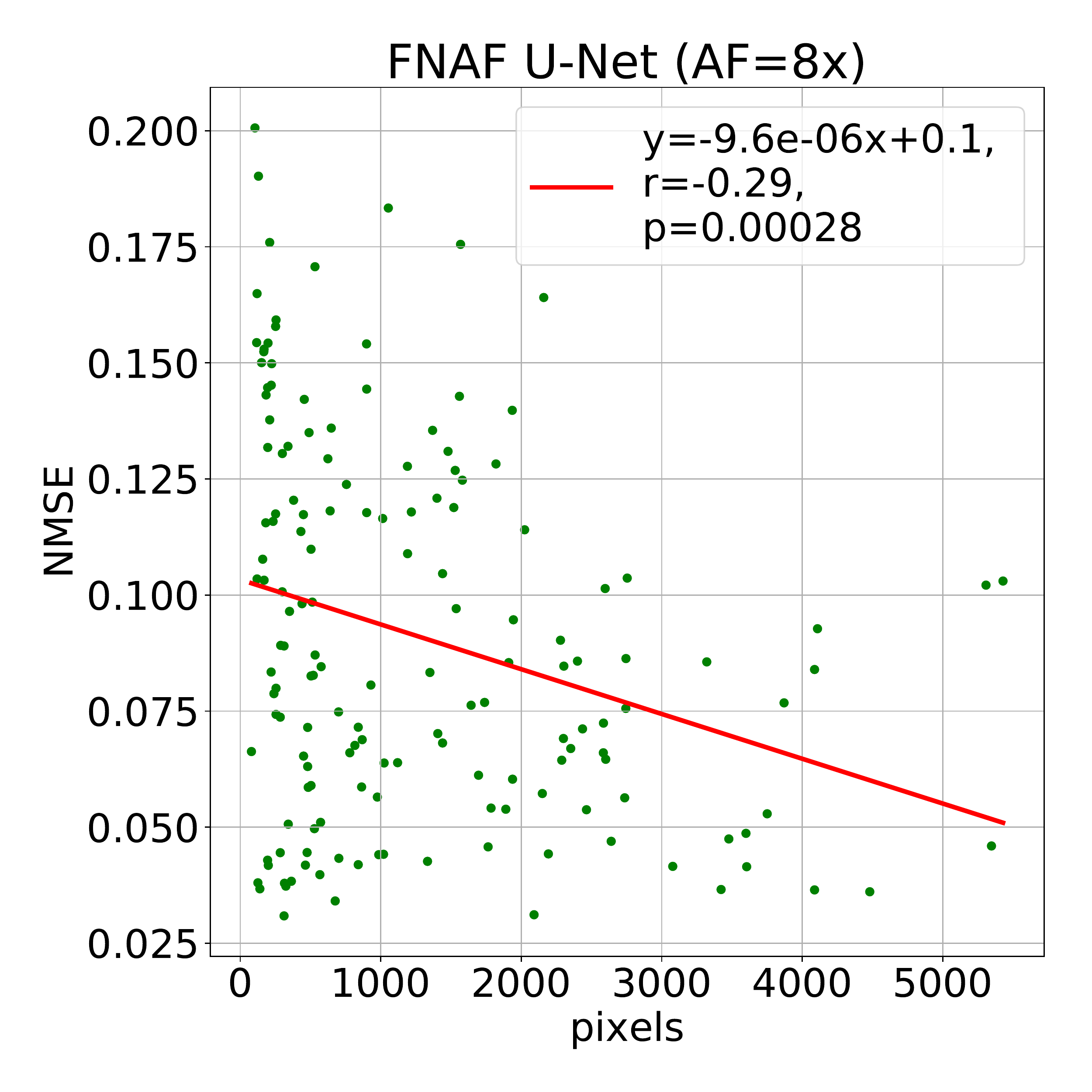} 
 \includegraphics[width=0.30\columnwidth]{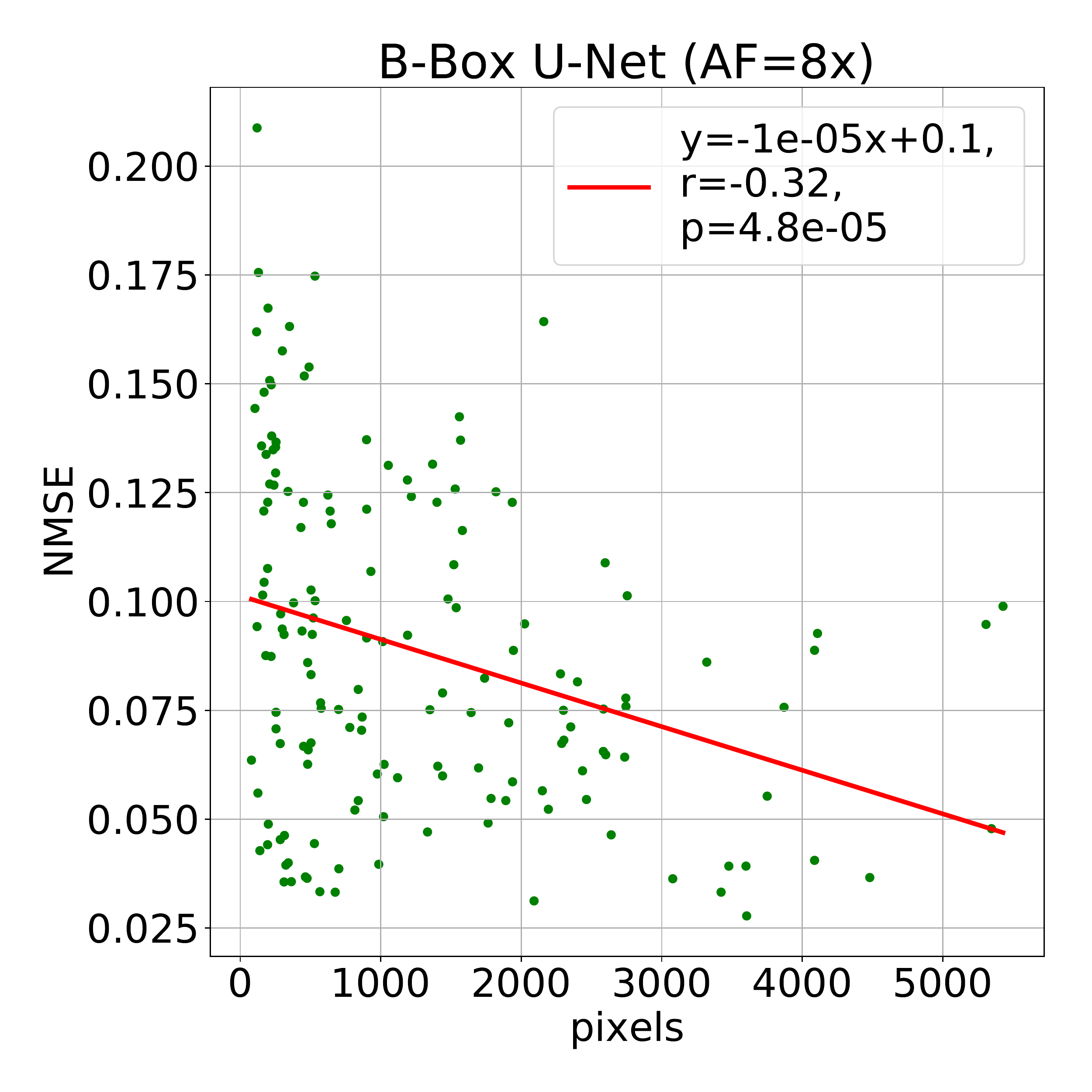} 

 \includegraphics[width=0.30\columnwidth]{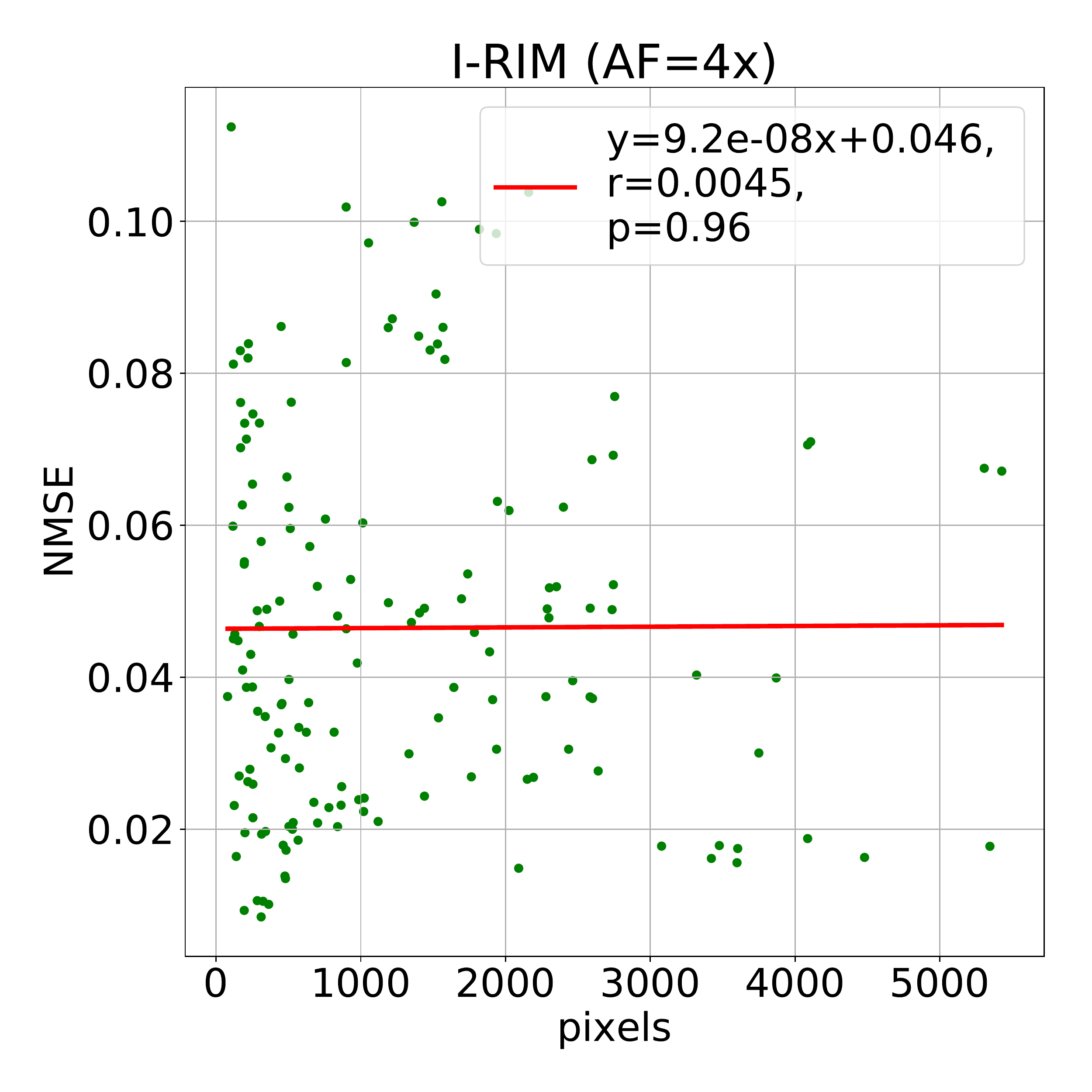} 
 \includegraphics[width=0.30\columnwidth]{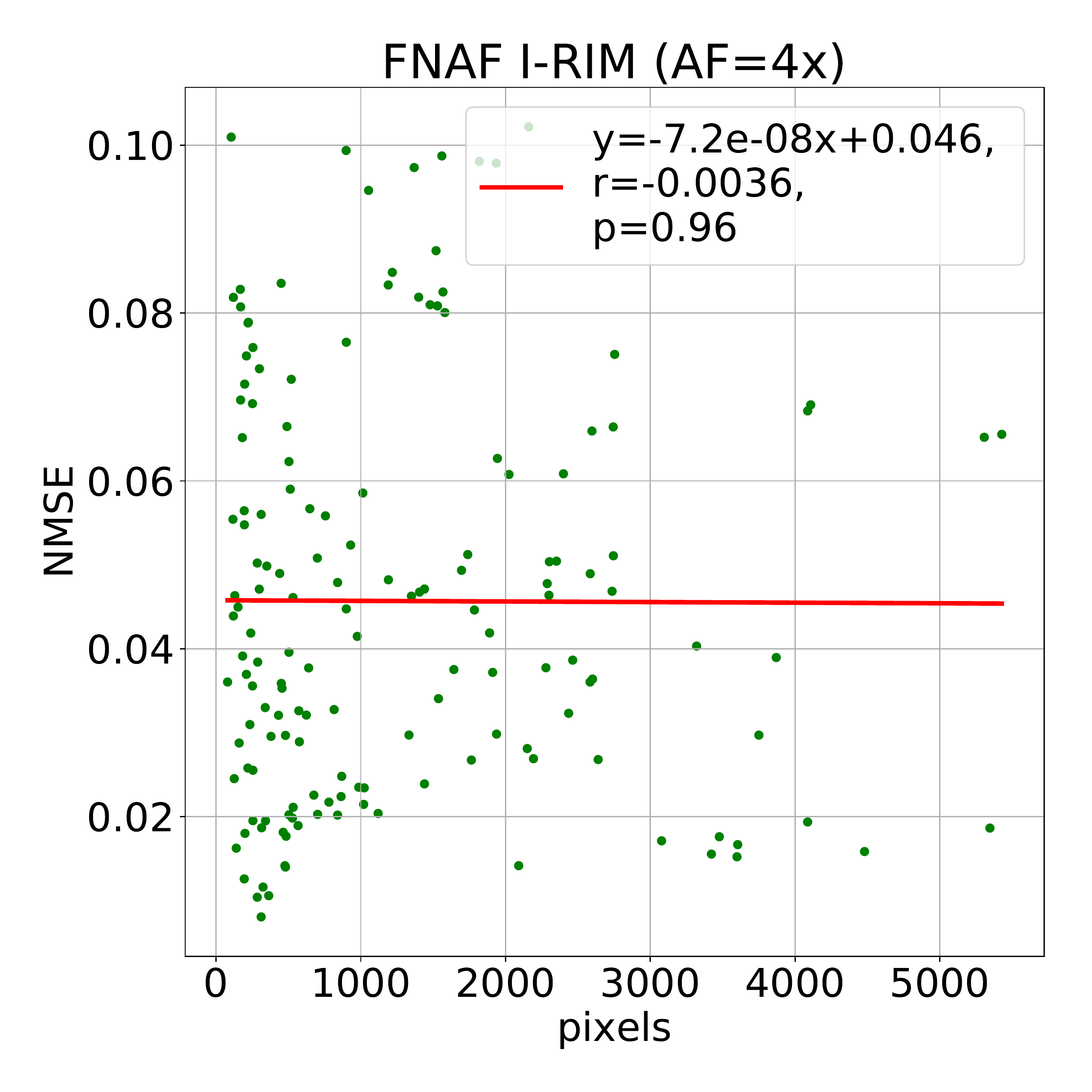}
 \includegraphics[width=0.30\columnwidth]{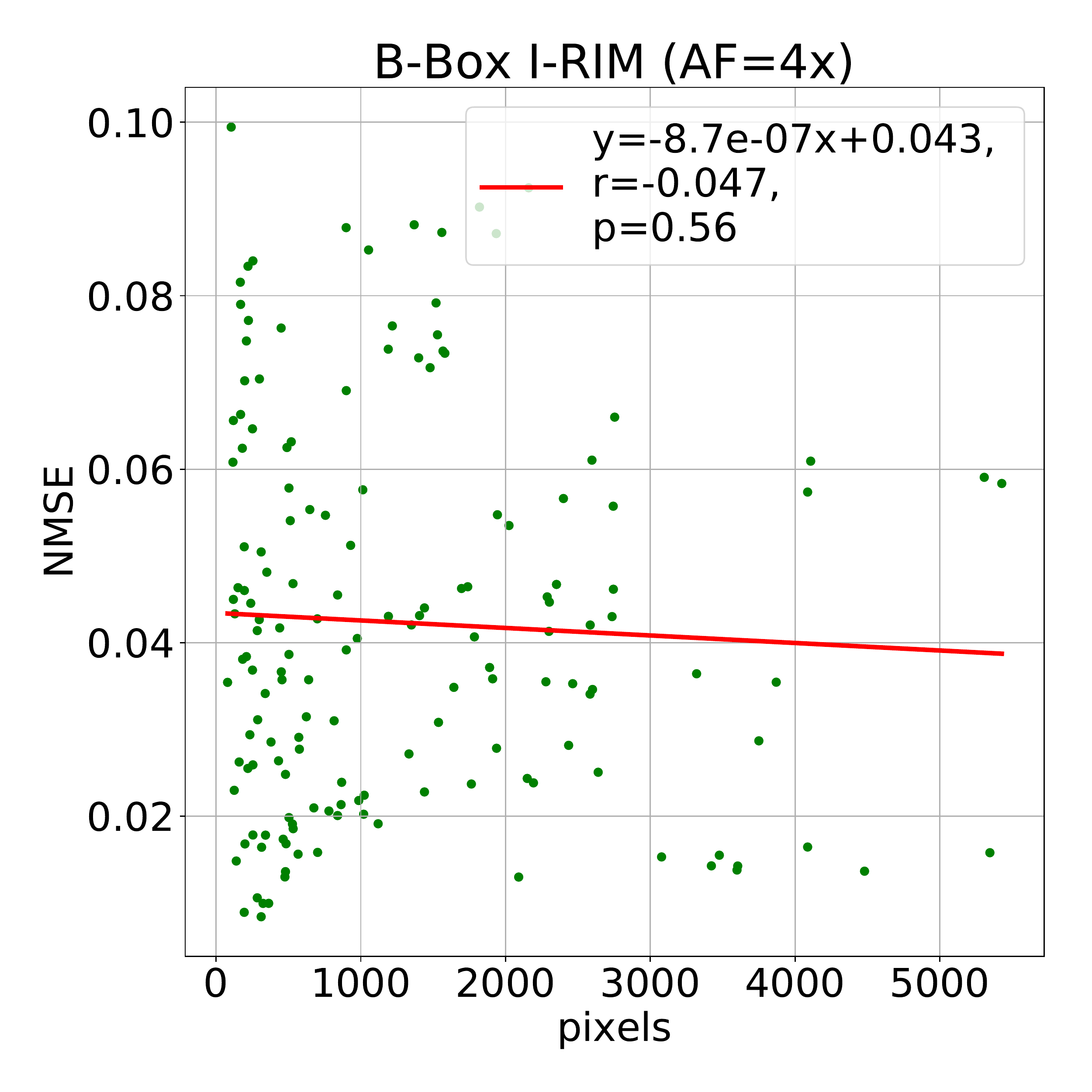} 

 \includegraphics[width=0.30\columnwidth]{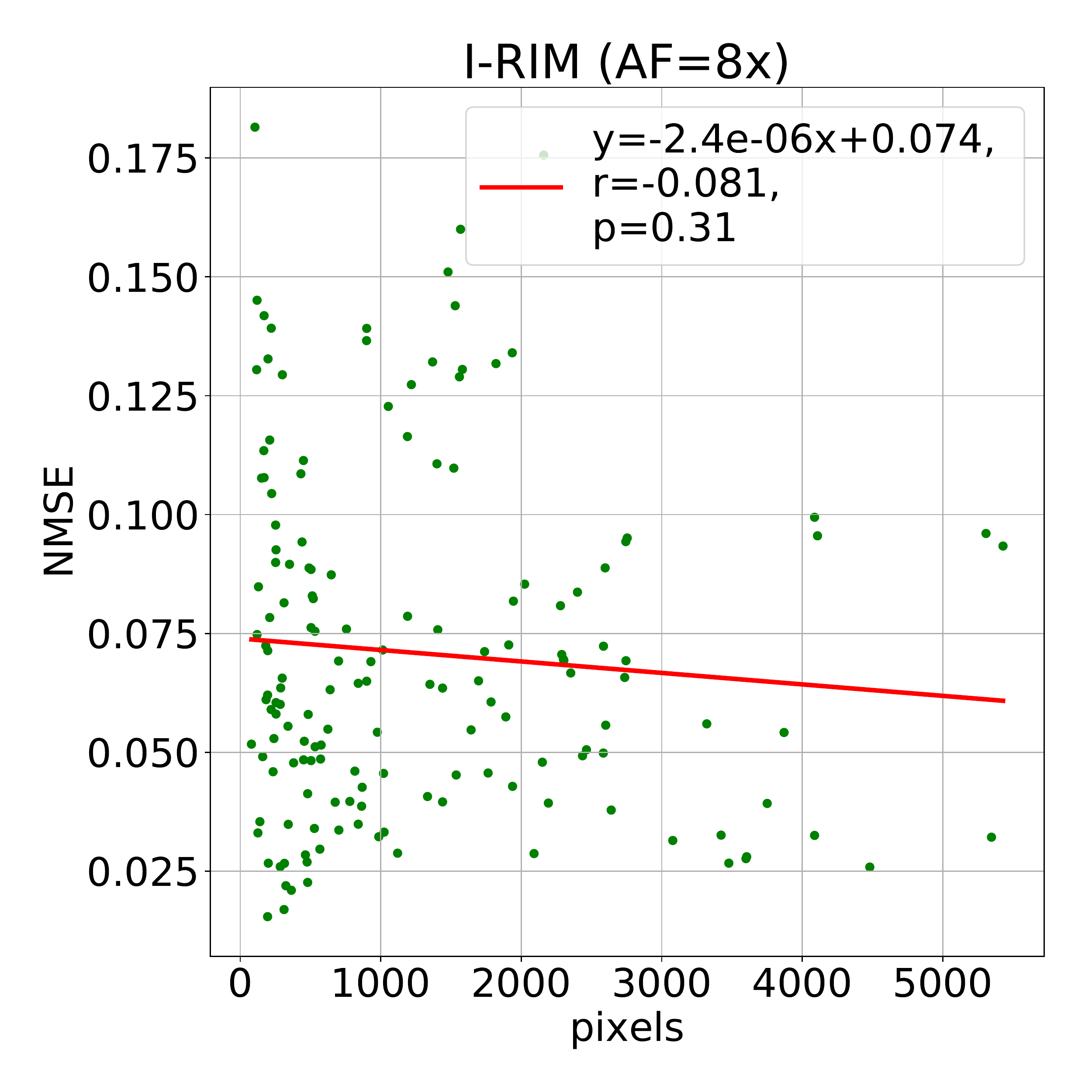} 
 \includegraphics[width=0.30\columnwidth]{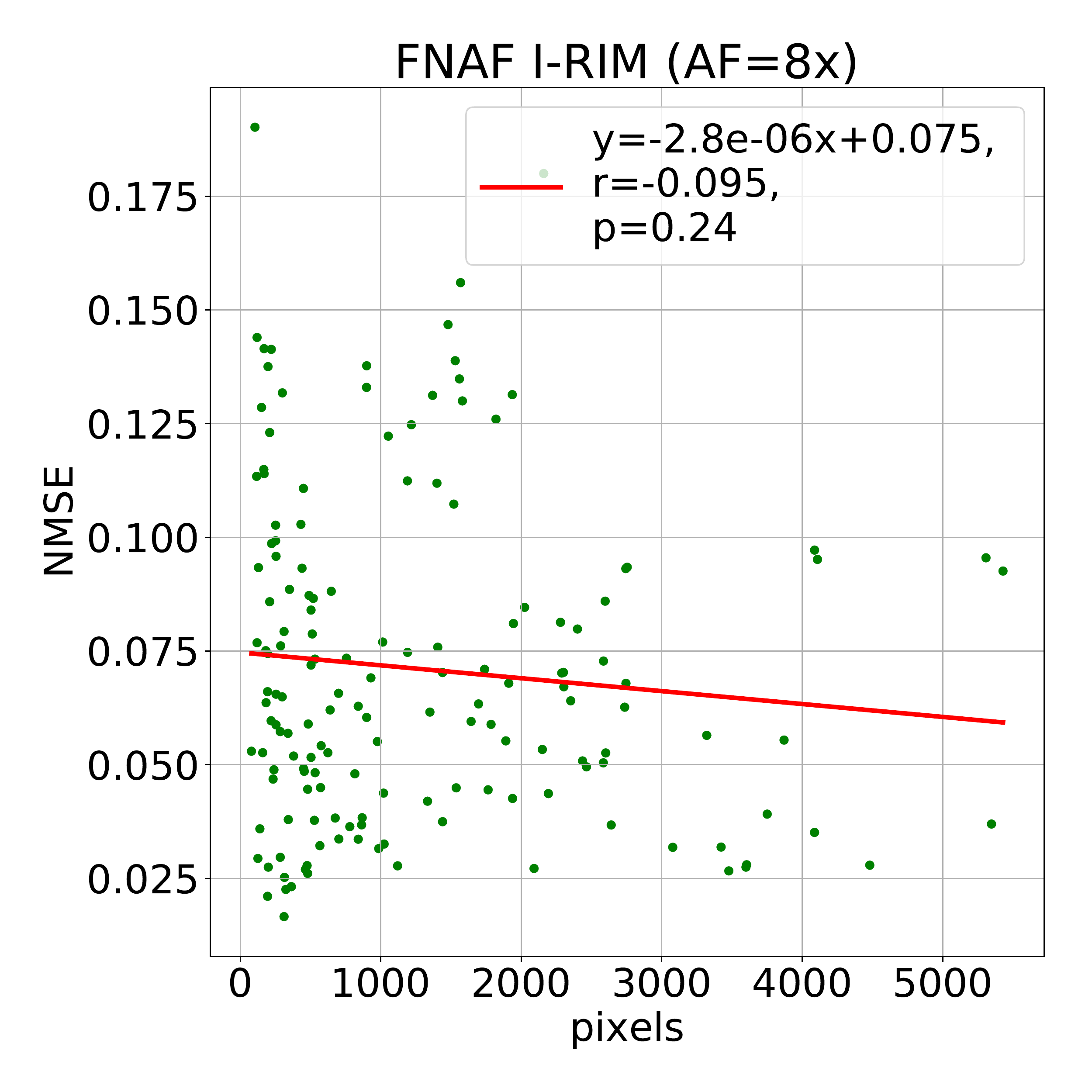} 
 \includegraphics[width=0.30\columnwidth]{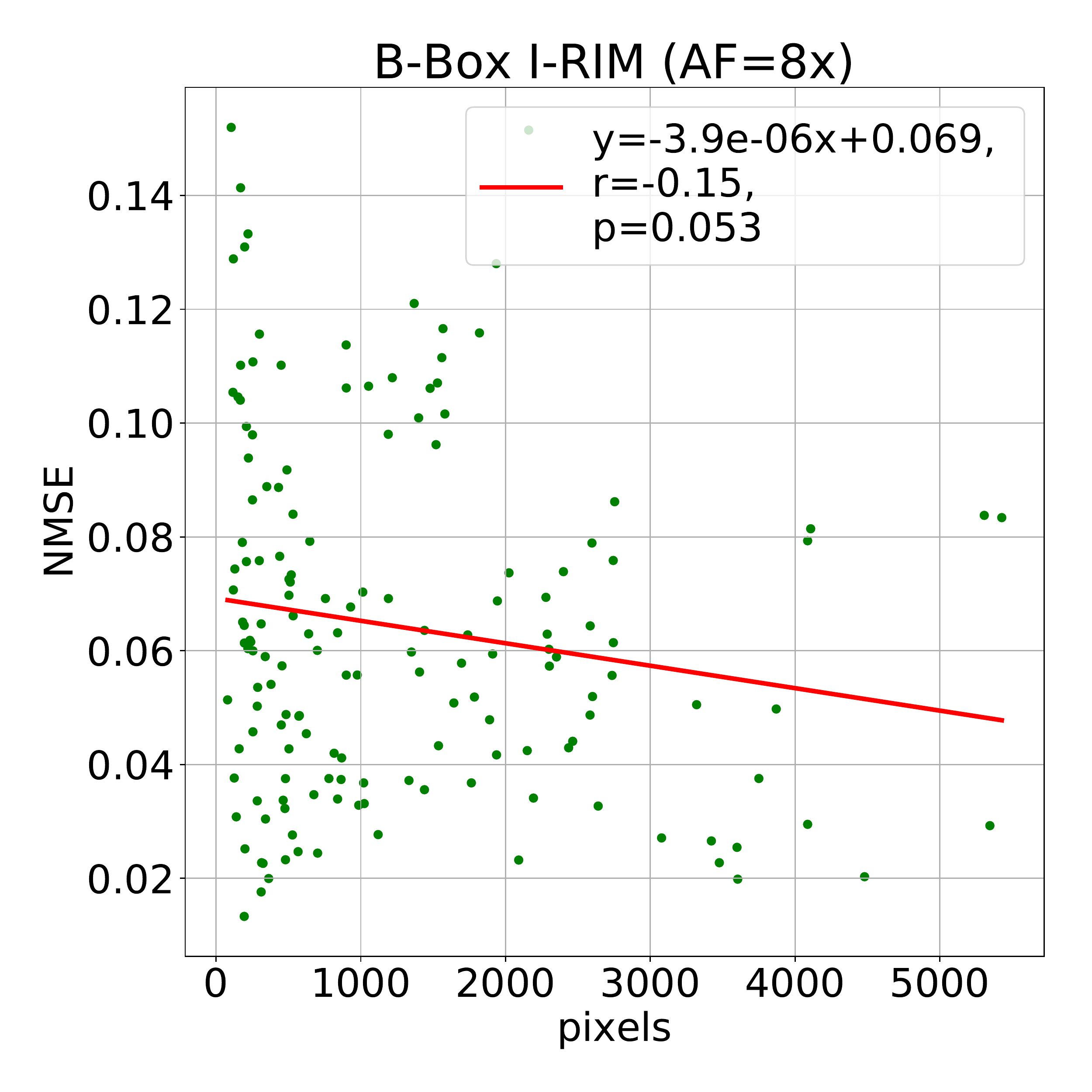} 
 
\caption{Correlation analysis between reconstruction NMSE and the size of abnormalities (measured in terms number of occupied pixels). Row 1 and 2: baseline U-Net and the robust trained versions with FNAF (FNAF U-Net) and real abnormalities (B-Box U-Net). Row 3 and 4: baseline I-RIM and the robust trained versions with FNAF (FNAF I-RIM) and real abnormalities (B-Box I-RIM). First and Third row: 4$\times$ undersampling factor. Second and Last row: 8$\times$ undersampling factor.}
\label{fig:loss_vs_size}
\end{figure}

A correlation analysis was conducted to investigate the effect of the abnormality size on the reconstruction quality. The abnormality size was measured by counting the number of pixels that abnormalities occupy. 
Fig.~\ref{fig:improvements_UNET} shows that for the 8$\times$ acceleration, the improvement of FNAF U-Net over the baseline increases as the size of bounding boxes decreases. This preliminarily validates that constraining FNAF's size during training leads to a better reconstruction of small abnormalities.
Nevertheless, both methods performed comparably in the 4$\times$ setting. 
In Fig.~\ref{fig:improvements_UNET}, the reconstruction performed with I-RIM showed no correlation with the abnormality size, arguably because relaxed constraints were used, when formulating the FNAF for I-RIM. B-BOX robust training showed the opposite correlation, which might be due to the presence of fewer abnormalities in the training set, some of which are larger than FNAF. 
Fig.~\ref{fig:loss_vs_size} shows that the reconstruction quality increases (\textit{i.e.} the loss decreases) as the size of abnormal region increases for U-Net but not I-RIM. This could demonstrate that I-RIM can reconstruct smaller abnormalities better than U-Net. 

\subsection*{Abnormality reconstruction analysis}
\begin{figure}[!tbh]
 \centering
 \includegraphics[width=0.45\columnwidth]{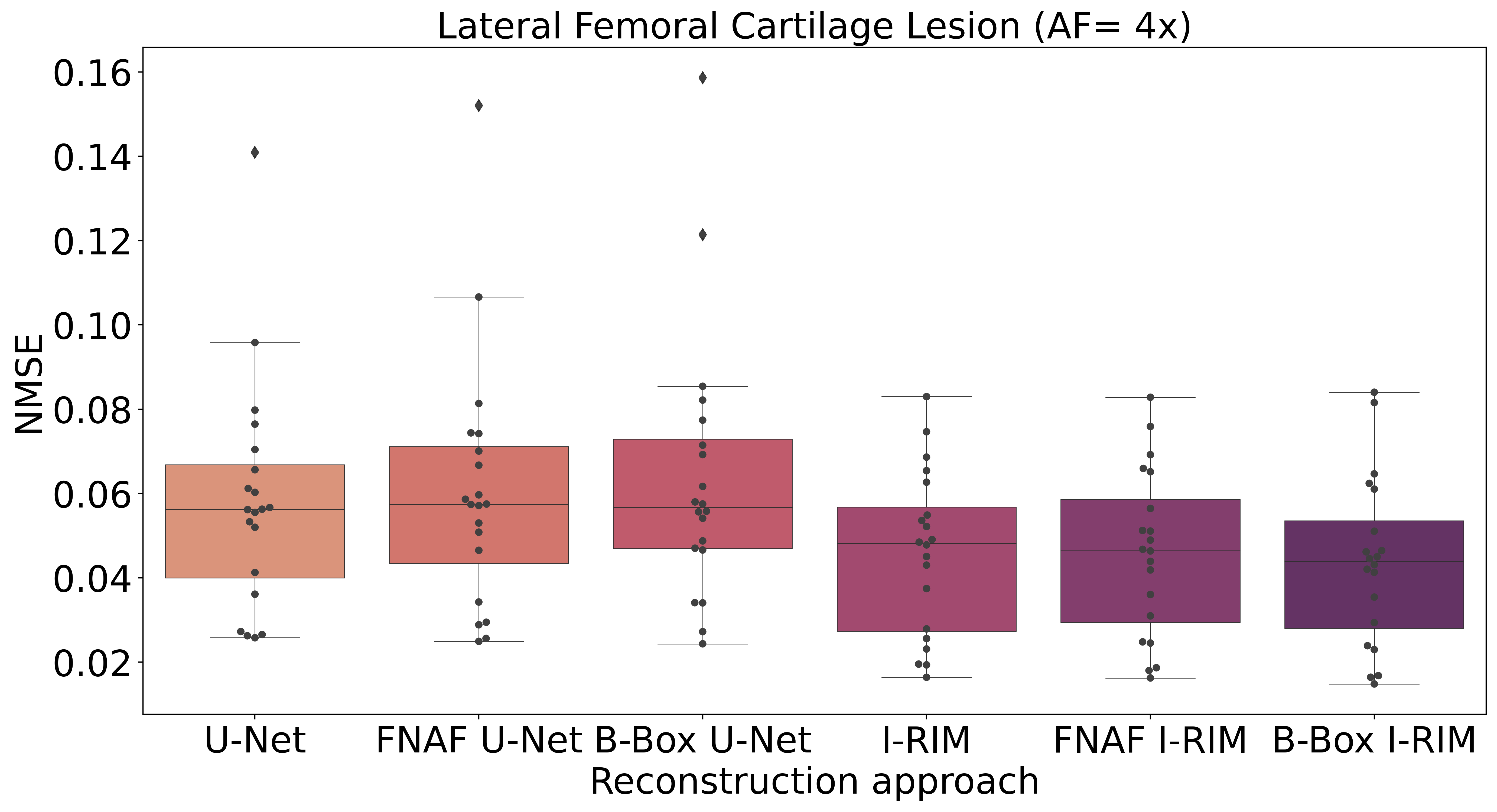}
 \includegraphics[width=0.45\columnwidth]{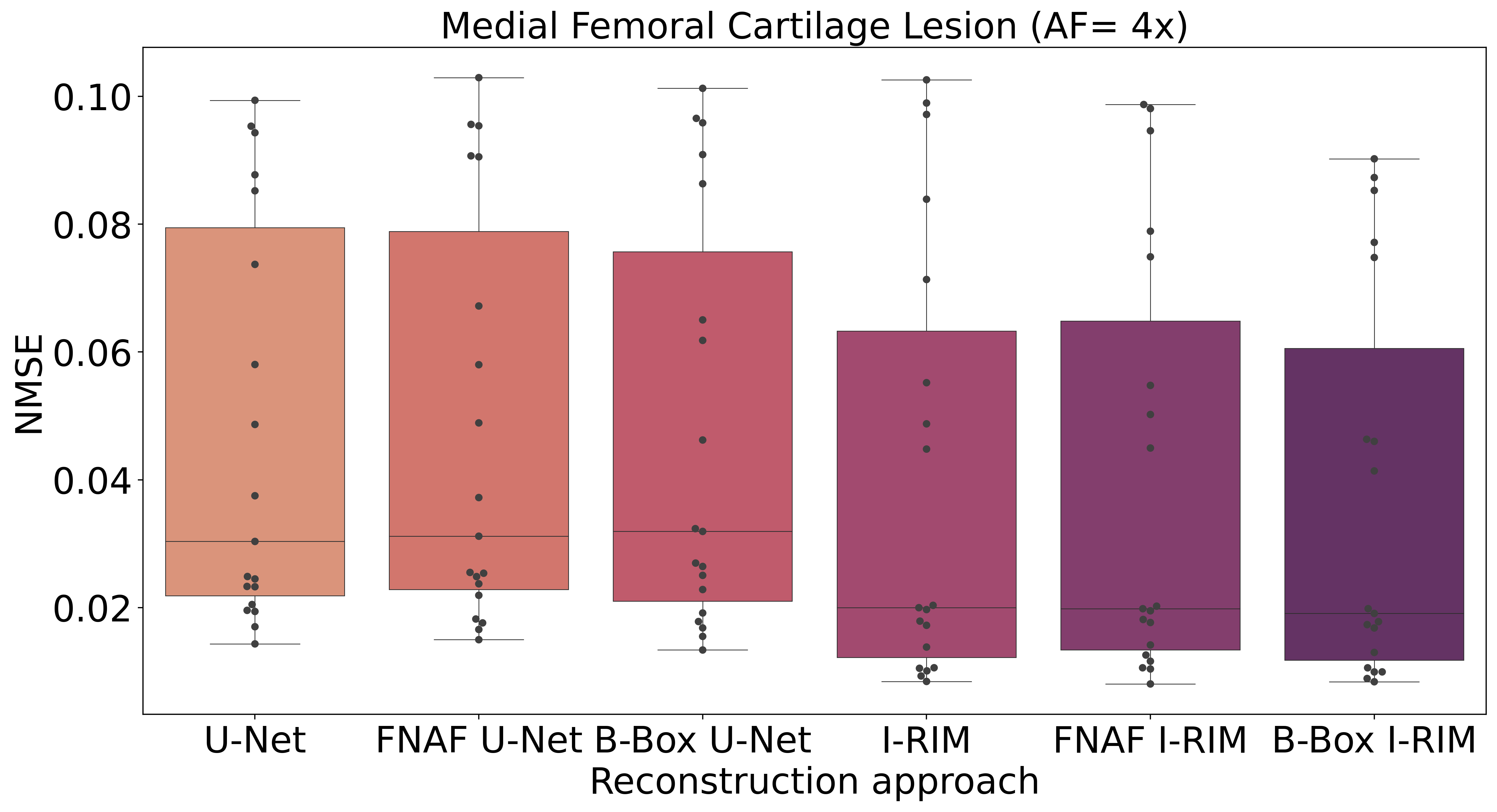}

 \includegraphics[width=0.45\columnwidth]{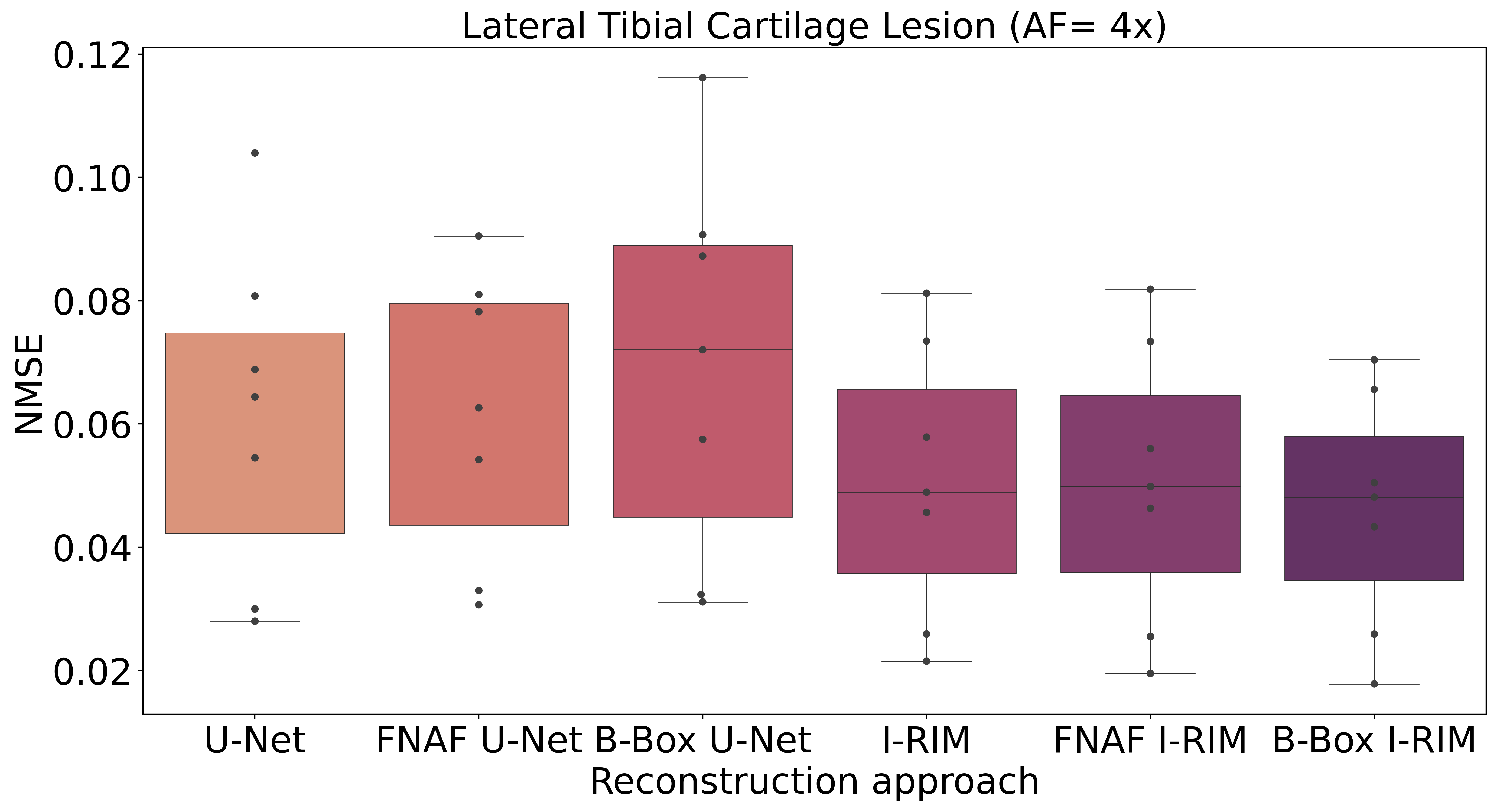}
 \includegraphics[width=0.45\columnwidth]{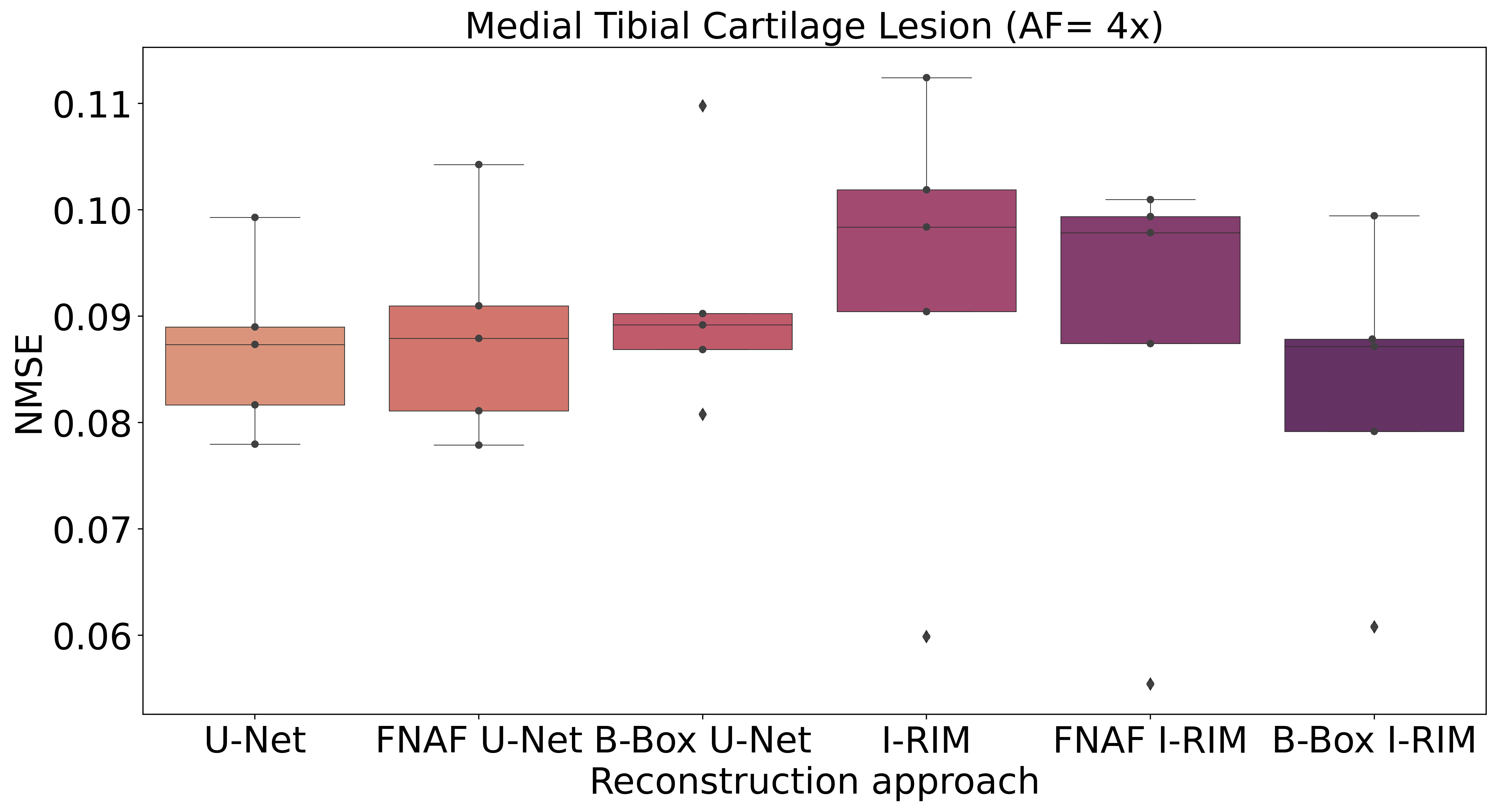}
 
  \includegraphics[width=0.45\columnwidth]{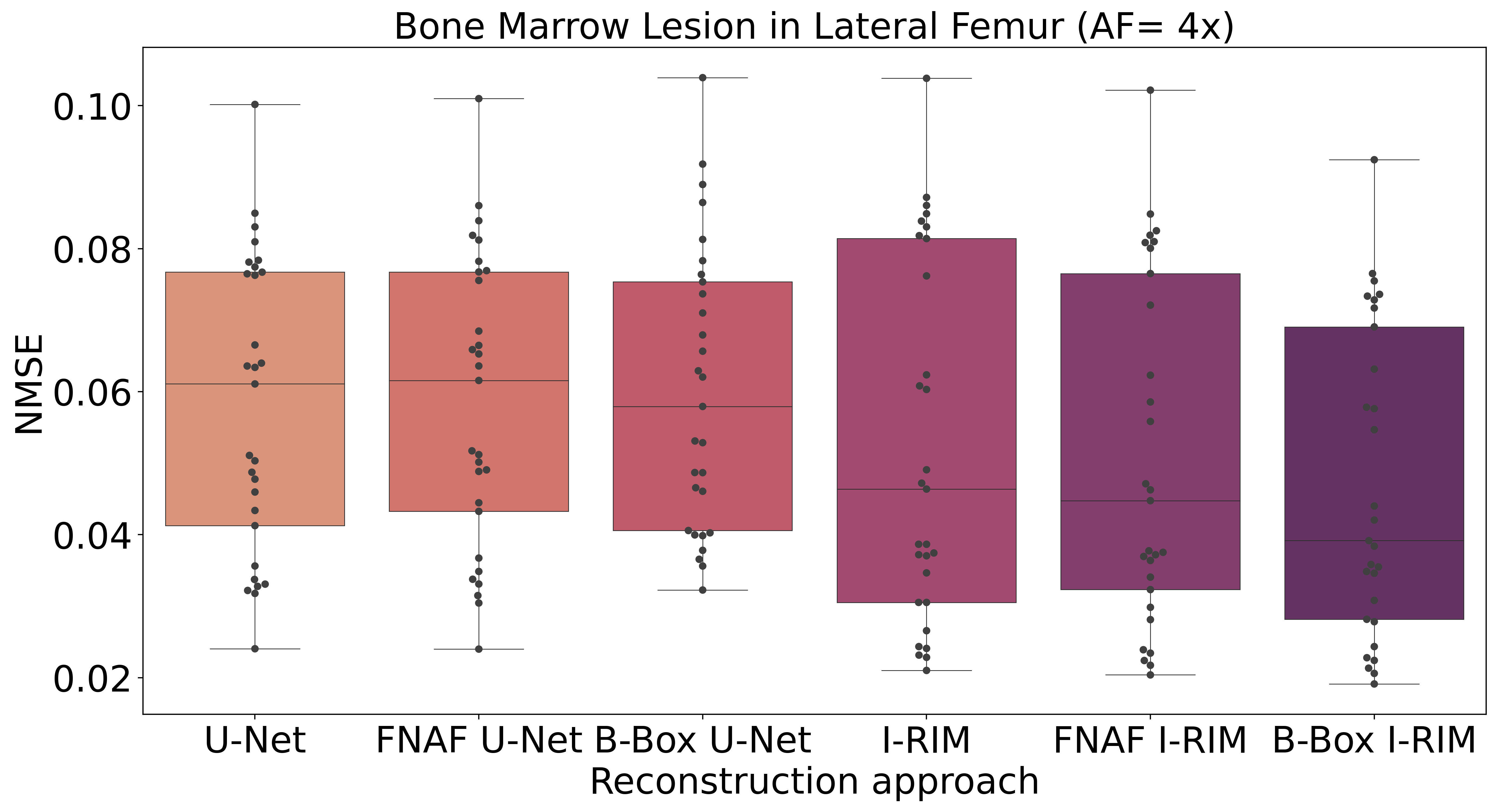}
 \includegraphics[width=0.45\columnwidth]{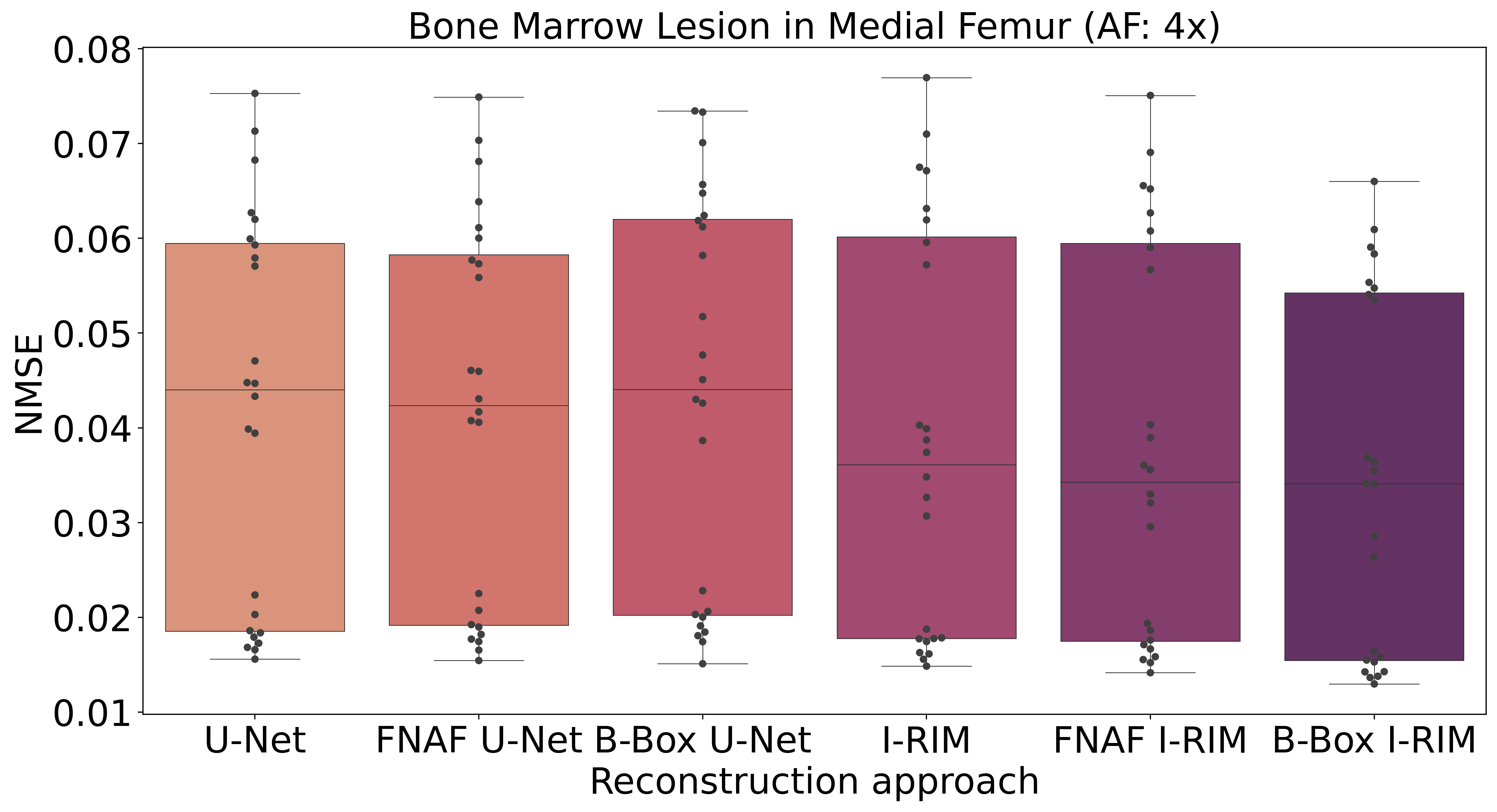}

 \includegraphics[width=0.45\columnwidth]{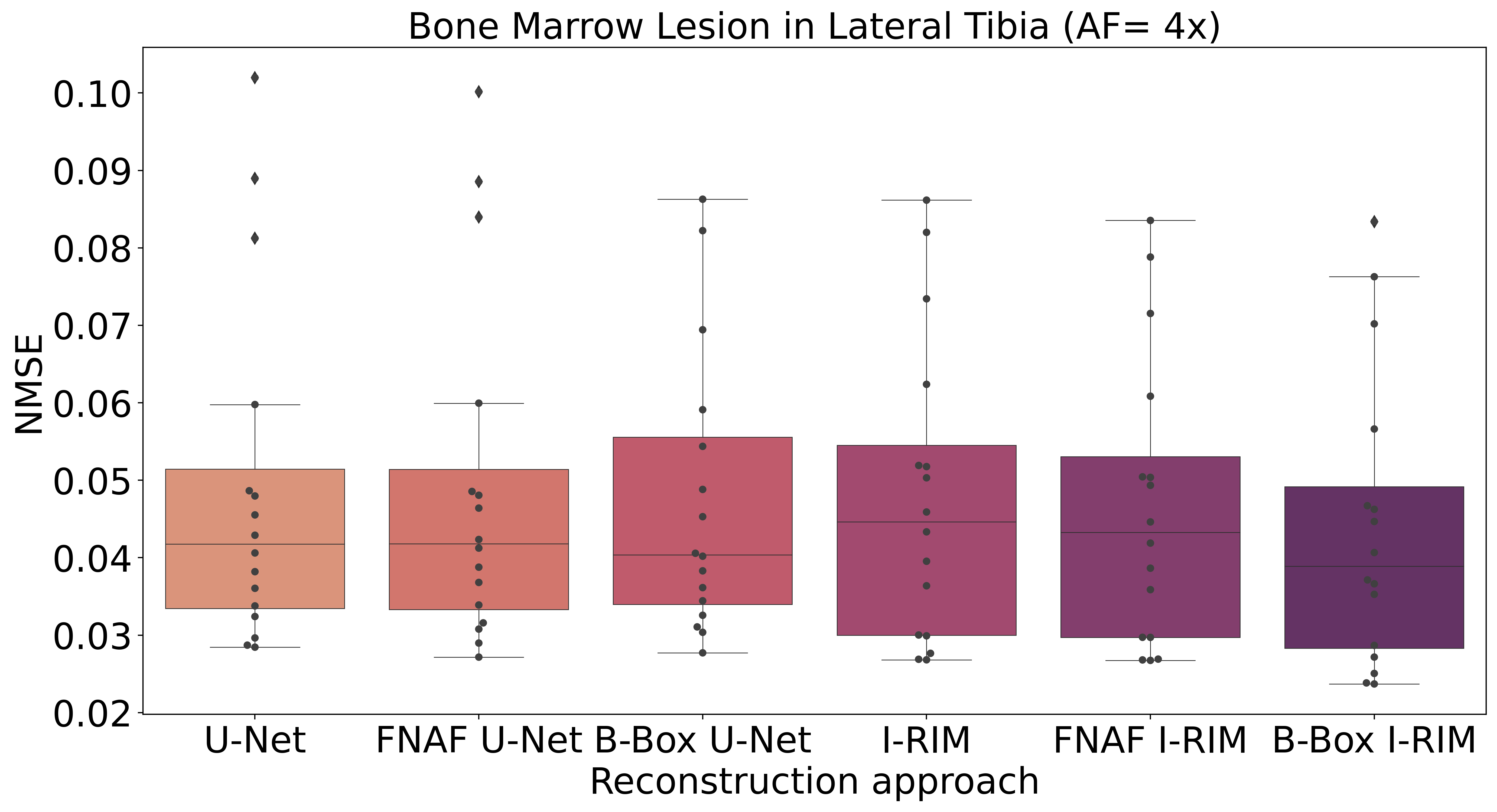}
 \includegraphics[width=0.45\columnwidth]{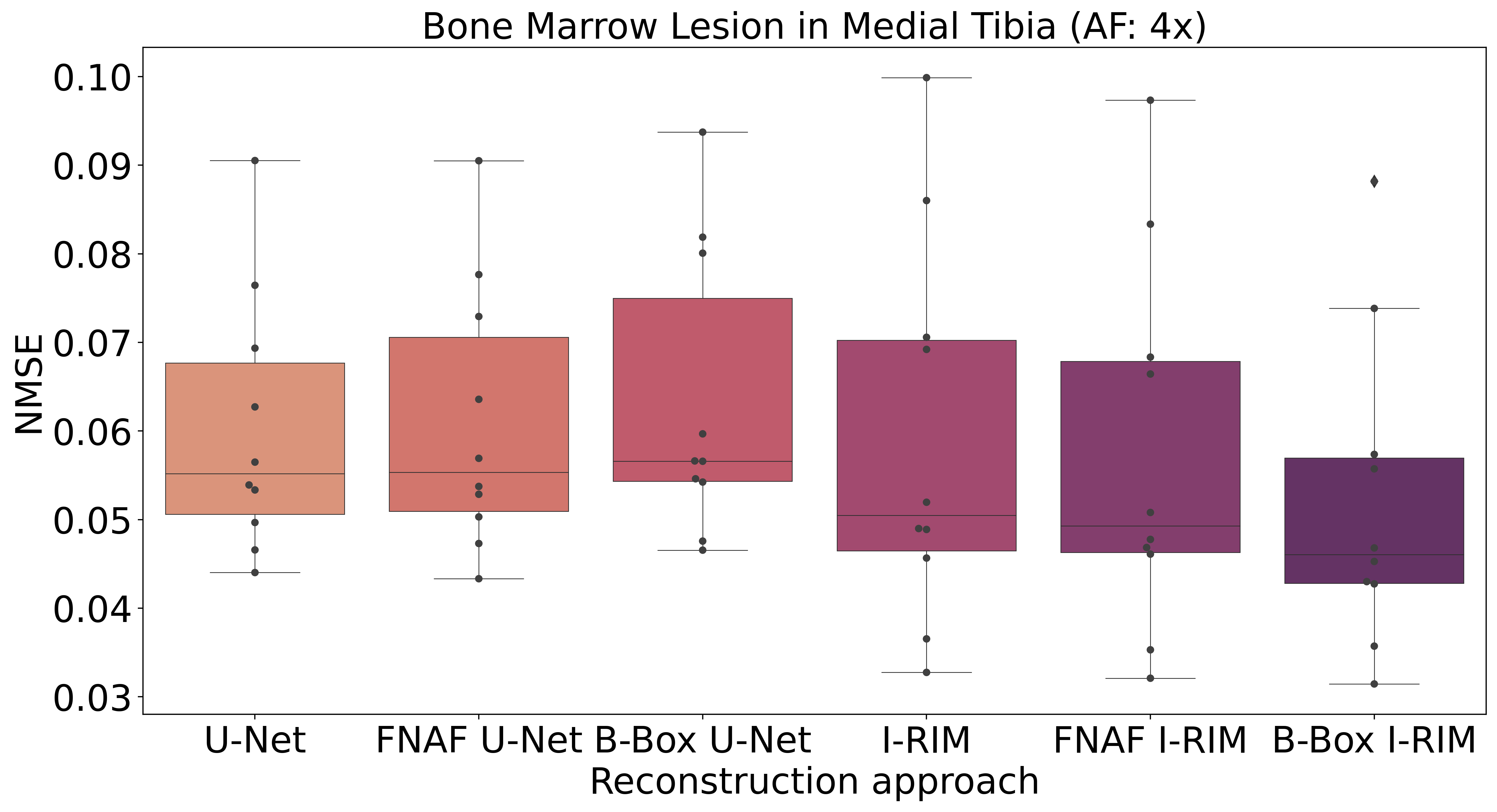}
 
 \includegraphics[width=0.45\columnwidth]{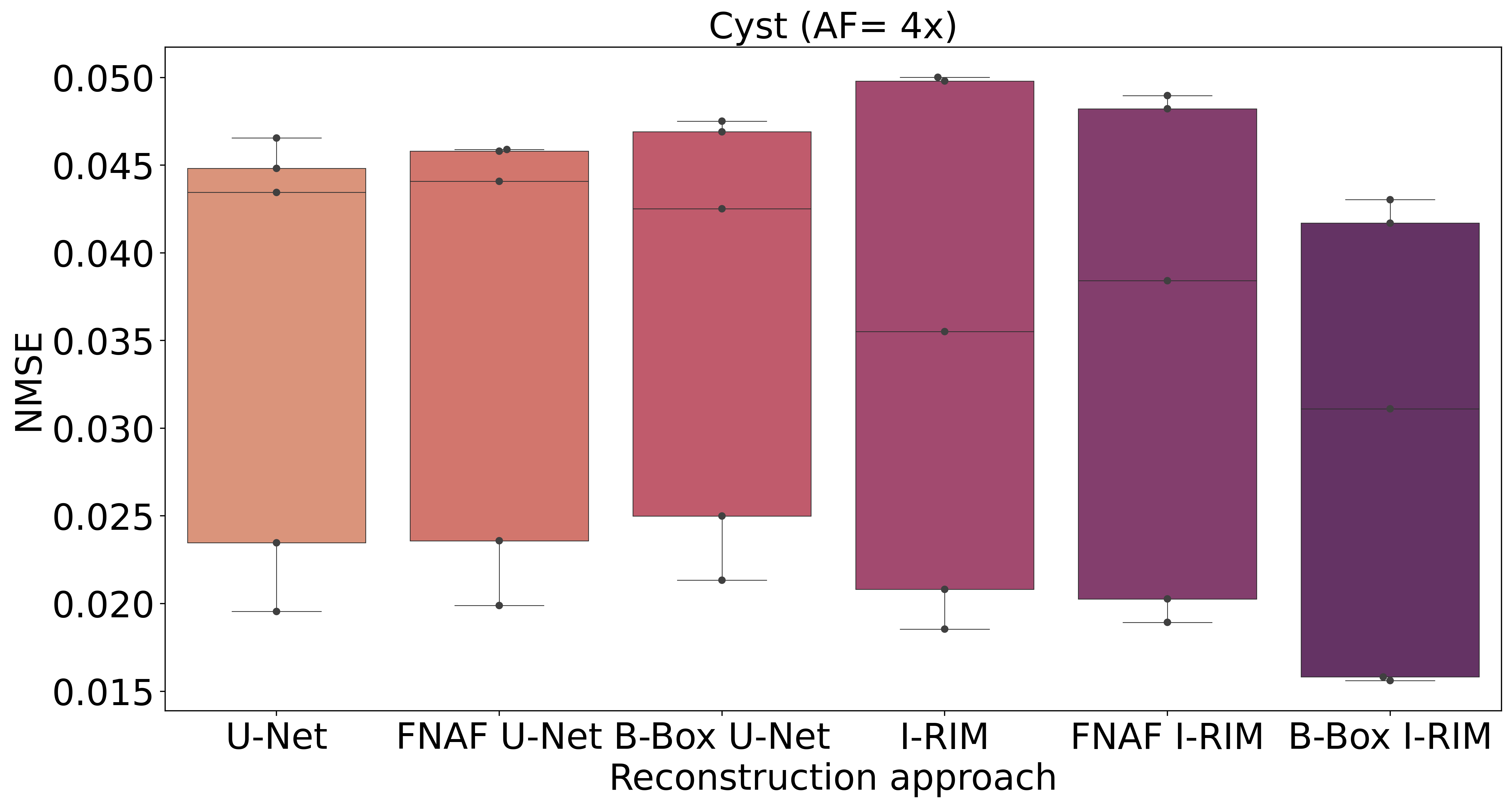}
\caption{Abnormality reconstruction performance summary reported in terms of normalized mean-squared error. The metric was computed within the manually annotated abnormality bounding-box regions. Prior to reconstruction, MRIs were undersampled with a $4\times$ AF factor. }
\label{fig:boxplot-4x}
\end{figure}

\begin{figure}[!tbh]
 \centering
 \includegraphics[width=0.45\columnwidth]{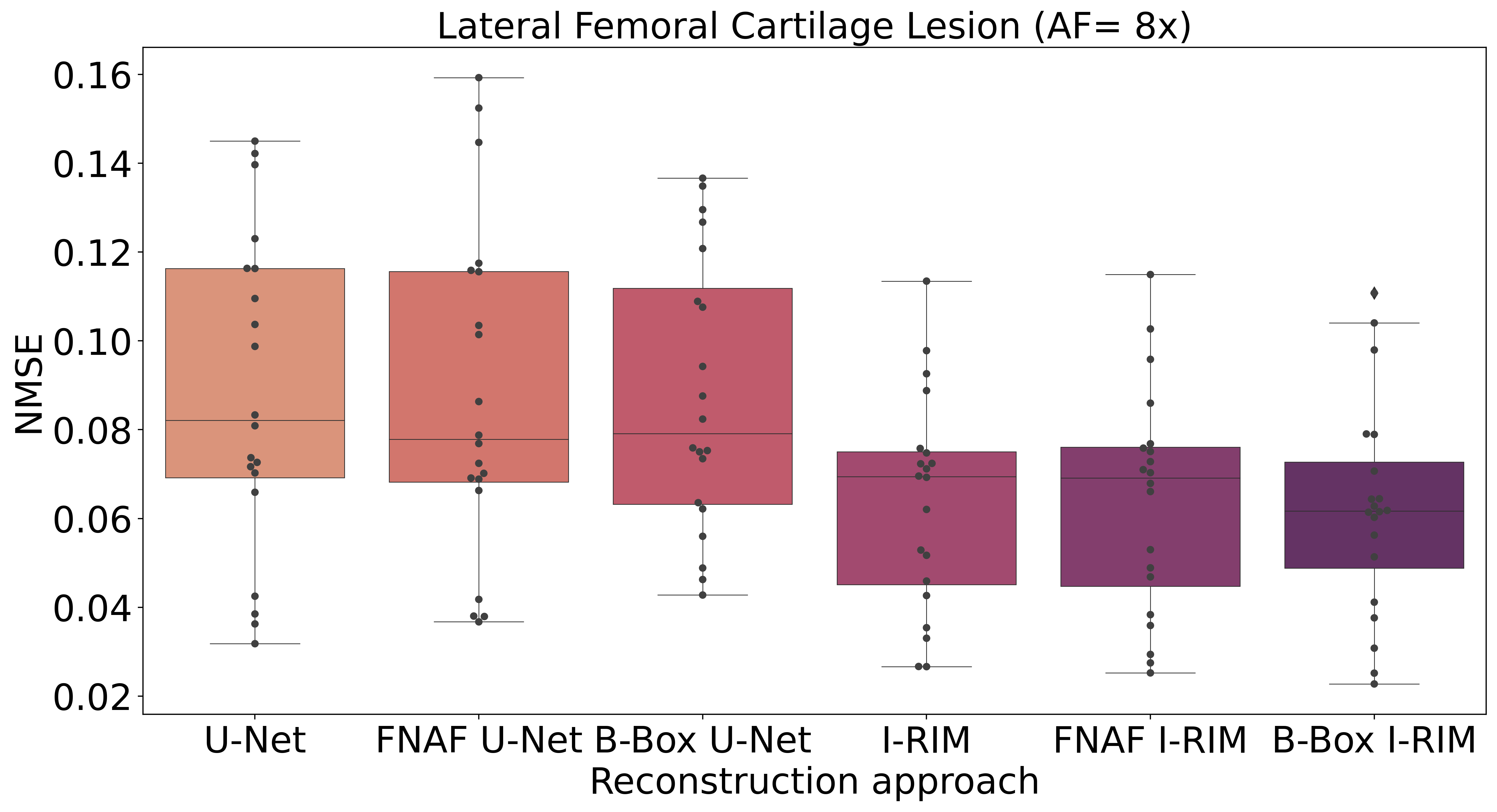}
 \includegraphics[width=0.45\columnwidth]{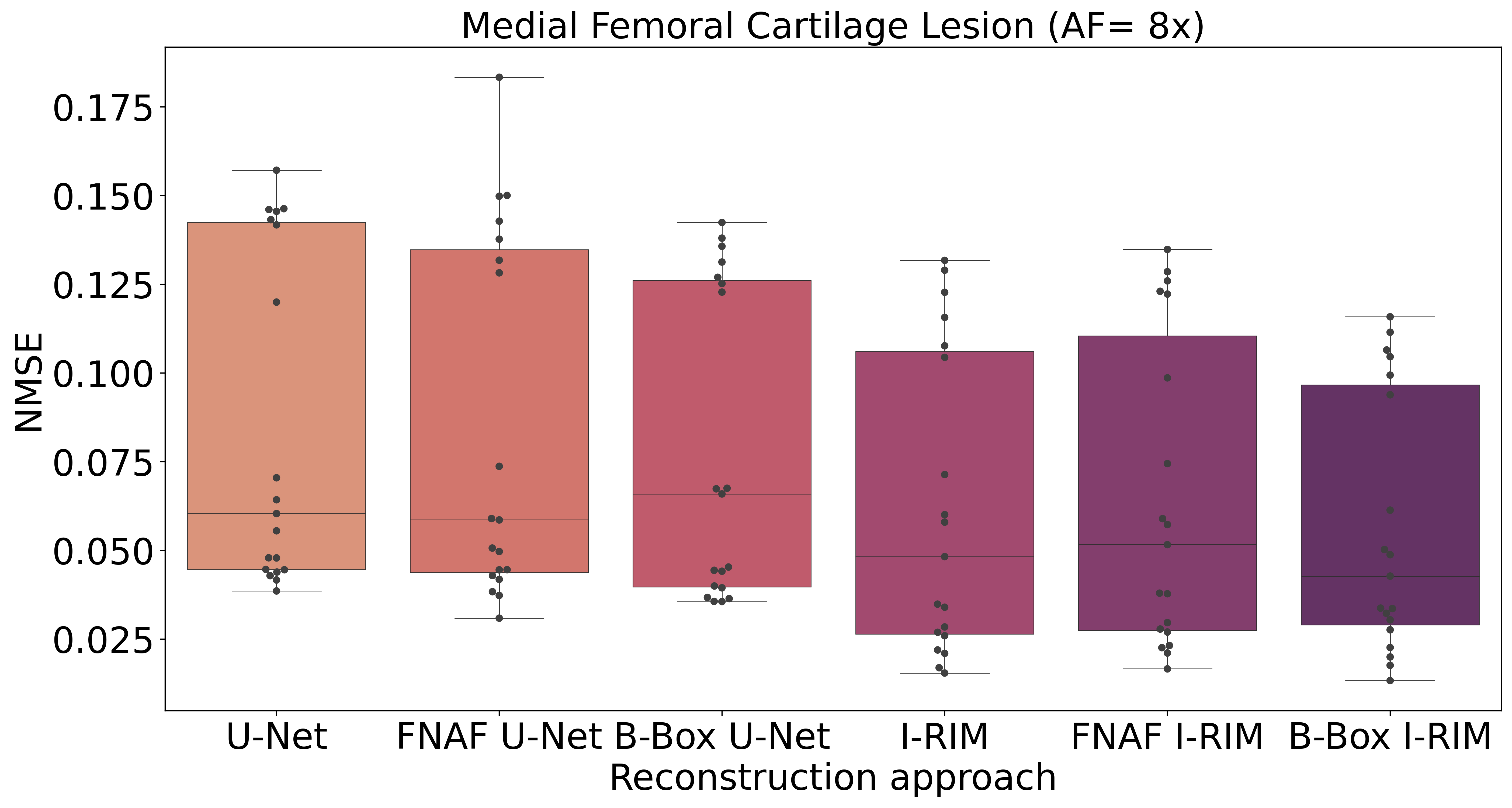}

 \includegraphics[width=0.45\columnwidth]{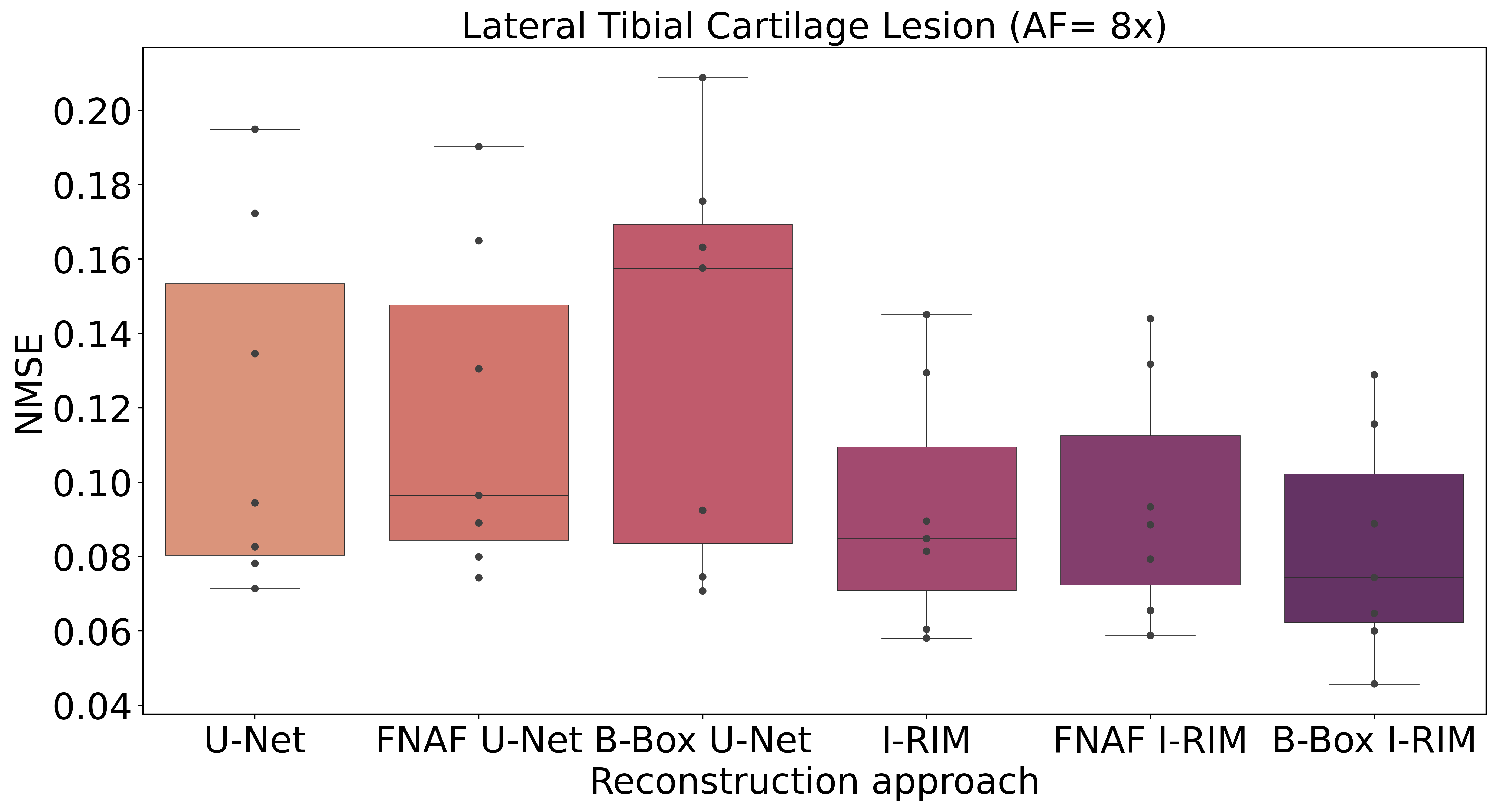}
 \includegraphics[width=0.45\columnwidth]{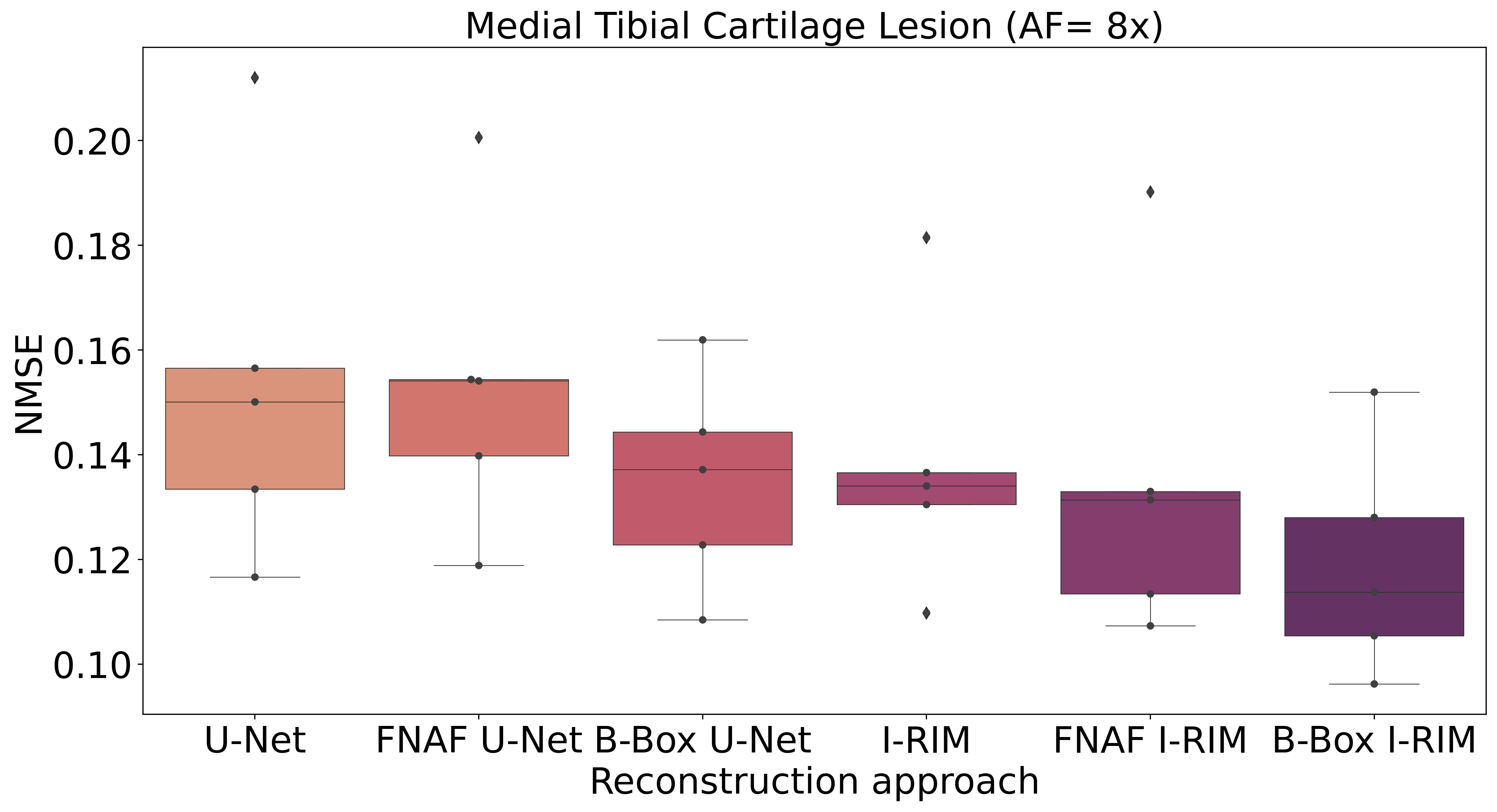}
 
  \includegraphics[width=0.45\columnwidth]{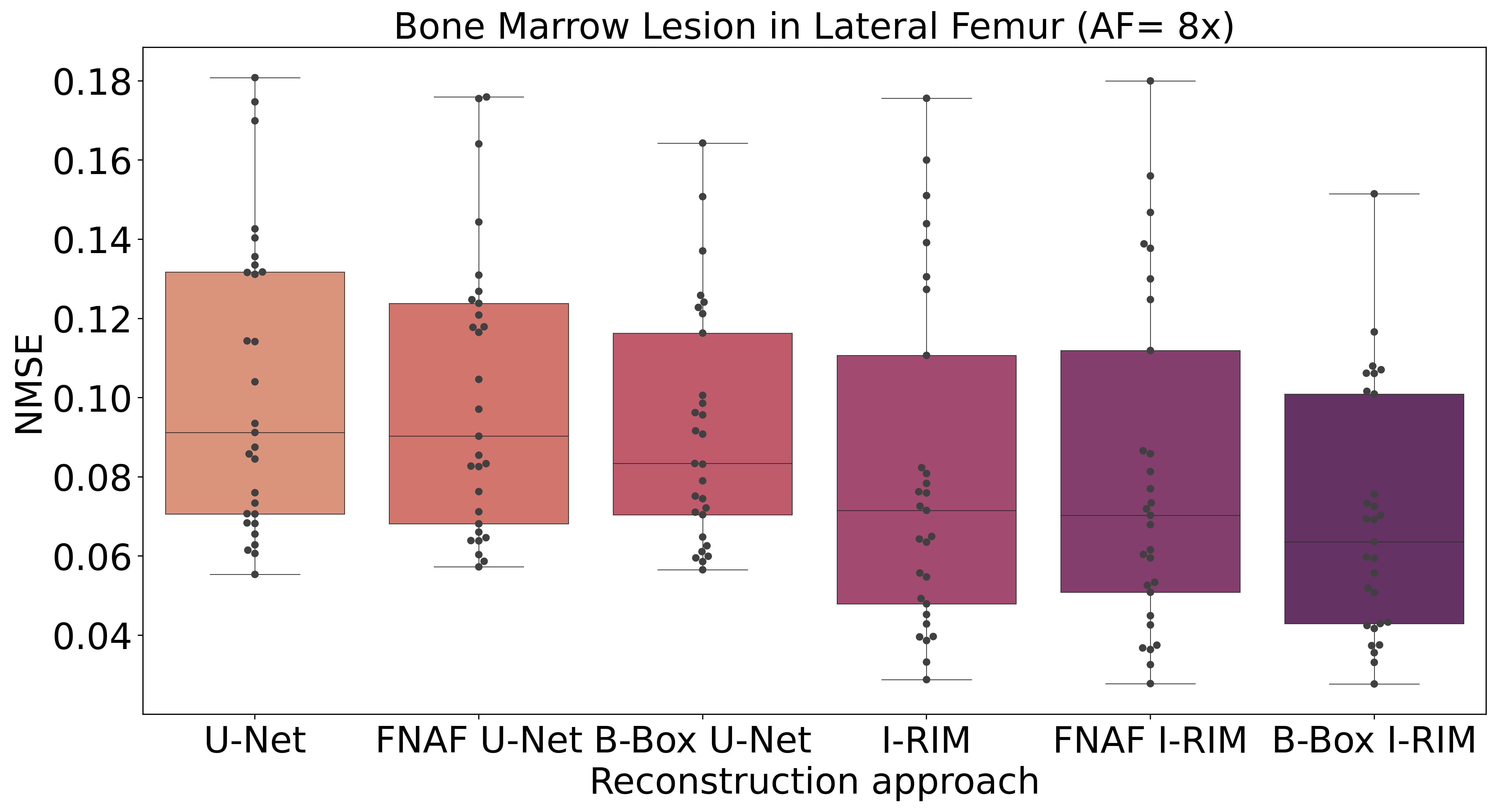}
 \includegraphics[width=0.45\columnwidth]{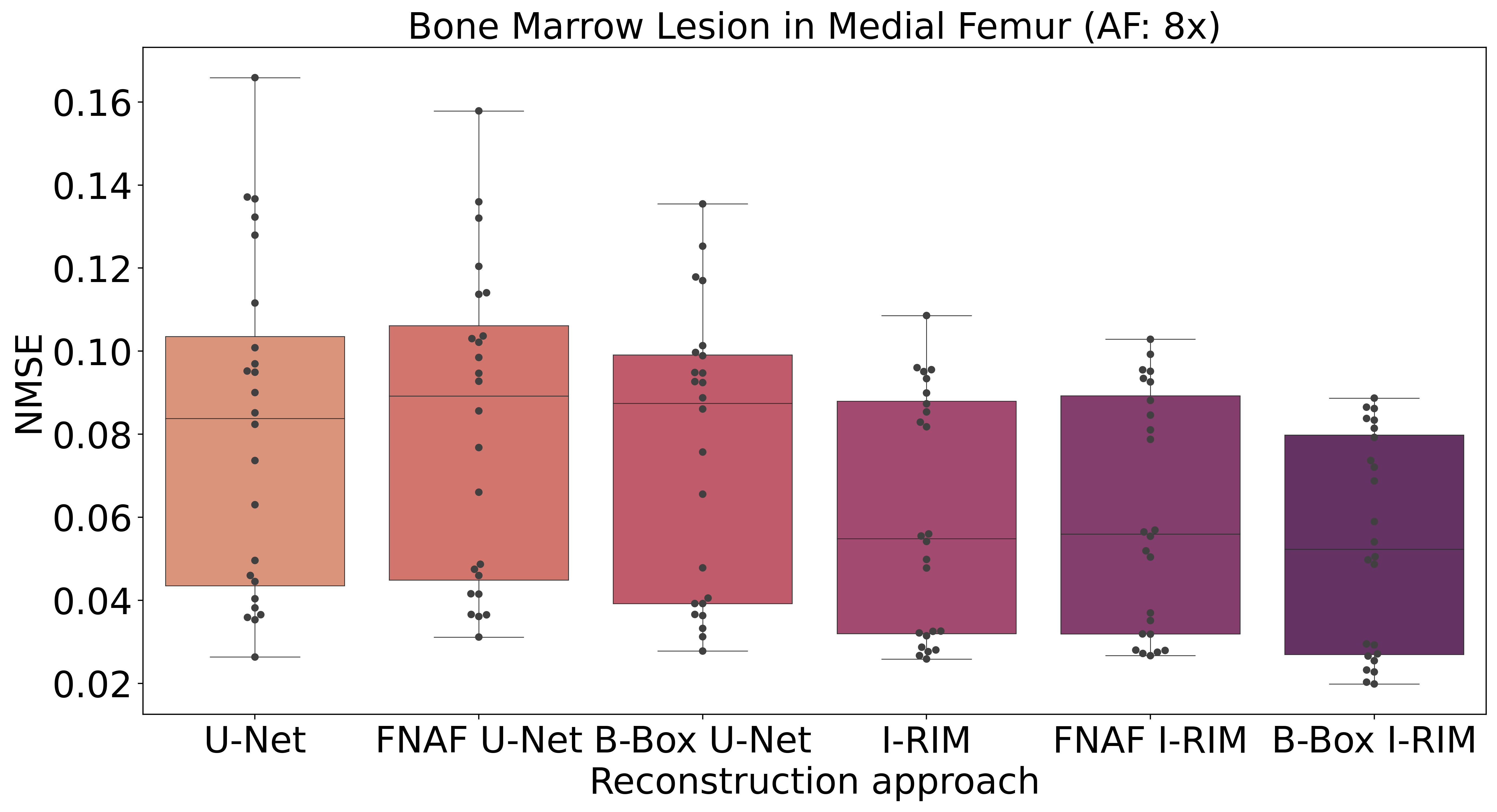}

 \includegraphics[width=0.45\columnwidth]{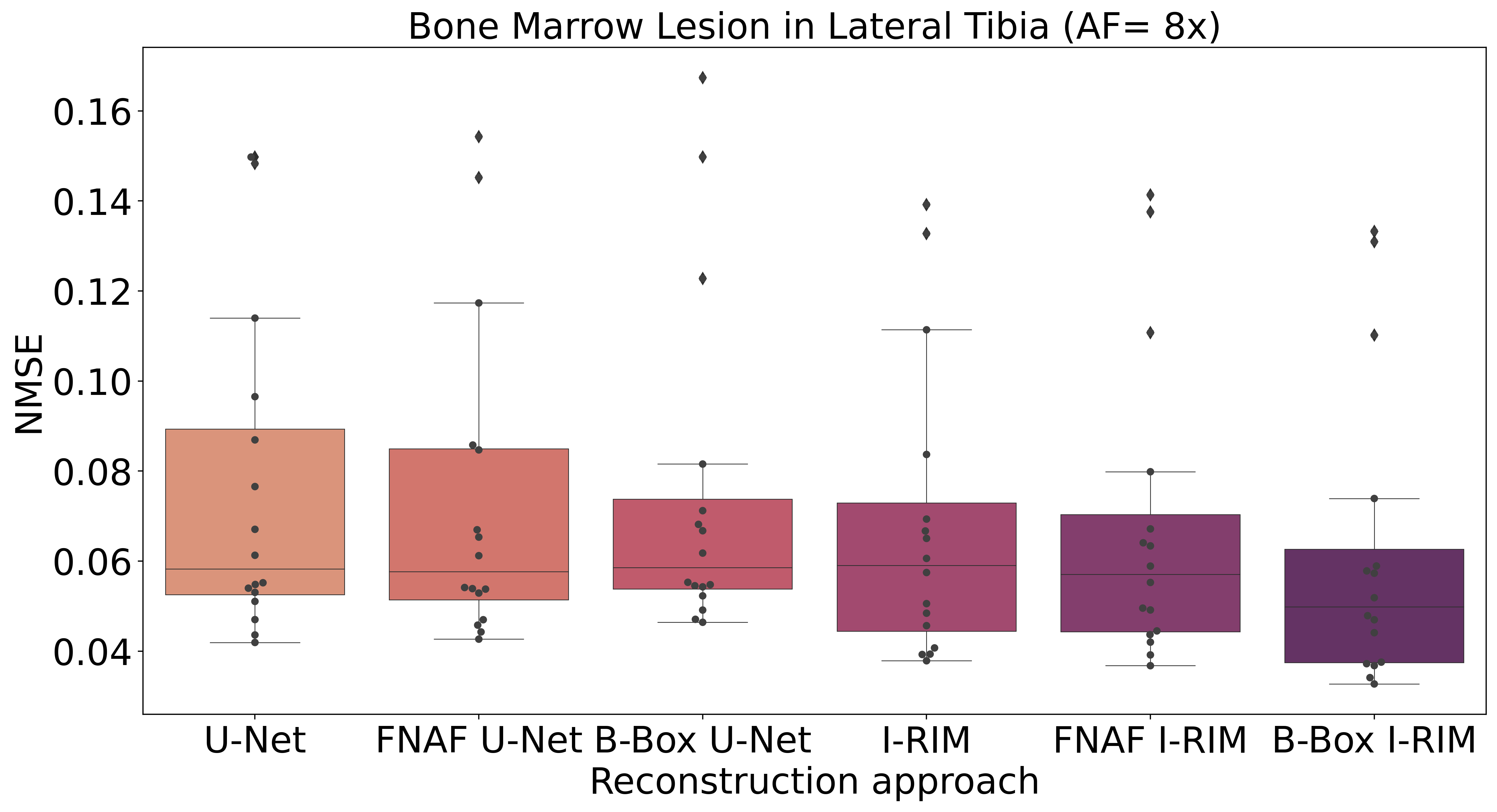}
 \includegraphics[width=0.45\columnwidth]{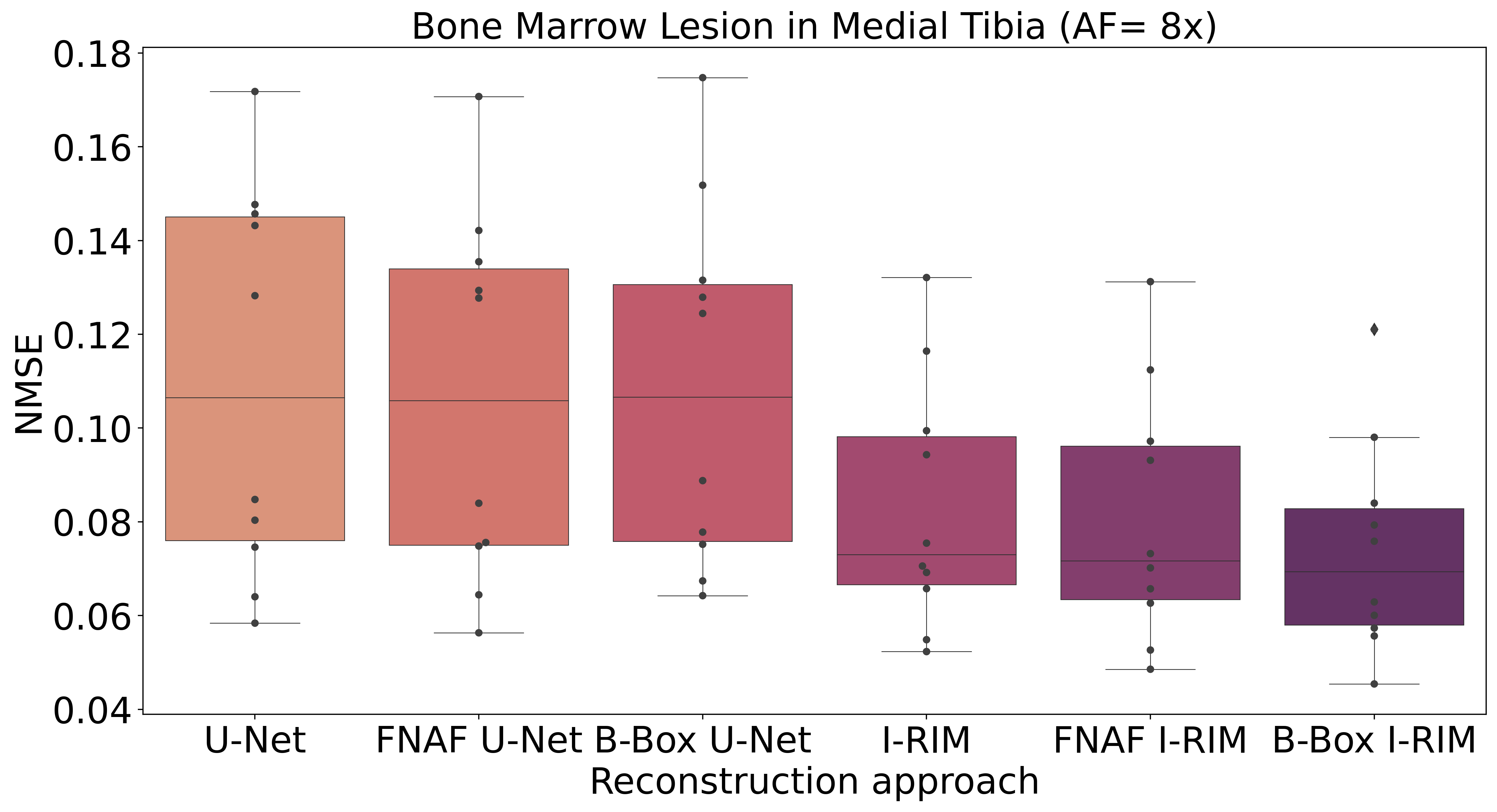}
 
 \includegraphics[width=0.45\columnwidth]{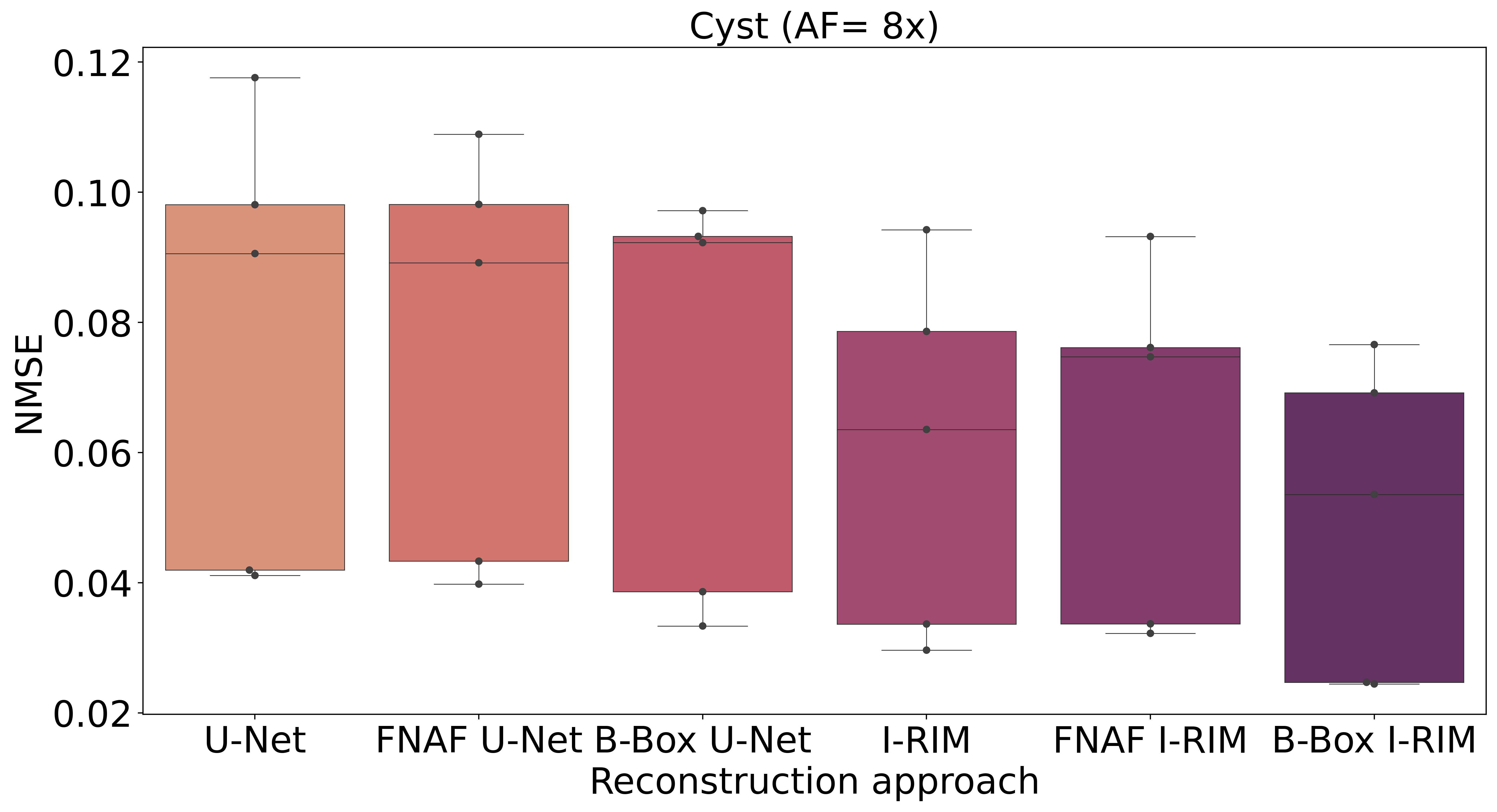}
\caption{Abnormality reconstruction performance summary reported in terms of normalized mean-squared error. The metric was computed within the manually annotated abnormality bounding-box regions. Prior to reconstruction, MRIs were undersampled with a $8\times$ AF factor.}
\label{fig:boxplot-8x}
\end{figure}

The quality of abnormality reconstruction was assessed in terms of normalized mean-squared error. Fig.~\ref{fig:boxplot-4x} depicts box-plots which summarize the reconstruction error in each manually annotated abnormality region. The metrics are reported per abnormality and organized per reconstruction approach. This is complementary to the results reported in Table~\ref{table: real_world_recon}. While in Fig.~\ref{fig:boxplot-4x}, the MRI data was previous undersampled with an 4-fold acceleration factor, in Fig.~\ref{fig:boxplot-8x} the undersampling factor was 8$\times$.

\noindent

\end{document}